%% file: main.tex
\documentclass[pdflatex,sn-mathphys-num]{sn-jnl}

\usepackage{graphicx}%
\usepackage{multirow}%
\usepackage{amsmath,amssymb,amsfonts}%
\usepackage{amsthm}%
\usepackage{mathrsfs}%
\usepackage[title]{appendix}%
\usepackage{xcolor}%
\usepackage{textcomp}%
\usepackage{manyfoot}%
\usepackage{booktabs}%
\usepackage{algorithm}%
\usepackage{algorithmicx}%
\usepackage{algpseudocode}%
\usepackage{listings}%

\usepackage{graphicx}
\usepackage{hyperref}
\usepackage{graphicx}
\usepackage{amsmath}
\usepackage{amsfonts}
\usepackage{amssymb}
\usepackage{booktabs}
\usepackage{comment}
\usepackage{xcolor}
\usepackage{subcaption}
\usepackage{amsmath}
\usepackage{algorithm}
\usepackage{algpseudocode}
\usepackage{breqn}
\usepackage{bm}

\theoremstyle{thmstyleone}%
\theoremstyle{thmstyletwo}%

\theoremstyle{thmstylethree}%

\begin{document}

\title[Article Title]{Domain-Decomposition Neural Surrogates for Scalable Decentralized Ensemble Kalman Filter Based Parameter Identification in High-Dimensional Stochastic PDEs}


\author*[1]{\fnm{Timm} \sur{G\"{o}dde}}\email{t.godde@utwente.nl}

\author[1,2]{\fnm{Bojana} \sur{Rosi\'{c}}}\email{bojana.rosic@tuwien.ac.at}


\affil*[1]{\orgdiv{Applied mechanics and data analysis (AMDA), MS3 Department}, \orgname{University of Twente}, \orgaddress{\street{Drienerlolaan 5}, \city{Enschede}, \postcode{7522NB}, \country{Netherlands}}}

\affil[2]{\orgdiv{Digital engineering group (DiGEN), Institute of Engineering Design and Product Development}, \orgname{Technische Universität Wien}, \orgaddress{\street{Lehargasse 6}, \city{Wien}, \postcode{1030}, \country{Austria}}}


\abstract{Ensemble Kalman filters (EnKF) provide an efficient framework for parameter identification of physics based laws from spatially distributed measurements. However, their forecast models often require a large number of samples to accurately represent apriori uncertainties, leading to high computational costs. To reduce this computational burden, a neural network (NN)-based surrogate model is introduced to replace the sample-based forecast model.
The proposed NN surrogate maps spatial coordinates and physics-based parameters to the forecasted observation. However, such maps typically require a large number of parameters for high-dimensional spatial domains. To overcome this limitation, a domain decomposition method (DDM) based on augmented Lagrange multipliers is developed, where local NN models are optimized independently before global communication and coupling. This approach reduces the number of NN parameters while improving local approximation accuracy and computational efficiency. Furthermore, a distributed and decentralized ensemble Kalman filter approach  based on DDM-NN surrogate model is investigated, where the parameter identification problem is decomposed into local subproblems. Each local estimator updates the material parameters using locally available information, while communication between neighboring subdomains enables the reconstruction of a consistent global estimate. This approach reduces the computational burden of the inversion while maintaining accurate uncertainty quantification of the identified parameters.
The proposed method is evaluated on a three-dimensional material parameter identification problem and compared with an ensemble Kalman filter (EnKF) and a Markov Chain Monte Carlo (MCMC) reference solution. The results show that the proposed DDM NN-based KF captures the posterior parameter distribution and approaches the solutions obtained with both EnKF and MCMC. While MCMC provides the most accurate representation of the posterior distribution, the proposed approach achieves comparable parameter estimates with reduced computational requirements for the forecast model.}




\maketitle

\input{formulas}

\input{Section/Introduction}
\input{Section/Problem}
\input{Section/Methodology}
\input{Section/Results}
\input{Section/Conclusion}
\input{Section/Acknowledgements}

\bibliography{sn-bibliography}

\end{document}

%% file: formulas.tex
\newcommand{\loss}{\mathcal{J}}
\newcommand{\Lagr}{\mathcal{L}}
\newcommand{\gebiet}{\mathcal{G}}
\newcommand{\ixa}{i}
\newcommand{\ixb}{j}
\newcommand{\ixc}{k}
\newcommand{\normal}{\eta}
\newcommand{\Q}{\boldsymbol{\mathcal{Q}}}
\newcommand{\wm}{\ws_{\ixa\ixb}}
\newcommand{\ls}{\boldsymbol{\lambda}}
\newcommand{\bound}{\varGamma_{\ixa\ixb}}
\newcommand{\point}{x}
\newcommand{\fm}{\fp_{\ws_{\ixa\ixb}}^{\textrm{int}}}
\newcommand{\li}{\boldsymbol{\lambda}_\ixa}
\newcommand{\Constr}{\boldsymbol{\mathcal{Q}}}
\newcommand{\Cwi}{\Constr(\wi)}
\newcommand{\id}{\vek{1}}
\newcommand{\real}{\mathbb{R}}
\newcommand{\fvar}{\boldsymbol u}
\newcommand{\Fvar}{A}
\newcommand{\fp}{\hat{\fvar}}
\newcommand{\fvi}{\fvar_{\ixa}}
\newcommand{\fwi}{\fp_{\wi}}
\newcommand{\fwj}{\fp_{\wj}}
\newcommand{\ws}{\boldsymbol{\theta}}
\newcommand{\wi}{\ws_{\ixa}}
\newcommand{\wii}{\ws_{\ixa}^{k}}
\newcommand{\wj}{\ws_{\ixb}}
\newcommand{\wij}{\ws_{\ixa\ixb}}
\newcommand{\w}{W}
\newcommand{\be}{b}
\newcommand{\xk}{\boldsymbol \point_\ixc}
\newcommand{\xoi}{\boldsymbol \point_\ixa}
\newcommand{\beq}{\begin{equation}}
\newcommand{\eeq}{\end{equation}}
\newcommand{\nat}{\mathbb{N}}
\newcommand{\pen}{\rho}
\newcommand{\dir}{\upsilon}
\newcommand{\bulk}{\kappa}
\newcommand{\neum}{\tau_n(x)}
\newcommand{\diri}{u_d(x)}
\newcommand{\probsp}{(S)}
\newcommand{\shear}{\mu}
\newcommand{\strain}{\epsilon_s}
\newcommand{\noise}{\varepsilon}
\newcommand{\paramsinit}{q_f}
\newcommand{\params}{\paramsinit}
\newcommand{\cmeas}{y_m} 
\newcommand{\fvarany}{u}
\newcommand{\expect}{\mathbb{E}}
\newcommand{\spatialvar}{x}
\newcommand{\stochasticvar}{q(\omega)}
\newcommand{\vars}{(\spatialvar, \stochasticvar)}
\newcommand{\probdist}{p}
\newcommand{\measf}{\varphi}
\newcommand{\paramskf}{\zeta}
\newcommand{\cov}{\Sigma}
\newcommand{\kalman}{K}
\newcommand{\n}{n}
\newcommand{\iter}{b}
\newcommand{\ncons}{N_{Q_i}}
\newcommand{\oq}{\omega}
\newcommand{\on}{\omega}
\newcommand{\Oq}{\varOmega}
\newcommand{\On}{\varOmega}

\newcommand{\y}{y}
\newcommand{\yf}{\y_\fc}
\newcommand{\ydisc}{\boldsymbol y}
\newcommand{\measo}{Y}
\newcommand{\mospace}{\mathcal{Y}}
\newcommand{\ymeas}{\y_m}
\newcommand{\phyop}{\mathcal{A}}
\newcommand{\extf}{f}
\newcommand{\fc}{f}
\newcommand{\ass}{a}
\newcommand{\idpara}{q}
\newcommand{\ev}{\omega}
\newcommand{\evpara}{\ev_\idpara}
\newcommand{\thism}{m}
\newcommand{\evmod}{\ev_\thism}
\newcommand{\disc}{p}
\newcommand{\Ndisc}{N_p}
\newcommand{\evdisc}{\ev_\disc}
\newcommand{\dims}{d}
\newcommand{\moderr}{\varepsilon}
\newcommand{\measerr}{\epsilon}
\newcommand{\fullerr}{\varepsilon_\thism}
\newcommand{\statef}{u}
\newcommand{\statefc}{\statef_\fc}
\newcommand{\paramsfc}{\idpara_\fc}
\newcommand{\paramsass}{\idpara_\ass}
\newcommand{\linmat}{H}
\newcommand{\linpara}{h}
\newcommand{\Hspace}{\mathcal{U}}

\newcommand{\linesearch}{\alpha}
\newcommand{\lsiter}{\linesearch^{(\iter)}}
\newcommand{\diter}{\dir^{(\iter)}}
\newcommand{\stepv}{s}
\newcommand{\graddiff}{r}
\newcommand{\siter}{\bold\stepv^{(\iter)}}
\newcommand{\giter}{\bold\graddiff^{(\iter)}}
\newcommand{\iden}{\mathbf{I}}
\newcommand{\invhes}{H}

%% file: Section/Introduction.tex
\section{Introduction}
One of the most common engineering challenges is to estimate parameters of a physics-based model given experimental measurements. In material science, this corresponds to the identification of material parameters entering the constitutive law, which is often interpreted as the construction of a so-called material passport. Assuming that the constitutive law is given in the form of a parametric family, the estimation problem consists of determining parameter values such that the discrepancy between the model response (typically obtained via finite element simulations) and the measured experimental data is minimized. This discrepancy is usually measured in a mean-squared sense leading to least-squares problem \cite{NoceWrig06}. Using the gradient of this error functional, the resulting optimization problem can be solved using iterative first-order or quasi-Newton methods, such as the limited-memory Broyden--Fletcher--Goldfarb--Shanno (L-BFGS) algorithm \cite{NoceWrig06}.

When formulated in a deterministic setting, the estimation problem may become ill-posed in the sense of Hadamard’s principles. In particular, the observational data are typically contaminated by noise, and the forward operator mapping the parameter space to the observation space may fail to be invertible or continuously stable. As a consequence, the estimation problem requires regularization to ensure well-posedness \cite{Stuart2010}. In a probabilistic setting, such regularization is naturally achieved through the incorporation of prior information into the inference problem. This prior knowledge may be derived from expertise in the field or from general mathematical properties of the chosen parametric family.
From a Bayesian perspective \cite{KF}, this means that the constitutive parameters are treated as unknown and hence modeled as random variables. 
Given observational data, the objective is to compute the posterior distribution of the parameters given likelihood and prior functions. 
A common approach for characterizing the posterior distribution is Markov Chain Monte Carlo (MCMC) method \cite{Metropolis1953}, which constructs a Markov chain whose invariant distribution coincides with the posterior. The main advantage of MCMC is that it does not require an explicit parametric representation of the posterior distribution, but instead provides asymptotically exact samples from the full (possibly high-dimensional and non-Gaussian) distribution. This makes it particularly suitable for inverse problems where the posterior has a complex structure that is not amenable to analytical approximation. However, MCMC is computationally expensive when combined with the Finite Element Method (FEM) predictions due to slow convergence. In the special case where the mapping between the parameter space and the observation space is linear and the prior distribution is Gaussian, the Bayesian posterior is Gaussian as well, and the posterior mean coincides with the best linear unbiased estimator, which is given by the Kalman filter (KF) estimate \cite{KF}. In this setting, the Kalman filter provides an exact solution to the Bayesian inference problem. In the KF ensemble or particle formulation \cite{Evensen2003}, however, the framework can be extended to nonlinear forward models and non-Gaussian distributions \cite{Nerger2021}, where it yields approximate representations of the posterior distribution.

Although filter-based approaches can reduce the number of samples required to estimate posterior moments, one may still encounter situations in which the forward FEM model is computationally expensive. In such cases, surrogate models are introduced to provide a computationally efficient approximation of the FEM-based forward model. In practice, surrogate-based models typically rely on regression approaches to approximate either the forward or inverse mapping. In the forward setting, the objective is to construct a minimally parameterized function that accurately approximates the mapping from unknown parameters to observable quantities. By learning this relationship, the surrogate model eliminates the need to repeatedly evaluate the computationally expensive finite element model whenever the parameter posterior distribution is estimated. A wide range of surrogate modeling strategies has been developed for this purpose. Classical approaches include polynomial chaos expansion methods \cite{wiener}, Gaussian process regression (kriging) \cite{Williams1995}, and reduced-order models based on projection techniques such as proper orthogonal or generalized decomposition \cite{Chatt2000}. More recently, machine learning-based methods such as feedforward neural networks (NN) \cite{Cybenko1989}, convolutional neural networks \cite{Hinton2012}, variational auto-encoders \cite{Kingma2019} and neural operators have emerged as powerful alternatives capable of approximating high-dimensional nonlinear mappings. These approaches differ in their assumptions, expressiveness, and data requirements, but all aim to provide efficient emulators of the underlying physics-based model. Besides these classical methods, physics-based NN-based surrogate models have emerged. They incorporate the governing physical laws \cite{Raissi2017,Basir2022,DRitz} and structural constraints directly into the learning process through the enforcement of differential equation residuals in the loss function.

More recently, learning-based approaches have emerged for direct approximation of the inverse map. 
In Kalman filtering sense the approximation of the direct inverse map is shown in KalmanNet \cite{Revach2022}. This model introduces a surrogate model that replaces the analytical computation of the Kalman gain with a recurrent neural network (RNN), thereby avoiding expensive covariance-related operations and matrix inversions. Similarly, Deep KF \cite{Chatto2023} extends this idea by replacing the complete Kalman update step with a NN to further reduce the computational effort required for state estimation.

Despite the significant progress of machine learning-based approaches for inverse problems, several fundamental limitations remain. Many of these methods, particularly deep NNs, variational autoencoders, and diffusion-based models, require large-scale training datasets generated from repeated evaluations of high-fidelity forward models. In computational mechanics, such data is typically obtained through finite element simulations, which are themselves computationally expensive, especially in high-dimensional, nonlinear, or time-dependent settings. As a result, the training phase can become prohibitively costly, both in terms of computational time and memory requirements.
Moreover, modern surrogate models often involve high-capacity architectures with millions of parameters, leading to substantial training times and significant demands on hardware resources. Even after training, uncertainty quantification or repeated inference in Bayesian settings may require large ensembles or multiple forward passes, further increasing computational costs. These challenges are larger, when the inverse problem must be solved repeatedly, for instance in design optimization, digital twin updating, or real-time monitoring applications.
Consequently, despite their expressiveness, purely data-driven approaches are often not computationally feasible without access to high-performance computing (HPC) resources. This motivates the development of scalable algorithms that exploit parallel architectures and distributed computing, as well as hybrid methods that reduce the reliance on large-scale training through physics-based constraints, reduced-order models, or more efficient probabilistic inference strategies.

Domain decomposition methods (DDMs) provide a natural framework for distributed and parallel computing by partitioning the computational domain into subdomains that can be solved independently on separate processors with only interface communication \cite{ToselliWidlund2005}.
In DDMs, e.g., a NN is decomposed into smaller sub-networks by partitioning the global feature domain into smaller, less complex local subdomains. Based on the splitting strategy DDMs can be classified into  overlapping \cite{Schwarz,Lions} or non-overlapping, i.e., sub-structured \cite{Lions2} methods.
Overlapping methods communicate between subdomains by applying a weighted average across the overlapping interface regions.
In contrast, sub-structured methods either update the subdomains sequentially, using the solution of the first updated subdomain as an interface constraint for the subsequent subdomains, or update them in parallel through the introduction of a coarse space.
The coarse space represents the interfaces between all subdomains and is updated independently of the local subdomain solutions.
In the parallel approach, the local subdomains are first updated simultaneously, followed by an update of the global coarse space that enforces consistency across the interfaces.
Furthermore, based on the loss-function used for training, DDMs for NNs are commonly classified as either data-driven or physics-driven. In data-driven methods \cite{Godde2026} either measurements/monitoring data or data coming from models, such as FEM, are used to generate the distributed surrogate model.
These methods can be further combined with reduced order modeling \cite{Xiao2019,Heaney2022}.
Although data-driven DDMs provide highly efficient surrogate models, their predictive accuracy is inherently limited by the quality and quantity of the available training data. Incorporating the underlying physical equations into the model can therefore improve generalization and enable reliable predictions beyond data-rich regions. A representative example is the deep domain decomposition method (D3M) \cite{KLi2019,WLi2020}, which introduces two overlapping DDM variants based on either the deep Ritz method \cite{DRitz} or the strong form of the governing equations, similar to PINNs \cite{Raissi2017}. In both approaches, interface conditions are enforced through additional penalty terms in the loss. The solution is obtained by alternating between local subdomain updates and interface updates until convergence. As the overlap decreases, more iterations are required and the accuracy deteriorates. A related approach, finite-basis PINNs (FBPINNs) \cite{Moseley}, combines local network outputs using window functions that smoothly restrict each network to its corresponding subdomain. Continuity is enforced implicitly through the overlapping regions while gradients are computed from the global loss. However, all overlapping approaches incur additional computational cost due to the overlap, and their performance generally degrades as the overlap is reduced. To reduce the computational overhead of overlapping methods, non-overlapping (substructuring) DDMs have been proposed. For PINNs, the extended PINN (XPINN) and conservative PINN (cPINN) \cite{Jagtap2020-1, Jagtap2020-2} enforce interface continuity through penalty terms on Dirichlet and Neumann conditions. While cPINN is limited to conservation laws, XPINN is applicable to general PDEs and offers greater flexibility in interface treatment. Both alternate between local subdomain updates and a global interface update, limiting parallel scalability. To address this, \cite{Jang} introduces a coarse interface space and augmented Lagrange multipliers, enabling global coupling while avoiding manual tuning of penalty weights. Another class of methods employs extreme learning machines (ELMs) \cite{Dong2021, LeeCO2025}, where only the final hidden layer is trained while deeper layers remain fixed after random initialization. These approaches enforce $C^k$ continuity across interfaces and determine the trainable weights via least squares.

Within the context of Kalman filtering, distributed approaches have been investigated to address the computational challenges associated with high-dimensional state estimation and parameter identification. These methods aim to exploit spatial decomposition and parallel computation by performing local updates while maintaining consistency between local estimates.  \cite{Shahid2014} proposed a distributed Kalman filtering framework in which local estimators update parameters based on subsets of available measurements, with a random gossip communication strategy \cite{Dimakis2010} employed to improve agreement between local estimates. Similarly,  \cite{Masooleh2022} introduced a communication strategy based on whale optimization to coordinate information exchange between subsystem estimators within a distributed Kalman filtering framework.

An alternative approach is domain localization, where the assimilation problem is decomposed into smaller subdomains with independent local updates \cite{Janjic2011}. While this reduces the computational burden of large-scale assimilation problems, independently updated subdomains may lead to inconsistencies at domain interfaces, requiring additional localization or regularization strategies to preserve smooth global solutions. To address scalability while maintaining accuracy, \cite{NinoRuiz2019} proposed a parallel ensemble Kalman filter implementation based on domain decomposition and sparse covariance estimation. Their method performs local covariance estimation and assimilation within subdomains, enabling parallel computation without significant loss of accuracy compared to established localized ensemble Kalman filter approaches. Similarly,  \cite{Zhang2024} developed a parallel ensemble Kalman method for large-scale field inversion by dividing the computational domain into non-overlapping subdomains and applying local ensemble updates with total variation regularization to improve consistency across interfaces.

The increasing interest in scalable Kalman filtering methods has also been summarized by  \cite{Li2026}, who categorize existing approaches into dimensionality reduction, ensemble and low-rank approximations, localization and sparsity methods, distributed computation, tensor-based formulations, and AI-augmented architectures. These developments highlight the ongoing need for efficient computational strategies that preserve uncertainty representation while enabling Kalman filtering for increasingly high-dimensional systems.

The novelty of this work lies in the development of an efficient Kalman-based inversion framework that combines domain decomposition, NN surrogate modeling, and iterative solution strategies for high-dimensional problems. While existing approaches commonly accelerate either the forecast model through surrogate approximations or the filtering step through modified Kalman formulations, the proposed approach addresses both computational bottlenecks. The expensive covariance-related operations in the Kalman update are avoided by reformulating the update as a linear system and solving it using a preconditioned Richardson iteration, while the computational cost of repeated forward model evaluations is reduced through locally trained NN surrogate models.

Unlike fully NN-based filtering approaches, the proposed method preserves the uncertainty interpretation of the Kalman framework by retaining the covariance-based estimation structure. Furthermore, the domain decomposition strategy enables parallel computation of local surrogate models while maintaining global consistency of the solution. This combination provides an efficient and scalable approach for uncertainty-aware parameter identification in high-dimensional physics-based models, where conventional Kalman filtering becomes computationally infeasible.

The paper is organized as follows: 
In Section 2 the inverse problem is introduced.
In Section 3 the surrogate model used to predict observations is presented. 
Then, the augmented Lagrange DDM with interface conditions is stated in Section 4 and the DDM NN Kalman filter is derived in Section 5.
Finally, in Section 6 numerical results for a three-dimensional material parameter estimation problem are shown and the results of the DDM NN and DDM NN KF are discussed.

%% file: Section/Problem.tex
\section{Inverse problem}
\label{Sec:forward}

Let there be a physical system described by 
\begin{equation}
    \phyop(\idpara,\statef(x)) = \extf(x)
    \label{Eq:PDE}
\end{equation}
on a bounded domain $\gebiet \subset \real^\dims$, where
$\phyop$ is an operator modeling the physics of the system, $\statef$ describes the system state in a Hilbert space $\Hspace$ and $\extf \in \Hspace^*$ (with $\Hspace^*$ being the dual space of $\statef$) describes an external influence on the system. The system is parametrized by a set of parameters $\idpara$, the values of which are assumed to be unknown.

The goal is to identify $\idpara$ given a measurement
\begin{equation}
    \y(x) = \measo(x, \idpara) \in \mospace,
\end{equation}
described by a possibly nonlinear operator $\measo$ mapping the state (and parameters) to the observation space
$\mospace$, which represents the space of admissible measurements. In practice, the observations are contaminated by measurement noise, and the observation is therefore given as:
\begin{equation}
    \ymeas(x) = \y(x) + \fullerr (x),
    \label{Eq:observation}
\end{equation}
in which $\fullerr(x)$ represents the measurement error. In the following, the observation operator is assumed to be local, meaning that
each observation depends on the solution in a limited spatial region. Therefore,
the observation vector can be interpreted as a collection of local measurements
associated with their corresponding spatial locations. This assumption is
satisfied for point-wise observations and localized measurement functionals, which
are commonly used in data assimilation problems. Under this assumption, the
observation operator does not introduce global coupling between distant regions
of the computational domain.

The identification of $\idpara$ given $\ymeas$ is an ill-posed problem due to Hadamard \cite{Hadamard1923}, and requires regularization. In a Bayesian setting the regularization can be achieved by adding  expert knowledge, also referred to as prior knowledge. In other words, $\idpara$ is assumed to be unknown, and hence uncertain. This means that $\idpara$ is modeled as a random variable $\idpara(\evpara)$ with finite second-order moments in  $\mathcal{L}_2(\varOmega_\idpara,\mathcal{F}_\idpara,\mathbb{P}_\idpara)$ in which $\varOmega_\idpara$ is the space of all events, $\mathcal{F}_\idpara$ is the sigma algebra, and $\mathbb{P}_\idpara$ is the probability measure. Given this description, one may predict the observation in Eq.~(\ref{Eq:observation}) as:
\begin{equation}
    \yf(x,\evpara, \evmod) = \measo(x, \paramsfc(\evpara)
    ) + \varepsilon_f(x,\evmod),
    \label{Eq:QOI}
\end{equation}
in which $\varepsilon_f$ is the measurement noise and $\statefc(x,\evpara)$ is computed by solving the stochastic extension of Eq.~(\ref{Eq:PDE}) including uncertainties in the parameters $\paramsfc(\evpara)$:
\begin{equation}
\phyop(\paramsfc(\evpara),\statefc(x,\evpara)) = \extf(x,\evpara), \quad \forall\evpara \in \varOmega_\idpara.
    \label{Eq:stpde}
\end{equation}
Furthermore, in Eq.~(\ref{Eq:QOI}) the measurement noise  $\varepsilon_f$ is modeled as a random variable/vector $\varepsilon_f(\evmod)$ in $\mathcal{L}_2(\varOmega_m,\mathcal{F}_m,\mathbb{P}_m)$,
independent of $q_f(\omega_q)$. This allows defining the joint probability space
\[
(\varOmega,\mathcal{F},\mathbb{P}) := (\varOmega_q \times \varOmega_m,\; \mathcal{F}_q \otimes \mathcal{F}_m,\; \mathbb{P}_q \otimes \mathbb{P}_m),
\]
with the joint event $\omega = (\omega_q,\omega_m) \in \varOmega$.

The objective is to update the prior knowledge of the parameters $\paramsfc(\omega)$ using the forecasted measurement data $y_f$ defined in Eq.~(\ref{Eq:QOI}), obtained from Eq.~(\ref{Eq:stpde}), by measuring the discrepancy between the predicted observations and the measurement data given in Eq.~(\ref{Eq:observation}). Given real measurements $\ymeas$, one may estimate a posterior probability density function of a parameter $q_f(\omega)$ by Bayes rule:
\begin{equation}
\probdist(\paramsfc | \ymeas)
= \frac{\probdist(\ymeas, \paramsfc)}{P(\ymeas)}
= \frac{\probdist(\ymeas | \paramsfc)\probdist(\paramsfc)}{P(\ymeas)}
\label{Eq:bayes}
\end{equation}
in which the normalizing constant in the denominator $P(\ymeas)$ is the evidence. 
The likelihood $\probdist(\ymeas | \paramsfc)$ describes how likely the data $\ymeas$ is, given the forecast parameters $\paramsfc$, and measures the agreement between the observations $\ymeas$ and the predicted measurements $y_f$.

Since evaluating the posterior distribution in Eq.~(\ref{Eq:bayes}) involves evaluating high-dimensional integrals, engineering practice often focuses on estimating its moments instead. In this paper we aim to compute the conditional expectation 
$\mathbb{E}[\paramsfc | \ymeas]$
which can be estimated via a projection-based approach following the methodology proposed in \cite{rosic2013}.  
Let $\mathcal{B}$ denote the sub-$\sigma$-algebra generated by the measurements $y_f$. The conditional expectation $\mathbb{E}[\paramsfc | y_f]$ is then defined as the orthogonal projection of $\paramsfc$ onto the sub-space of $\mathcal{L}_2(\Omega, \mathcal{B}, \mathbb{P})$, i.e.:
\begin{equation}
\mathbb{E}[\paramsfc | y_f] 
= \operatorname*{arg\,min}_{\hat{q}_f \in \mathcal{L}_2(\Omega, \mathcal{B}, \mathbb{P})} \mathbb{E}(\| \paramsfc - \hat{q}_f \|^2),
\label{Eq:exp_cond_proj}
\end{equation}
and is measurable with respect to $\mathcal{B}$. 
Therefore, according to the Doob–Dynkin lemma \cite{Bobrowski2005} there is a measurable function $\measf(\cdot)$ parameterized by $\theta$ such that:
\begin{equation}
    \expect(\paramsfc| y_f) = \measf(y_f;\theta)
\end{equation}
holds. 
Substituting the last equation in Eq.~(\ref{Eq:exp_cond_proj}), one obtains:
\begin{equation}
  \theta^* = \operatorname*{arg\,min}_{\theta} \mathbb{E}(\| \paramsfc - \measf(\yf;\theta)\|^2)
    \label{Eq:exp_cond_doob}
\end{equation}
with $ \expect(\paramsfc| y_f) = \measf(y_f;\theta^*)$.
Under the assumption that $\measf(\cdot)$ is a linear function parameterized by $\linmat$ and $\linpara$, the previous equation can be written as:
\begin{equation}
    (\kalman,b) = \operatorname*{arg\,min}_{\linmat,\linpara} \mathbb{E}(\| \paramsfc - (\linmat \yf + \linpara)\|^2) 
    \label{Eq:exp_cond_lin}
\end{equation}
leading to $\mathbb{E}[\paramsfc |y_f]=K y_f+b$. The optimal parameters $(K,b)$ are determined by solving optimality conditions leading to the Kalman gain:
\begin{equation}
    \kalman
    =
    \operatorname{Cov}(\paramsfc,\yf)
    \left(
    \operatorname{Cov}(\yf,\yf)
    +
    \operatorname{Cov}(\varepsilon_f,\varepsilon_f)
    \right)^{-1}.
    \label{Eq:gain}
\end{equation}
where $\operatorname{Cov}(\paramsfc,\yf)$ denotes the cross-covariance matrix between the parameter forecast  $\paramsfc$ and the measurement forecast $\yf$, defined as:
\[
   \operatorname{Cov}(\paramsfc,\yf) := \mathbb{E}\!\left[(\paramsfc-\mathbb{E}[\paramsfc])\otimes (\yf-\mathbb{E}[\yf])\right].
\]

The Kalman gain further involves the covariance matrix $ \operatorname{Cov}(\varepsilon_f,\varepsilon_f)$ of the measurement noise and the covariance matrix of the observations:
\[
    \operatorname{Cov}(\yf,\yf) := \mathbb{E}\!\left[(\yf-\mathbb{E}[\yf])\otimes(\yf-\mathbb{E}[\yf])\right],
\]
which are defined analogously to $\operatorname{Cov}(\paramsfc,\yf)$.

Given the optimal mapping, the prior knowledge can be further updated to the posterior $q_a(\omega)$ using the measurement data through a linear correction of the prior distribution:
\begin{equation}
    q_a(\omega) = q_f(\omega) + \Delta q_f(\omega),
\end{equation}
where the discrepancy $\Delta q(\omega)$ is defined as
\[
    \Delta q_f(\omega) := \mathbb{E}[\paramsfc |\ymeas] - \mathbb{E}[\paramsfc | y_f].
\]
This leads to the update equation:
\begin{equation}
    q_a(\omega) =q_f(\omega) + \mathbb{E}[\paramsfc |\ymeas] - \mathbb{E}[\paramsfc | y_f],
\end{equation}
which in a linear case reads:
\begin{equation}
    \paramsass(\ev) = \paramsfc(\ev) + \kalman(\ymeas - \yf(\ev)),
    \label{Eq:kalman}
\end{equation}
and is known as the Gauss-Markov-Kalman filter (GMKF) \cite{KF}. For general nonlinear mapping $\varphi(\cdot)$ the filter can be expressed as:
\begin{equation}
    \paramsass(\ev) = \paramsfc(\ev) + \varphi(\ymeas;\theta^*) - \varphi(\yf(\ev);\theta^*),
    \label{Eq:kalmanGEN}
\end{equation}
and can cover nonlinear observation and forward operators as further discussed in \cite{Matthies2016,Rosic2019}. 

The update in Eq.~(\ref{Eq:kalman}) is formulated in continuous form and depends on both the spatial variables and the stochastic components of the system. 
To solve the inverse problem numerically, the forward problem in
Eq.~(\ref{Eq:stpde}) must first be discretized in both the spatial and
stochastic dimensions. This discretization enables the construction of a
computable approximation of the observation operator and allows the discrete
form of the Kalman update in Eq.~(\ref{Eq:kalman}) to be evaluated, yielding the
analysis and parameter update.

The forward problem and stochastic PDE in Eq.~(\ref{Eq:stpde}) can be approximated using the finite element method (FEM). Let $\{\phi_i(x)\}_{i=1}^{N_h}$ denote the spatial finite element basis functions over the domain. Then the solution $\statefc(\omega)$ can be expressed as:
\begin{equation}
\label{fem}
    \statefc(x,\omega) \approx u_{f,h}(x,\omega) = \sum_{i=1}^{N_h} {\statefc}_i(\omega) \, \phi_i(x), \quad \forall \omega \in \varOmega,
\end{equation}
where ${\statefc}_i(\omega)$ are the stochastic coefficients corresponding to each basis function.  Substituting this expansion into Eq.~(\ref{Eq:stpde}) and applying the Galerkin procedure yields the FEM system:
\begin{equation}
    \bm{R}(\bm{q}_f(\omega), \bm{u}_f(\omega)) = \bm{\mathit{0}}, \quad \forall \omega \in \varOmega,
\end{equation}
where $\bm{u}_f(\omega) = [{\statefc}_1(\omega), \dots, {\statefc}_{N_h}(\omega)]^\top$ collects the unknown coefficients, and $\bm{R}$ is the residual vector. Note that this system is not necessarily linear in $\bm{u}_f$, reflecting the potential nonlinearity of the original PDE with respect to $\bm{u}_f$ and/or $\bm{q}_f$. Here, $\bm{q}_f(\omega)$ denotes  discretized parameter set.
Once spatially discretized we may forecast the discretized measurement:
\begin{equation}
 \bm{y}_f(\omega)  = \measo_h(\bm{q}_f(\omega),\bm{u}_f(\omega))+\bm{\varepsilon}_h(\omega)+\bm{\varepsilon}_f(\omega)
\end{equation}
in which $\measo_h$ is observation operator $\measo$ as a function of FEM coefficients, and $\bm{\varepsilon}_h(\omega)$ is the associated FEM discretization error.
After spatial discretization, the update in Eq.~(\ref{Eq:kalman}) reads:
\begin{equation}
    \bm{q}_a(\omega) = \bm{q}_f(\omega) + \bm{\kalman}_h(\bm{y}_m - \bm{y}_f(\omega)), 
    \label{Eq:kalman_spatial}
\end{equation}
and is characterized by the Kalman gain
\begin{equation}
    \bm{K}_h
    =
    \operatorname{Cov}\!\left(
    \bm q_f,\bm y_f
    \right)
    \left(
    \operatorname{Cov}\!\left(
    \bm y_f,\bm y_f
    \right)
    +
    \operatorname{Cov}\!\left(
    \bm\varepsilon_h+\bm\varepsilon_f
    \right)
    \right)^{-1},
    \label{Eq:gain_spatial}
\end{equation}
that accounts for the discretization error \(\bm{\varepsilon}_h(\omega)\) that is either known a priori (e.g.~for elliptic problems), or can be estimated \cite{
Harlim2017}, as discussed later in the text.

The previous update equation is continuous in the stochastic space and thus must subsequently be approximated. The simplest approach is to use Monte Carlo sampling \cite{Chan_2013} leading to:
\begin{equation}
    \bm{q}_a(\omega_i)= \bm{q}_f(\omega_i)+ \bm{\kalman}_h^s(\bm{y}_m - \bm{y}_f(\omega_i)), 
    \quad i=1,\dots,N_s.
\end{equation}
This ensemble approximation \cite{Evensen2003} enables the computation of statistics such as the mean and covariance of the updated parameters, and relies on the statistical estimation of the Kalman gain \(\bm{\kalman}_h^s\) that for large sample size asymptotically converges to \(\bm{\kalman}_h\). 
As a consequence, the predicted measurements $\bm{y}_f(\omega)$ and state $\bm{u}_f(\omega)$ must also be sampled, which requires multiple evaluations of the stochastic forward problem defined in Eqs.~(\ref{Eq:stpde}) and (\ref{Eq:QOI}). Performing these evaluations directly can be computationally expensive, particularly for high-dimensional or nonlinear models. To mitigate this cost, we instead collect the prediction data and construct a surrogate model, which provides an efficient approximation of the forward mapping while preserving the essential statistical properties of the system.

\section{Surrogate model for forward prediction}
\label{Sec:forwardpred}
To reduce the number of samples required for predicting $\bm{u}_f$, a neural network surrogate model is introduced. 
In particular, a feedforward neural network is used to approximate the solution $u_f$ by learning the mapping from the input parameters $\bm{q}_f(\omega)$ to the FEM-discretized solution $\bm{u}_f(\omega)$ as:
\begin{equation}
    \bm{u}_{f}(\omega) \approx \hat{\bm{u}}_{f,\xi}(\bm{q}_{f}(\omega)) = F_{\xi}(q_{f}(\omega))=
    (\Fvar_{L} \circ \boldsymbol{a} \circ \Fvar_{L-1} \circ ... \circ \boldsymbol{a} \circ \Fvar_{1})(\bm{q}_{f}(\omega)),
    \label{Eq:compositionnn}
\end{equation}
in which $A_k:\real^{m_{k-1}} \rightarrow \real^{m_{k}}$ is an affine map, with $m_k \in \mathbb{N}$, defined as:
\beq
    x^{(k)} \mapsto \w^{(k)}x^{(k-1)} + b^{(k)},
\eeq
and $\boldsymbol{a}$ is a vector-valued function expressed as:
\begin{equation}
    \boldsymbol a (\boldsymbol h) := \left[a(h_1), a(h_2), \dots, a(h_{m_k})\right], \quad \boldsymbol h \in \mathbb{R}^{m_k},
\end{equation}
which acts element-wise on the hidden states $h_i$ of the network, i.e., the outputs of the affine transformation to the nonlinear activation function. Here, $\w^{(k)}$ and $b^{(k)}$ denote the trainable weights and biases of the network, which are collected into the parameter vector $\xi := \left\{ \w^{(k)}, \be^{(k)} \right\}^L_{k=1}$.  To obtain the network weights, one collects the training data:
\[
\mathcal{D} =
\left\{
\big(\bm{q}_f(\omega_i),\, \bm{u}_f(\omega_i)\big)
\;:\;
i=1,\dots,N_s
\right\},
\]
and defines the empirical loss function:
\[
\mathcal{J}(\ws)
\approx
\frac{1}{N_s}
\sum_{i=1}^{N_s}
\left\|
\bm{u}_f(\omega_i)
-
\hat{\bm{u}}_{f,\xi}(\bm{q}_{f}(\omega_i))
\right\|_2^2 
\]
that is further minimized by gradient-descent algorithms \cite{NoceWrig06}.

The previously described neural network approximates the mapping from stochastic input parameters to the finite element nodal coefficients. A continuous surrogate solution is then obtained by reconstructing the finite element function using the associated basis expansion
\[
\hat{u}_{f,h}(x;\bm{q}_f) = \sum_{i=1}^{N_h} F_\xi^i(\bm{q}_f)\, \phi_i(x),
\]
where \(F_\xi^i(\bm{q}_f)\) represents the $i$th nodal value obtained by a neural network approximation and $\{\phi_i(x)\}_{i=1}^{N_h}$ denote the finite element basis functions. Since the reconstruction is performed in the finite element space, the resulting surrogate solution $\hat{u}_{f,h}(\cdot;\bm{q}_f)$ is a piecewise continuous function, inheriting spatial regularity from the discretization while the neural network learns the parameter-to-coefficient map.

The previously described mapping is useful when the objective is solely to learn a relationship between the stochastic input space and a quantity of interest. However, the observation also depends explicitly on the spatial position within the computational domain, i.e.~it is a spatially distributed field $u(x, {q}_f(\omega))$, and the parameterization can be naturally extended by including the spatial coordinate $x$. Therefore, to account for full parameterization one requires a neural network representation with augmented inputs, enabling the approximation of the solution as a continuous function over the entire domain rather than at isolated points. 
This formulation yields a surrogate model that can be evaluated at arbitrary points in the domain directly and promotes efficient information sharing across space, which is particularly advantageous for high-dimensional stochastic PDE problems. 

Let the parameter set be augmented by the spatial coordinate $(x, \bm{q}_f(\omega))$, then a feedforward neural network with an extended input representation can be defined as:
\begin{equation}
\label{main_mapping}
    {u}_f(x,\omega) \approx \hat{{u}}_{f,\ws} (x, \bm{q}_f(\omega))= F_\ws(x, \bm{q}_f(\omega)).
\end{equation}
In particular, one has:
\begin{equation}
    \bm{u}_f(\omega) := \big(\bm{u}_f(x_l,\omega)\big)_{l=1}^{N_l} 
    \approx \hat{\bm{u}}_f(\omega):=\big(F_\ws(x_l, \bm{q}_f(\omega))\big)_{l=1}^{N_l}
     \label{Eq:compositionnn}
\end{equation}
which further generates the dataset used to train the neural network:
\[
\mathcal{D} =
\left\{
\big((x_l, \bm{q}_f(\omega_i)),\, \bm{u}_f(x_l, \omega_i)\big)
\;:\;
l=1,\dots,N_l,\;
i=1,\dots,N_s
\right\}.
\]
Given the data, the unknown parameters $\ws$ are evaluated by minimizing the empirical loss:
\[
\mathcal{J}(\ws) \approx \frac{1}{N_s} \frac{1}{N_l} \sum_{i=1}^{N_s} \sum_{l=1}^{N_l} 
\big\| \bm{u}_f(x_l, \omega_i) - \hat{\bm{u}}_{f,\ws}(x_l, \bm{q}_f(\omega_i)) \big\|_2^2.
\]
by gradient based algorithms \cite{NoceWrig06}.

In a direct mapping approach, the neural network approximates the solution field directly as a function of both spatial coordinates and stochastic input parameters. In contrast to FEM-based surrogate models, no intermediate finite element representation is used (besides for training), and the spatial dependence is learned implicitly by the network rather than being enforced through basis functions. The neural network therefore acts as a global function approximator over the joint space $\varOmega \times \mathcal{G}$, where $\varOmega$ denotes the parameter space and $\mathcal{G}$ the physical domain.

The previous approximation introduces the modeling error that can be decomposed into the finite element discretization error and the neural network approximation error:
\[
\varepsilon_u(x,q):=u(x,\bm{q}) - \hat{u}_h(x,\bm{q})
=
\underbrace{u(x,\bm{q}) - u_h(x,\bm{q})}_{\text{finite element discretization error}}
+
\underbrace{u_h(x,\bm{q}) - \hat{{u}}_{f,\ws}(\omega)}_{\text{neural network approximation error}}.
\]
This reveals a hierarchical two-level approximation structure, where the finite element method induces a projection onto a finite-dimensional function space and the neural network approximates the resulting parameter-to-coefficient mapping. These errors should be accounted for in the update procedure, as they may introduce bias into the estimated state.

Once the solution to the forward problem has been approximated by the neural network surrogate, the prediction of the observation becomes:
\[
y_f(x,\omega)
=
\hat{y}_f(x,\omega)
+
\varepsilon_h(x,\omega)
+
\varepsilon_{\mathrm{NN}}(x,\omega), \quad \hat{y}_f(x,\omega)=Y_h\big(\hat{u}_{f,\theta}(x,\bm{q}_f(\omega))\big) + \varepsilon_f(x,\omega)
\]
where $Y_h$ denotes the discrete observation operator associated with the finite element approximation, $\hat{u}_{f,\theta}$ is the neural network surrogate of the finite element solution, and $\varepsilon_f$ represents the measurement noise. 
Here,
\[
\varepsilon_h(x,\omega)
:=
Y\!\left(u_f(x,\omega)\right)
-
Y_h\!\left(u_{f,h}(x,\omega)\right)
\]
is the observation error induced by the finite element discretization, and
\[
\varepsilon_{\mathrm{NN}}(x,\omega)
:=
Y_h\!\left(u_{f,h}(x,\omega)\right)
-
Y_h\!\left(\hat{u}_{f,\theta}(x,\bm{q}_f(\omega))\right)
\]
is the neural network surrogate approximation error.

If the discrete observation operator $Y_h(\cdot)$ is further approximated by a neural network surrogate $G_{\boldsymbol{\zeta}}(\cdot)$, the predicted observation can be expressed as
\[
\hat{y}_f(x,\omega)
=
G_{\boldsymbol{\zeta}}\!\left(\hat{u}_{f,\theta}(x,\bm{q}_f(\omega))\right)
=
\left(\phi^{(L)} \circ \phi^{(L-1)} \circ \cdots \circ \phi^{(1)}\right)
\hat{u}_{f,\theta}(x,\bm{q}_f(\omega)),
\]
where each hidden layer is defined by
\[
\phi^{(k)}(\mathbf{z})
=
\sigma\!\left(V^{(k)}\mathbf{z}+s^{(k)}\right),
\qquad k=1,\ldots,L-1,
\]
and the output layer is
\[
\phi^{(L)}(\mathbf{z})
=
V^{(L)}\mathbf{z}+s^{(L)}.
\]
Here, $V^{(k)}$ and $s^{(k)}$ denote the weight matrices and bias vectors, respectively, $\sigma$ is a componentwise activation function, and
\[
\boldsymbol{\zeta}
=
\{V^{(k)},s^{(k)}\}_{k=1}^{L}
\]
collects all trainable parameters of the observation surrogate.

Following Eq.~(\ref{Eq:compositionnn}), the observation prediction is given by:
\[
y_f(x,\omega) \approx \hat{y}_f(x,\omega)
= \big(G_{\boldsymbol{\zeta}} \circ F_{\theta}\big)(x,\bm{q}_f(\omega))+\varepsilon_f(x,\omega)
\]
i.e., the mapping:
\[
(x, \bm{q}_f(\omega)) \;\;\mapsto\;\; F_{\theta}(x, \bm{q}_f(\omega)) \;\;\mapsto\;\; G_{\boldsymbol{\zeta}}\big(F_{\theta}(x,\bm{q}_f(\omega))\big)
\]
defines an approximate observation operator that depends on both the forward neural network parameters \(\theta\) and the observation neural network parameters \(\boldsymbol{\zeta}\).

When the observation network $G_{\boldsymbol{\zeta}}(\cdot)$ is trained separately to approximate the discrete observation operator $Y_h(\cdot)$, the training problem is
\[
\min_{\boldsymbol{\zeta}} \mathcal{J}(\boldsymbol{\zeta})
=
\frac{1}{N_s}
\sum_{i=1}^{N_s}
\left\|
\bm{y}_{f,h}(\omega_i)
-
\hat{\bm{y}}_{f,h}(\omega_i)
\right\|_2^2,
\]
where
\[
\bm{y}_{f,h}(\omega_i)
=
\big(
Y_h(u_{f,h}(x_1,\omega_i)),
\dots,
Y_h(u_{f,h}(x_{N_l},\omega_i))
\big)^\top,
\]
and
\[
\hat{\bm{y}}_{f,h}(\omega_i)
=
\big(
G_{\boldsymbol{\zeta}}(u_{f,h}(x_1,\omega_i)),
\dots,
G_{\boldsymbol{\zeta}}(u_{f,h}(x_{N_l},\omega_i))
\big)^\top.
\]

The resulting observation operator approximation error is
\[
\varepsilon_G(x,\omega)
:=
Y_h(u_{f,h}(x,\omega))
-
G_{\boldsymbol{\zeta}}(u_{f,h}(x,\omega)).
\]

During the prediction however, the observation network receives the output of the forward surrogate rather than the finite element solution, which introduces the propagated forward error
\[
\varepsilon_{\mathrm{prop}}(x,\omega)
:=
G_{\boldsymbol{\zeta}}(u_{f,h}(x,\omega))
-
G_{\boldsymbol{\zeta}}(F_\theta(x,\bm{q}_f(\omega))).
\]
Since $G_{\boldsymbol{\zeta}}(\cdot)$ is generally nonlinear, $\varepsilon_{\mathrm{prop}}$ is a nonlinear transformation of the forward approximation error
\[
\varepsilon_{\mathrm{forward}}(x,\omega)
:=
u_{f,h}(x,\omega)
-
F_\theta(x,\bm{q}_f(\omega)).
\]

The exact observation can therefore be written as
\[
\begin{aligned}
y_f(x,\omega)
&=
G_{\boldsymbol{\zeta}}(F_\theta(x,\bm{q}_f(\omega)))
+
\varepsilon_h(x,\omega)
+
\varepsilon_G(x,\omega)
+
\varepsilon_{\mathrm{prop}}(x,\omega)
+
\varepsilon_f(x,\omega),
\end{aligned}
\]
where
\[
\varepsilon_h(x,\omega)
:=
Y(u_f(x,\omega))
-
Y_h(u_{f,h}(x,\omega))
\]
denotes the discretization error of the observation operator and $\varepsilon_f(x,\omega)$ is the measurement noise. Hence, the total modelling error becomes:
\[
\varepsilon_{\mathrm{model}}(x,\omega)=\varepsilon_h(x,\omega)
+
\varepsilon_G(x,\omega)
+
\varepsilon_{\mathrm{prop}}(x,\omega).
\]

Alternatively, the forward surrogate and the observation network may be trained jointly by solving
\[
\min_{(\theta,\boldsymbol{\zeta})}
\mathcal{J}(\theta,\boldsymbol{\zeta})
=
\frac{1}{N_s}
\sum_{i=1}^{N_s}
\left\|
\bm{y}_f(\omega_i)
-
\hat{\bm{y}}_f(\omega_i)
\right\|_2^2,
\]
where
\[
\hat{\bm{y}}_f(\omega_i)
=
\big(
G_{\boldsymbol{\zeta}}(F_\theta(x_1,\omega_i)),
\dots,
G_{\boldsymbol{\zeta}}(F_\theta(x_{N_l},\omega_i))
\big)^\top.
\]

In this case, the composite mapping $G_{\boldsymbol{\zeta}}\circ F_\theta$ is optimized directly, and the propagated forward error is implicitly accounted for during training. Consequently, the total modeling error is
\[
\varepsilon_{\mathrm{model}}(x,\omega)
:=
Y(u_f(x,\omega))
-
G_{\boldsymbol{\zeta}}(F_\theta(x,\bm{q}_f(\omega))),
\]
and the observation model is written as
\[
y_f(x,\omega)
=
G_{\boldsymbol{\zeta}}(F_\theta(x,\bm{q}_f(\omega)))
+
\varepsilon_{\mathrm{model}}(x,\omega)
+
\varepsilon_f(x,\omega).
\]

Introducing
\begin{equation}
    \bm{y}_f(\omega)
    :=
    \big(y_f(x_l,\omega)\big)_{l=1}^{N_l},
    \qquad
    \hat{\bm{y}}_f(\omega)
    :=
    \big(G_\xi(F_\theta(x_l,\bm q_f(\omega)))\big)_{l=1}^{N_l},
\end{equation}
\[
    \bm{\varepsilon}_{\mathrm{model}}(\omega)
    :=
    \big(\varepsilon_{\mathrm{model}}(x_l,\omega)\big)_{l=1}^{N_l},
    \qquad
    \bm{\varepsilon}_f(\omega)
    :=
    \big(\varepsilon_f(x_l,\omega)\big)_{l=1}^{N_l},
\]
and substituting into the update equation, Eq.~(\ref{Eq:kalman}), we obtain
\begin{equation}
    \bm q_a(\omega)
    =
    \bm q_f(\omega)
    +
    \mathbf K_{\mathrm{NN}}
    \big(
    \bm y_m
    -
    \hat{\bm y}_f(\omega)
    \big).
    \label{Eq:kalmanNN}
\end{equation}

Denoting
\[
\bm{\varepsilon}_{fm}
:=
\bm{\varepsilon}_{\mathrm{model}}
+
\bm{\varepsilon}_f,
\]
the Kalman gain is given by
\begin{equation}
    \mathbf K_{\mathrm{NN}}
    =
    \operatorname{Cov}(\bm q_f,\hat{\bm y}_f)
    \left(
    \operatorname{Cov}(\hat{\bm y}_f,\hat{\bm y}_f)
    +
    \operatorname{Cov}(\bm{\varepsilon}_{fm})
    \right)^{-1}.
    \label{Eq:gainNN}
\end{equation}
and, due to the presence of modeling errors, generally leads to a higher uncertainty in the posterior compared to the classical Kalman gain computed without modeling error.

In practice, $\varepsilon_{\mathrm{model}}(x,\omega)$ is typically treated as an additional stochastic term with a suitable covariance, which can be estimated empirically from validation data, for example by Monte Carlo simulations. In this paper we approximate the distribution of the error by the normal distribution, and estimate its moments empirically from validation data. The modeling error at each spatial point $x_i$ and for each random realization $\omega_j$ is computed as
\begin{equation}
\varepsilon_{\mathrm{model}}(x_i,\omega_j)
=
y_f(x_i,\omega_j)
-
\hat{y}_f(x_i,\omega_j),
\qquad
i=1,\ldots,N_l,\;
j=1,\ldots,N_m,
\label{Eq:mod_err}
\end{equation}
where $N_l$ denotes the number of spatial locations and $N_m$ the number of validation samples.

The sample mean of the modeling error at each spatial point is
\[
\bar{\varepsilon}_{\mathrm{model}}(x_i)
=
\frac{1}{N_m}
\sum_{j=1}^{N_m}
\varepsilon_{\mathrm{model}}(x_i,\omega_j).
\]

The corresponding sample covariance (variance) is estimated by
\[
\operatorname{Cov}_{\varepsilon}(x_i)
=
\frac{1}{N_m-1}
\sum_{j=1}^{N_m}
\left(
\varepsilon_{\mathrm{model}}(x_i,\omega_j)
-
\bar{\varepsilon}_{\mathrm{model}}(x_i)
\right)^2.
\]

The estimates of the mean and covariance are based on a finite number of samples and therefore contain statistical uncertainty. This uncertainty arises from sampling variability, finite sample size, and the stochastic nature of the surrogate modeling error.
An alternative approach is to augment the model error with the state vector in the data assimilation procedure and estimate it jointly with the unknown parameters. In this setting, the model error can be represented, for example, by an auto-regressive (AR) process or by a low-dimensional neural network whose parameters are updated during the assimilation procedure \cite{mortada2025}.

%% file: Section/Methodology.tex
\section{Domain decomposition}
\label{Sec:method}

Although the surrogate model in Eq.~(\ref{Eq:kalmanNN}) significantly reduces the computational cost of the online update compared with classical ensemble filtering methods, its construction may still be computationally expensive due to the large number of trainable parameters required to approximate high-dimensional mappings. Consequently, mesh refinement can substantially increase the offline training time, particularly for large-scale three-dimensional problems. Next to this, a global neural network as presented in Eq.~(\ref{Eq:kalmanNN}) is trained by minimizing a loss function defined over the entire computational domain. As a result, the optimization primarily controls the average approximation error rather than the local error. Consequently, regions with complex solution features, such as steep gradients or localized nonlinearities, may exhibit significantly larger approximation errors, even when the global loss is small. This non-uniform distribution of the approximation error motivates the use of domain decomposition, where local networks can better resolve spatially localized solution features.

In this paper we decompose the spatial domain in the  mapping in Eq.~(\ref{main_mapping}) into smaller subdomains, each approximated by a dedicated NN, to reduce the computational cost of solving the forward problem. 
This decomposition enables parallel and distributed computation, as each subdomain can be processed independently or with limited communication. 
Such an approach also increases flexibility, allowing the use of locally optimized network architectures, training strategies, and mesh resolutions. 
The DDM, as described in the following, is employed to implement this framework efficiently.

Let the spatial domain $\gebiet$ be partitioned into $M$ non-overlapping subdomains $\gebiet_i$:
\[
\gebiet = \bigcup_{i=1}^{M} \gebiet_i, 
\qquad 
\gebiet_i \cap \gebiet_j = \emptyset \quad \text{for } i \neq j.
\]
The global approximation introduced in Eq.~(\ref{Eq:compositionnn}) is then replaced by a collection of $M$ local neural networks, each defined on its respective subdomain. Specifically, the approximation on subdomain $\gebiet_i \quad \forall x \in \gebiet_i,\ \forall \omega \in \varOmega$ is given by
\begin{equation}
    u_{f,i}(x,\omega) \approx \hat{u}_{f,\wi} (x, \bm{q}_f(\omega))= F_{\wi}(x, \bm{q}_f(\omega))=\big(F_{\wi}(x_l, \bm{q}_f(\omega))\big)_{l=1}^{N_l}.
     \label{Eq:splitnn}
\end{equation}
where the vector-valued activation function $\boldsymbol a^{(i)}$ and the affine maps $ \Fvar_{L_i}^{(i)}$ are defined in a similar manner analogous to Eq.~(\ref{Eq:compositionnn}).
In particular, at the nodes:
\begin{equation}
    \bm{u}_{f,i}(\omega) := \big(\bm{u}_{f,\wi}(x_l,\omega)\big)_{l=1}^{N_l} 
    \approx \hat{\bm{u}}_{f,\wi}(\omega):=\big(F_{\wi}(x_l, \bm{q}_f(\omega))\big)_{l=1}^{N_l}.
     \label{Eq:compositionnn_sep}
\end{equation}

Given the local neural network representations, the objective is to determine the unknown parameter vectors $\wi$, $i=1,\dots,M$, by minimizing the local mean-squared error loss on each subdomain
\begin{align}
    \wi^{*}
    &= \arg\min_{\wi} \loss_{\ixa}(\wi), \nonumber \\
    \loss_{\ixa}(\wi)
    &= \mathbb{E}
    \left[
        \left\|
            \bm{u}_{f,i}(\omega) - \hat{\bm{u}}_{f,\wi}(\bm{q}_f(\omega))
        \right\|_2^2
    \right], \quad \forall \omega \in \varOmega.
    \label{Eq:unconst_eq_loc}
\end{align}
In this approach, the solution $\bm{u}_f(x,\omega)$ is approximated separately in each subdomain. The interfaces between subdomains, however, may not be continuous if treated independently. Even if ample data exists, NNs often predict badly when queried in an extrapolatory regime, such as on the interfaces \cite{Swiler2020}. To prevent this, we enforce $C^1$ continuity constraints at the interfaces, ensuring that both the network output and its first derivative are smooth across subdomains. This yields a globally consistent and smooth approximation across the entire domain.

By enforcing continuity of both the solution and its normal derivative, the global approximation problem in Eq.~(\ref{Eq:unconst_eq_loc}) can be reformulated as a set of constrained local optimization problems:
\begin{align}
    &\wi^* = \arg\min_{\wi} \loss_i(\wi),
    && i = 1, \dots, M,
    \label{Eq:unconst_eq_loc1} \\
    &\text{s.t.} \quad
    \hat{u}_{f,\wi}(x,\bm{q}_f(\omega)) = \hat{u}_{f,\wij}(x,\bm{q}_f(\omega)),
    && \forall x \in \bound, \quad \forall \omega \in \varOmega,
    \label{Eq:unconst_eq_loc2} \\
    &\text{s.t.} \quad
    \partial_\normal \hat{u}_{f,\wi}(x,\bm{q}_f(\omega))\big|_{\normal}
    =
    \partial_\normal \hat{u}_{f,\wij}(x,\bm{q}_f(\omega))\big|_{\normal},
    && \forall x \in \bound, \quad \forall \omega \in \varOmega,
    \label{Eq:unconst_eq_loc3}
\end{align}
where $\normal$ denotes the unit normal vector to the interface and $\hat{u}_{f,\wij}$ is the approximation of the interface solution defined on \(\bound = \gebiet_{\ixa} \cap \gebiet_{\ixb}, \ \ixa \neq \ixb\). The latter is approximated by a neural network:
\begin{equation}
    \bm{u}_{f,\wij}(\omega) := \big(u_{f,\wij}(x_l,\omega)\big)_{l=1}^{N_{int}} 
    \approx \hat{\bm{u}}_{f,\wij}(\omega):=\big(F_{\wij}(x_l, \bm{q}_f(\omega))\big)_{l=1}^{N_{int}}.
     \label{Eq:compositionInt}
\end{equation}
the weights of which are learned by minimizing the loss
\begin{equation}
    \wij^*
    =
    \arg\min_{\wij}
    \mathcal{J}_{\ixa\ixb}(\wij),
\end{equation}
where
\begin{equation}
\begin{aligned}
    \mathcal{J}_{\ixa\ixb}(\wij)
    &=
    \sum_{j \in \{\ixa,\ixb\}}
    \mathbb{E} \Big[
        \|\fp_{f,\wj}^{\textrm{int}}(\omega) - \fp_{f,\wij}^{\textrm{int}}(\omega)\|_2^2
    \Big] \\
    &\quad +
    \sum_{j \in \{\ixa,\ixb\}}
    \mathbb{E} \Big[
        \|\partial_\normal \fp_{f,\wj}^{\textrm{int}}(\omega)\big|_{\normal}
        -
        \partial_\normal \fp_{f,\wij}^{\textrm{int}}(\omega)\big|_{\normal}\|_2^2
    \Big],
    \quad \forall \omega \in \varOmega.
\end{aligned}
\label{Eq:inbetween}
\end{equation}

An efficient numerical treatment of the interface constraints in Eq.~(\ref{Eq:unconst_eq_loc}) is achieved by incorporating them directly into the local objective functions using Lagrange multipliers.
In this manner the constrained optimization problems are reformulated as unconstrained by the method of augmented Lagrange multipliers \cite{NoceWrig06,Godde2026}:
\begin{equation}
    \Lagr_\ixa (\wi ,\li) = 
    \loss_{\ixa}(\wi)
    +\li^T \Cwi+\pen\|\Cwi\|^2,
    \label{Eq:main_eq2}
\end{equation}
where
\begin{equation}
    \Cwi
    :=
    \begin{bmatrix}
        \fp_{f, \wi}^{\textrm{int}}(\omega) - \fp_{f,ij}^{\textrm{int}}((\omega) = 0\\
        \partial_\normal \fp_{f,i}^{\textrm{int}}(\omega)|_\normal -  \partial_\normal \fp_{f,\wij}^{\textrm{int}}(\omega)|_\normal = 0\\
    \end{bmatrix}, \qquad \forall\omega \in \varOmega.
    \label{Eq:const_mat2}
\end{equation}
Here, \(\boldsymbol{\lambda}_i\) are the Lagrange multipliers and $\fwi^{\textrm{int}}(\omega)$ and $\fm(\omega)$ are nodal approximations on the interface coming from subdomain network and the interface network.
The term \( \|\Cwi\|^2 \) represents an additional regularization term of the sum of squares type, which penalizes the growth of the constraints in Eq.~(\ref{Eq:main_eq2}), while the penalty factor \( \pen \) adjusts its weight. This ensures that each subdomain satisfies its local constraints.

For simplicity reasons, the formulation is restricted to interfaces between pairs of adjacent local domains. This approach extends naturally to configurations with multiple intersecting subdomain interfaces. 
Equation~\eqref{Eq:inbetween} provides a mechanism for computing a global interface approximation, 
which facilitates information exchange between distant subdomains. 
This approximation defines a global coarse space that captures the dominant, large-scale behavior 
of the solution across the entire domain. 
Meanwhile, each subdomain can perform detailed local computations independently, 
without continuously comparing with neighboring subdomains, 
and periodically incorporate the global interface information to maintain consistency. 
This strategy balances efficient local computation with effective global coupling, 
thereby accelerating convergence.

To numerically solve the previous optimization problem given in Eq.~(\ref{Eq:main_eq2}), a primal-dual update algorithm \cite{Godde2026} based on the dual ascent method \cite{Boyd2010} is used. The algorithm is based on an alternating scheme in which the primal variables \( \wi \) are updated by minimizing the Lagrangian, 
followed by an update of the dual variables \( \li \) through adjustment of the Lagrange multipliers in response to constraint violations:
\begin{align}
\wi^{(\iter+1)} &:= \arg\min_{\wi} \, \Lagr_i\left(\wi, \li^{(\iter)}\right), \label{eq:update_wi} \\
\li^{(\iter+1)} &:= \li^{(\iter)} + \pen \Constr_i\left(\wi^{(\iter+1)}\right), \quad \iter=1, \dots, D_m. \label{eq:update_li}
\end{align}
While the update for \( \li \) is performed using the gradient ascent algorithm, the minimization of the Lagrangian for \( \wi \) is carried out using the Limited-memory Broyden–Fletcher –Goldfarb–Shanno (LBFGS) method \cite{lbfgs}. 
By keeping the Lagrange multipliers $\li$ constant, the LBFGS algorithm updates the network parameters $\wi$ according to:
\begin{equation}
    \wi^{(\iter+1)} = \wi^{(\iter)} - \lsiter \bm{v}^{(b)},
\end{equation}
where:
\begin{equation}
   \bm{v}^{(b)} = \invhes^\iter \nabla \Lagr(\wi^\iter, \li).
\end{equation}
The step size \( \lsiter \) is determined via a line search, defined as
\[ \lsiter = \arg\min_{\alpha} \Lagr(\wi^{(\iter)} + \alpha \bm{v}^{(b)}, \li). \]
The matrix \( \invhes^{(\iter)} \) approximates the inverse Hessian and is initialized as the identity matrix, making the first LBFGS iteration equivalent to a gradient descent step.
Subsequent updates of the inverse Hessian are performed as follows:
\begin{equation}
    \invhes^{(\iter+1)} = \Big(\iden - \frac{\siter {\giter}^T}{{\giter}^T \siter}\Big) 
    \invhes^{(\iter)} 
    \Big(\iden - \frac{\giter {\siter}^T}{{\giter}^T \siter}\Big) 
    + \frac{{\siter} {\siter}^T}{{\giter}^T {\siter}},
\end{equation}
where $\siter := \lsiter \bm{v}^{(b)}$ and $\giter := \nabla \Lagr(\wi^{(\iter+1)}, \li) - \nabla \Lagr(\wi^\iter, \li)$ denote the update step and the change in gradients, respectively. Due to the non-convex nature of the problem, the previous algorithm has the convergence to a stationary point. Nevertheless, LBFGS typically provides fast local convergence through progressively improved inverse Hessian approximations \cite{Zocco2020, niu2023}. 

Following the process of solving the local minimization problems, a global update of the interface solution $\fm$ is performed by solving the optimization problem in Eq.~(\ref{Eq:compositionInt}). Using the augmented Lagrangian method, which adds a quadratic penalty on the constraint violations, stabilizes the updates and improves convergence by smoothing the dual landscape and coupling local subproblems more effectively. While global optimality cannot be guaranteed, this approach typically leads to convergence to stationary points that satisfy the constraints approximately \cite{Boyd2010}. For a detailed explanation of the algorithm we refer to \cite{Godde2026}.

After convergence of the algorithm, the global approximation of the solution $u_f$ is reconstructed as:
\begin{equation}
  \hat{u}_{f}(\point, \paramsfc(\omega))=\sum^{N_i}_i \hat{u}_{f,\wi}(\point, \paramsfc(\omega)) 1_{\gebiet_i}(\point)  + \sum_{i<j}\hat{u}_{f,\wij}(\point, \paramsfc(\omega)) 1_{\mathcal{G}_{ij}}^T (\point) \, ,
  \label{eq:global_solu}
\end{equation}
with the interface indexing \[
\mathcal{I} = \{ (i,j) \mid 1 \le i < j \le N_i, \; \gebiet_i \cap \gebiet_j \neq \emptyset \}
\]
and indicator functions
\begin{equation}
1_{\gebiet_i}(\point)=\begin{cases}
    1, & \text{if } \point \in \gebiet_i, \\
    0, & \text{if } \point \notin \gebiet_i. \\
\end{cases}.
\label{Eq:indi}
\end{equation}

In contrast to classical domain decomposition methods, where the approximation error mainly originates from the spatial discretization and the treatment of interfaces between subdomains, neural-network-based domain decomposition introduces an additional surrogate approximation error. The total error can therefore be interpreted as the combination of the finite element discretization error, the domain decomposition error associated with the coupling of local solutions, and the neural network approximation error resulting from finite model capacity, optimization, and generalization limitations. While classical domain decomposition reduces the computational cost of solving the discretized system, NN-based approaches replace the repeated numerical solution of local problems by learned approximations, introducing a trade-off between computational efficiency and surrogate accuracy.

Given the forward solution DDM NN approximation, the local predicted observation is defined as
\begin{equation}
    \hat{y}_{f,i}(\omega)
    =
    Y_{h,i}
    \left(
    \hat{u}_{f,\wi}(\cdot,\paramsfc(\omega))
    \right),
    \label{eq:local_observation}
\end{equation}
where $Y_{h,i}$ denotes the observation operator restricted to the subdomain 
$\mathcal{G}_i$. We assume that the observation operator is compatible with the domain
decomposition, such that each measurement depends only on the solution within
the corresponding subdomain. Under this assumption, the DDM approximation of the forward solution naturally
induces a corresponding decomposition of the observation operator, allowing
local observation surrogates to be constructed independently on each subdomain.
The global predicted observation is then reconstructed from the local observations as
\begin{equation}
    \hat{y}_f(\omega)
    =
    \sum_{i=1}^{N_i}
    \hat{y}_{f,i}(\omega)
    +
    \sum_{i<j}
    \hat{y}_{f,ij}(\omega),
    \label{eq:global_observation}
\end{equation}
where $\hat{y}_{f,ij}$ denotes the contribution associated with the interface 
regions $\mathcal{G}_{ij}$. This formulation then leads to the modeling error in the observation space defined as:
\begin{equation}
    \bm{\varepsilon}_{\mathrm{model}}(\omega)
    :=
    \bm{y}_f(\omega)
    -
    \hat{\bm{y}}_f(\omega)  =
    \sum_{i=1}^{N_i}
    \left(
    \bm{y}_{f,i}(\omega)
    -
    \hat{\bm{y}}_{f,i}(\omega)
    \right)
    +
    \sum_{i<j}
    \left(
    \bm{y}_{f,ij}(\omega)
    -
    \hat{\bm{y}}_{f,ij}(\omega)
    \right),
    \label{eq:model_error_observation}
\end{equation}

where $\bm{y}_{f,i}$ and $\hat{\bm{y}}_{f,i}$ denote the exact and DDM NN 
predicted observations on the subdomain $\Omega_i$, respectively, and 
$\bm{y}_{f,ij}$ and $\hat{\bm{y}}_{f,ij}$ denote the corresponding interface 
contributions.

Given predicted observations, one may define Kalman gain:
\begin{equation}
    \mathbf K_{\mathrm{DDMNN}}
    =
    \operatorname{Cov}
    \big(
    \bm q_f,\hat{\bm y}_f
    \big)
    \left(
    \operatorname{Cov}
    \big(
    \hat{\bm y}_f,\hat{\bm y}_f
    \big)
    +
    \operatorname{Cov}
    \big(
    \bm\varepsilon_{\mathrm{model}}
    +
    \bm\varepsilon_f
    \big)
    \right)^{-1}.
    \label{eq:gain_ddmnn}
\end{equation}
and hence the update equation 
\begin{equation}
    \bm q_a(\omega)
    =
    \bm q_f(\omega)
    +
    \mathbf K_{\mathrm{DDMNN}}
    \left(
    \bm y_m-\bm y_f(\omega)
    \right),
    \label{eq:kalmanDD}
\end{equation}
based on the domain decomposition approximation.

Here, $\varepsilon_{\textrm{model}}(\omega)$ denotes the modeling error resulting from the neural network approximations and the domain decomposition approach. This error can also be computed following the procedure described for the modeling error in the previous section. Consequently, the Kalman gain $\kalman_{DDMNN}$ has a form similar to that in Eq.~(\ref{Eq:gainNN}).

 \section{Domain decomposition based ensemble Gauss-Markov-Kalman filter}

In the previous section we have explained how to decompose the solution $u_f(x,\omega)$ into local approximations. However, using this approximation in Eq.~(\ref{eq:kalmanDD}) may still be computationally expensive as one has to compute the Kalman gain on full domain which may be represented by high-resolution finite element discretization. 
The main computational bottleneck is the inversion of the observation covariance 
matrix \(\operatorname{Cov}
    \big(
    \hat{\bm y}_f,\hat{\bm y}_f
    \big)\), whose size scales with the number of observations $N_y$. This
requires the inversion of an $N_y\times N_y$ matrix, with a computational cost 
of $\mathcal{O}(N_y^3)$ for a direct solver. Since the number of observations can 
increase rapidly with mesh refinement, this becomes a limiting factor for large-scale problems. Therefore, localizing the update by computing Kalman gains on 
individual subdomains reduces the dimension of the covariance matrices and 
enables parallel computation of the local updates.

 The global posterior estimate can be recovered by introducing a correction increment $\Delta \bm{q}_f(\omega)$ to the prior $\bm{q}_f(\omega)$ such that
\begin{equation}
\bm{q}_a(\omega) = \bm{q}_f(\omega) +\Delta \bm{q}_f(\omega), \quad \Delta \bm{q}_f(\omega)
=
\bm{K}_{DDMNN}\,\bm{r}_f
=
\operatorname{Cov}\!\left(\bm{q}_f,\bm{y}_f\right)
\left(
\operatorname{Cov}\!\left(\bm{y}_f,\bm{y}_f\right)\right)^{-1}
\bm{r}_f
\label{decouple}
\end{equation}
holds. Here, $\bm{r}_f=\bm{y}_m-\bm{y}_f$ denotes the residual (or innovation) between the measurements and the observation predictions computed using the DDM NN surrogate, see Eq.~(\ref{eq:global_observation}). The evaluation of the Kalman gain constitutes the main computational bottleneck, as it requires the inversion of the observation covariance matrix,
\(
\operatorname{Cov}\!\left(\bm{y}_f,\bm{y}_f\right),
\)
whose dimension grows with the number of observations. To alleviate this computational burden and enable efficient parallelization, the update is reformulated in terms of local subdomain contributions.

We introduce an auxiliary variable $\bm{v}(\omega)$ defined by
\begin{equation}
\bm{v}(\omega) := \operatorname{Cov}\!\left(\bm{y}_f,\bm{y}_f\right)^{-1}  \, \bm{r}_f(\omega),
\end{equation}
and aim to solve the linear system
\begin{equation}
\label{linearsystem}
\operatorname{Cov}\!\left(\bm{y}_f,\bm{y}_f\right) \bm{v}(\omega)=  \bm{r}_f(\omega),
\end{equation}
after which the update equation can be written as:
\begin{equation}
\bm{q}_a(\omega) = \bm{q}_f(\omega) +\operatorname{Cov}\!\left(\bm{q}_f,\bm{y}_f\right) \, \bm{v}(\omega).
\label{Eq:enKF}
\end{equation}

For large-scale problems, the direct solution of system in Eq.~(\ref{linearsystem}) becomes
computationally expensive due to the size of the observation covariance matrix.
Therefore, a preconditioned Richardson iteration \cite{Richardson1911,Stotsky2023} is introduced. The
preconditioner is constructed using the domain decomposition and is defined as a
block-diagonal approximation of the covariance matrix
\begin{equation}
\tilde{\operatorname{Cov}}_{\bm{y}_f\bm{y}_f}
=
\begin{bmatrix}
\operatorname{Cov}(\bm{y}_{f,1},\bm{y}_{f,1}) & 0 & \cdots & 0\\
0 & \operatorname{Cov}(\bm{y}_{f,2},\bm{y}_{f,2}) & \cdots & 0\\
\vdots & \vdots & \ddots & \vdots\\
0 & 0 & \cdots & \operatorname{Cov}(\bm{y}_{f,M},\bm{y}_{f,M})
\end{bmatrix},
\label{eq:block_covariance}
\end{equation}
where the off-diagonal covariance blocks
\(
\operatorname{Cov}(\bm{y}_{f,i},\bm{y}_{f,j}),
\quad i\neq j,
\)
representing correlations between different subdomains, are neglected.
Using this block-diagonal approximation as a preconditioner, the Richardson
iteration reads
\begin{equation}
\bm{v}^{(k+1)}(\omega)
=
\bm{v}^{(k)}(\omega)
+
\alpha \tilde{\operatorname{Cov}}_{\bm{y}_f\bm{y}_f}^{-1}
\left(
\bm{r}_f(\omega)
-
\operatorname{Cov}(\bm{y}_f,\bm{y}_f)
\bm{v}^{(k)}(\omega)
\right),
\label{richardson}
\end{equation}
where $\alpha$ represents the relaxation parameter.
The iterative solution of the covariance system can be expressed directly in
terms of the Kalman correction in the parameter space. Introducing
\[
\Delta \bm{q}_f^{(k)}(\omega)
=
\operatorname{Cov}(\bm{q}_f,\bm{y}_f)
\bm{v}^{(k)}(\omega),
\]
the Richardson iteration can be transformed into an update for the parameter
correction. Multiplying Eq.~(\ref{richardson}) by
$\operatorname{Cov}(\bm{q}_f,\bm{y}_f)$ yields
\begin{equation}
\begin{aligned}
\Delta \bm{q}_f^{(k+1)}(\omega)
=
\Delta \bm{q}_f^{(k)}(\omega)
+
\alpha \operatorname{Cov}(\bm{q}_f,\bm{y}_f)
\tilde{\operatorname{Cov}}_{\bm{y}_f\bm{y}_f}^{-1}
\big(
\bm{r}_f(\omega)
-
\operatorname{Cov}(\bm{y}_f,\bm{y}_f)
\bm{v}^{(k)}(\omega)
\big).
\end{aligned}
\label{eq:parameter_richardson}
\end{equation}

Using the relation between the forecast and analysis parameters,
\[
\bm{q}_a^{(k)}(\omega)
=
\bm{q}_f(\omega)
+
\Delta \bm{q}_f^{(k)}(\omega),
\]
the iterative update can equivalently be written as
\begin{equation}
\begin{aligned}
\bm{q}_a^{(k+1)}(\omega)
=
\bm{q}_a^{(k)}(\omega)
+
\alpha \operatorname{Cov}(\bm{q}_f,\bm{y}_f)
\tilde{\operatorname{Cov}}_{\bm{y}_f\bm{y}_f}^{-1}
\big(
\bm{r}_f(\omega)
-
\operatorname{Cov}(\bm{y}_f,\bm{y}_f)
\bm{v}^{(k)}(\omega)
\big).
\end{aligned}
\label{eq:q_richardson}
\end{equation}
The block-diagonal preconditioner allows the operator appearing in the
last equation to be expressed in terms of local Kalman gains. By
partitioning the observation vector according to the domain decomposition,
\[
\bm{y}_f=
\begin{bmatrix}
\bm{y}_{f,1}\\
\vdots\\
\bm{y}_{f,M}
\end{bmatrix},
\]
one may write
\[
\operatorname{Cov}(\bm{q}_f,\bm{y}_f)
\tilde{\operatorname{Cov}}_{\bm{y}_f\bm{y}_f}^{-1}
=
\begin{bmatrix}
\operatorname{Cov}(\bm{q}_f,\bm{y}_{f,1})
\operatorname{Cov}(\bm{y}_{f,1},\bm{y}_{f,1})^{-1}
&
\cdots
&
\operatorname{Cov}(\bm{q}_f,\bm{y}_{f,M})
\operatorname{Cov}(\bm{y}_{f,M},\bm{y}_{f,M})^{-1}
\end{bmatrix}.
\]

Introducing the local Kalman gains
\[
\bm{K}_{i}^{\mathrm{local}}
=
\operatorname{Cov}(\bm{q}_f,\bm{y}_{f,i})
\operatorname{Cov}(\bm{y}_{f,i},\bm{y}_{f,i})^{-1},
\qquad i=1,\ldots,M,
\]
the preconditioned Kalman operator becomes
\[
\operatorname{Cov}(\bm{q}_f,\bm{y}_f)
\tilde{\operatorname{Cov}}_{\bm{y}_f\bm{y}_f}^{-1}
=
\begin{bmatrix}
\bm{K}_{1}^{\mathrm{local}}
&
\cdots
&
\bm{K}_{M}^{\mathrm{local}}
\end{bmatrix}.
\]

Therefore, the preconditioned Richardson update for the parameter vector can
be written as
\[
\bm{q}_a^{(k+1)}(\omega)
=
\bm{q}_a^{(k)}(\omega)
+
\alpha \begin{bmatrix}
\bm{K}_{1}^{\mathrm{local}}
&
\cdots
&
\bm{K}_{M}^{\mathrm{local}}
\end{bmatrix}
\left(
\bm{r}_f(\omega)
-
\operatorname{Cov}(\bm{y}_f,\bm{y}_f)
\bm{v}^{(k)}(\omega)
\right).
\]

Here, the local Kalman gains define the block-diagonal preconditioner, while
the residual term contains the full covariance matrix and therefore accounts
for the neglected cross-correlations between different subdomains. Hence, the
iterative procedure converges towards the global Kalman update without
explicitly forming the inverse of the full observation covariance matrix.

By decomposing the observation vector into $M$ subdomain contributions, where
each subdomain contains approximately $N_y/M$ observations, the inversion is
replaced by $M$ independent local covariance inversions. The resulting
computational cost scales as
\[
\sum_{i=1}^{M}
\mathcal{O}\left(\left(\frac{N_y}{M}\right)^3\right)
=
\mathcal{O}\left(
M\left(\frac{N_y}{M}\right)^3
\right)
=
\mathcal{O}\left(\frac{N_y^3}{M^2}\right),
\]
neglecting the additional cost associated with the interface contributions.
Hence, the ideal reduction in computational complexity is by a factor of
approximately $M^2$ compared to the direct global inversion.

In practice, the total computational gain is influenced by the size of the
interface blocks, the communication overhead between subdomains, and the number
of iterations required by the iterative solver to recover the global covariance
coupling. Nevertheless, the block-diagonal preconditioning enables efficient
parallel computation while avoiding the explicit inversion of the full
observation covariance matrix.

%% file: Section/Results.tex
\section{Numerical results}
\label{Sec:results}
\begin{figure}[h]
    \begin{subfigure}[b]{0.4\textwidth}  
    \centering 
    \includegraphics[width=0.6\textwidth]{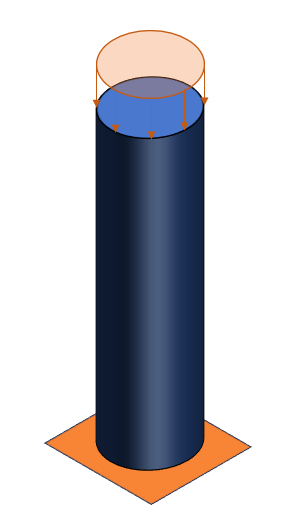}
    \caption{}
    \label{Fig:3dcyl2}
    \end{subfigure}
    \begin{subfigure}[b]{0.6\textwidth} 
    \centering 
    \includegraphics[width=1\textwidth]{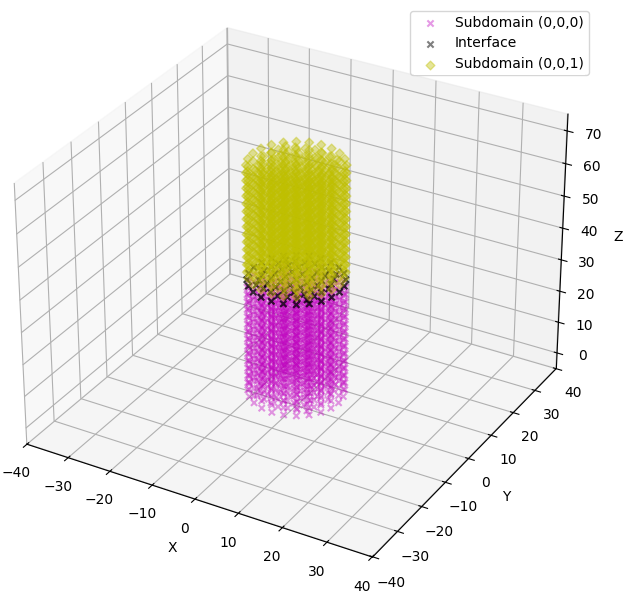}
    \caption{}
    \label{Fig:3dcyl}
     \end{subfigure}
    \caption{a) Schematic of the 3D Cylinder with boundary conditions. b) Data points on the cylinder split into two domains.} 
    \label{Fig:problem}
\end{figure}

As a numerical example of the identification procedure, we consider a three-dimensional cylinder representing a homogenized battery cell subjected to a compressive load of \(20{,}000\,\mathrm{N}\) applied at the top, while the bottom surface is fully clamped (Fig.~\ref{Fig:problem}). Assuming linear elastic material behavior, the cylinder is characterized by its bulk modulus \(\kappa\) and shear moduli \(\mu\), which are treated as the unknown material parameters to be identified from full-field displacement measurements on the cylinder surface. The measurement data are generated synthetically by solving the forward problem using the true parameter values, \(\kappa_{\mathrm{true}}=160{,}419\) MPa and \(\mu_{\mathrm{true}}=80{,}707\)MPa, and subsequently contaminating the resulting displacement field with additive Gaussian noise. Specifically, the measurement error is modeled as a realization of the random variable
\(
\varepsilon_f \sim \mathcal{N}(0, c_f u_{\textrm{true}} I),
\)
where \(c_f\) denotes the noise level corresponding to a coefficient of variation of \(1\%\) of the true measurement $u_{\textrm{true}}$.

\begin{figure}[t]
    \centering
    \begin{subfigure}{0.45\textwidth}
        \centering
        \includegraphics[width=\textwidth]{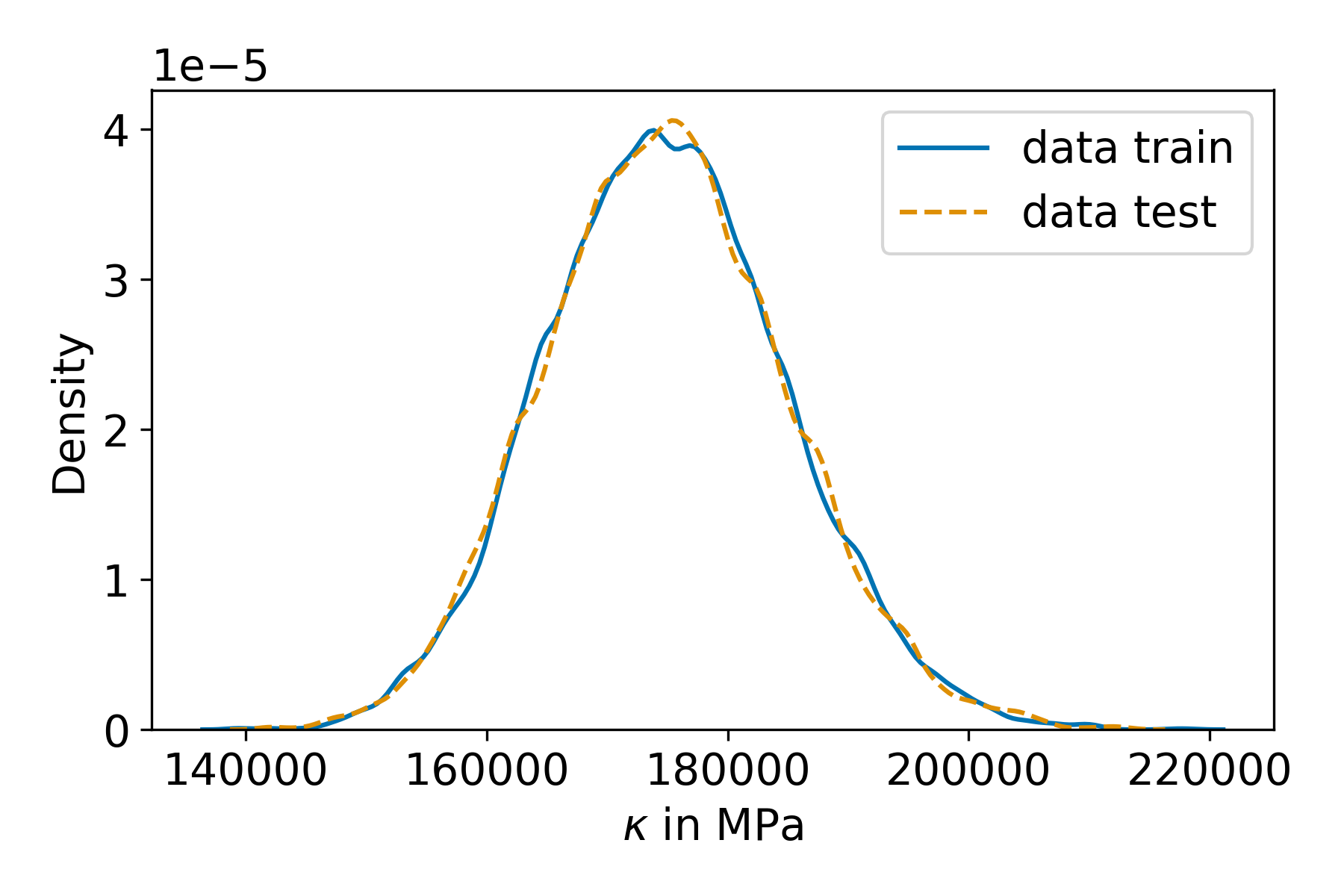}
        \caption{Bulk modulus.}    
        \label{Fig:input_dist_kappa}
    \end{subfigure}
    \hfill
    \begin{subfigure}{0.45\textwidth}  
        \centering 
        \includegraphics[width=\textwidth]{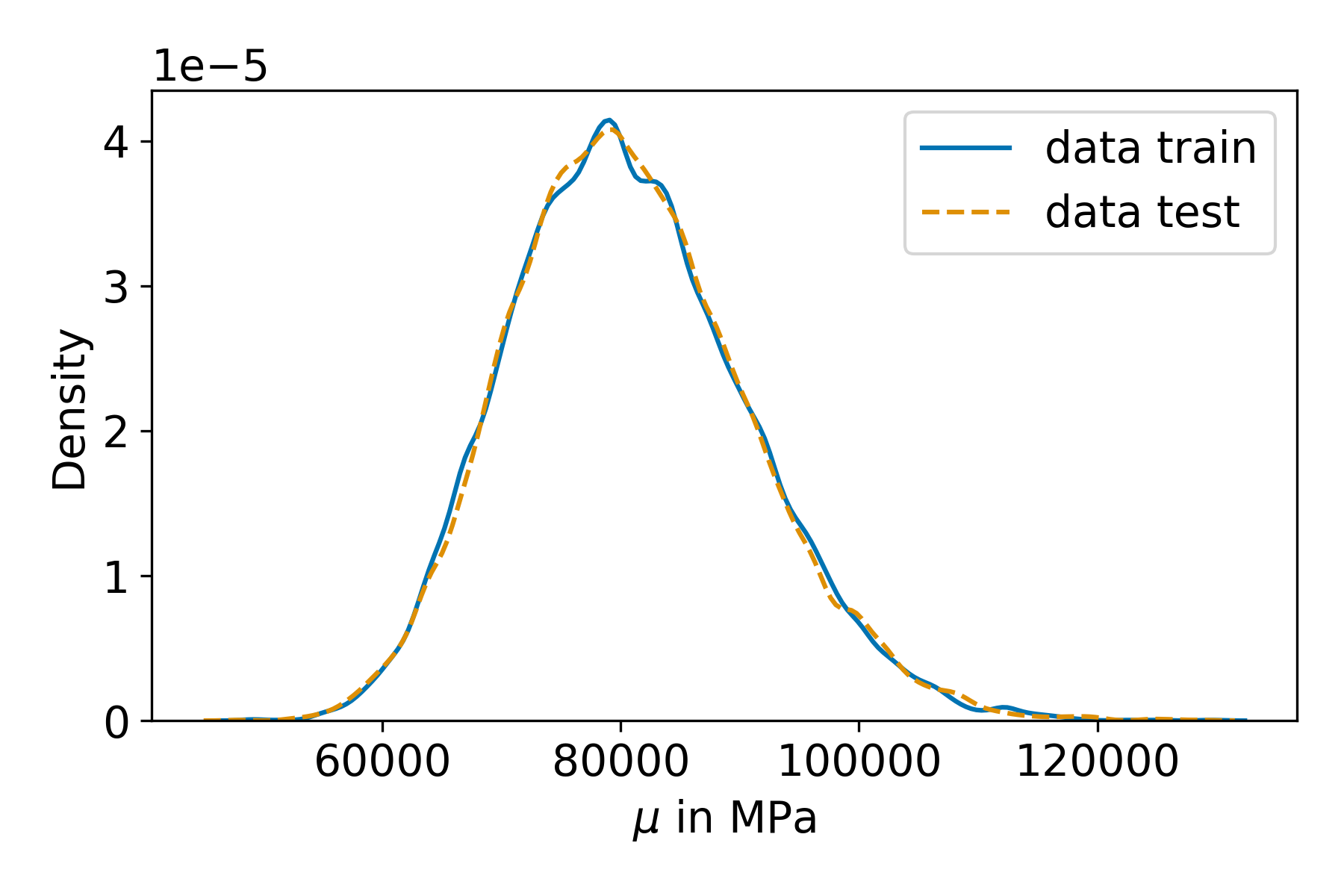}
        \caption{Shear modulus.}     
        \label{Fig:input_dist_mu}
    \end{subfigure}
    \caption{Input parameter kernel density figures. The input distributions of the bulk $\kappa$ and shear $\mu$ modulus of the training samples are shown against new, test samples from the lognormal distributions.} 
    \label{Fig:input_dists}
\end{figure}

The unknown material parameters are treated as uncertain and are therefore modeled a priori as the random vector
\(
\params(\ev) := [\bulk(\ev),\, \shear(\ev)]^{T},
\)
where \(\ev \in \varOmega\) denotes the stochastic event and the components of \(\params\) are assumed to be independent lognormally distributed random variables with the statistical properties summarized in Tab.~\ref{Tab:3dcy}. Given this prior characterization, the corresponding quasi-static boundary value problem is formulated as follows:
\begin{align}
\nabla \cdot \sigma(\point, \ev) &= f(\point), &\forall \point& \in \gebiet,\forall\ev \in \Oq,\label{Eq:PDEs1}\\
    u(\point, \ev) &= \diri, &\forall \point& \in \partial\gebiet_D,\forall \ev \in \Oq,\label{Eq:PDEs2}\\
    \sigma(\point, \ev) \cdot n &= \neum, &\forall \point& \in \partial\gebiet_N, \forall \ev \in \Oq.
    \label{Eq:PDEs3}
\end{align}
where $\sigma \in \Lagr_{2}(\gebiet \times \varOmega, \textrm{Sym}(\mathbb{R}^{d}))$ is the stress tensor, $u \in \mathcal{U} \times \varOmega$ is the displacement field and $f \in \mathcal{U}^*$ describes the internal forces, 
$\tau_n$ denote Neumann boundary conditions on $\partial\gebiet_N$ and $u_D$ is the Dirichlet boundary condition on $\partial\gebiet_D$.
The body is made of material described by Hooke's constitutive law:
\begin{equation}
    \sigma(\ev) = C(\ev):\strain(\ev),
    \label{Eq:Const_stoch}
\end{equation}
in which $C := C(\bulk(\ev),\shear(\ev)) \in \mathcal{L}(Sym(\real^{d})) \otimes \mathcal{S}$ denotes the $4^{th}$ order symmetric, bonded, measurable and point-wise stable elasticity tensor $C$. The linear strains $\strain \in \mathcal{S} \otimes L_2(\gebiet,Sym(\real^+))$ are denoted as: 
\begin{equation}
    \strain = \frac{1}{2}(\nabla_s u +(\nabla_s u)^T),
    \label{Eq:linela}
\end{equation}
in which the linear mapping $\nabla_s$ between strains $\strain$ and displacements $u$ is defined as:
\begin{equation}
\nabla_s: u_1(x)u_2(\oq) \rightarrow (\nabla_su_1(x))u_2(\oq).
\end{equation}
The weak form of the boundary value problem is discretized using the Galerkin finite element method and solved in the \textsc{LS-DYNA} \cite{LSDYNA2024} software package using 1,800 finite elements of type eight-point hexahedron. The stochastic response is obtained by means of Monte Carlo sampling in the parameter space. The full-field displacement, which also serves as the observation in the inverse problem, is selected as the quantity of interest and approximated using the proposed DDM NN surrogate model.
The dataset is generated by performing \(20{,}000\) finite element simulations, of which \(10{,}000\) are used for training and the remaining \(10{,}000\) for testing. The distributions of the sampled bulk and shear moduli are shown in Fig.~\ref{Fig:input_dists}, where the training and test samples are represented by the solid blue and dotted yellow curves, respectively.
All DDM NN computations are carried out on two Linux cluster nodes, each equipped with \(64\) Intel\textsuperscript{\textregistered} Xeon\textsuperscript{\textregistered} Silver 4216 processors operating at \(2.10\,\mathrm{GHz}\) and \(140\,\mathrm{GB}\) of memory.
Once the displacement field has been approximated, the predicted observations are used to estimate the unknown material parameters from the measurement data using the EnKF update given by Eq.~(\ref{Eq:enKF}).

\begin{table}[h]
    \centering
    \caption{Cylinder properties for the finite element simulations.}
    \begin{tabular}{llllllllllll}
    \toprule
        \multicolumn{3}{c}{Dimensions [mm$^3$]} & \multicolumn{2}{c}{Bulk mod. [GPa]} & \multicolumn{2}{c}{Shear mod. [GPa]} & \multicolumn{1}{c}{Element} & \multicolumn{2}{c}{Number of}\\
        \cmidrule(r){1-3} \cmidrule(r){4-5} \cmidrule(r){6-7} \cmidrule(r){9-10}
         \multicolumn{1}{c}{diameter} & x & \multicolumn{1}{c}{height} & \multicolumn{1}{c}{mean} & \multicolumn{1}{c}{variance} & \multicolumn{1}{c}{mean} & \multicolumn{1}{c}{variance} & \multicolumn{1}{c}{type} & \multicolumn{1}{c}{elements} &\multicolumn{1}{c}{nodes}\\
    \midrule
        20 & x & 70 & 175 & 10 & 81 & 10 & hexahedron & 1800 & 31 x 73\\
    \bottomrule
    \end{tabular}
    
    \label{Tab:3dcy}
\end{table}

\begin{table}[h] 
    \centering
    \caption{NN parameter 3D cylinder for a DDM NN split into two equally sized vertically stacked domains along the axis of the cylinder, in which $NN_1$ denotes the lower half.}
    \label{Tab:nn_params_3d}
    
    \footnotesize                 
    \setlength{\tabcolsep}{3.5pt} 
    
    \begin{tabular}{lllrllllrrl}
    \toprule
        \multicolumn{1}{c}{Name} &
        \multicolumn{1}{c}{Inputs} &
        \multicolumn{1}{c}{Outputs} &
        \multicolumn{3}{c}{Architecture} & 
        \multicolumn{1}{c}{Training} &
        \multicolumn{1}{c}{Activation} &
        \multicolumn{1}{c}{Optimizer} &
        \multicolumn{1}{c}{Initializer} &
        \multicolumn{1}{c}{Penalty} \\
        \cmidrule{4-6}
          &  &  &
        \multicolumn{1}{c}{Width} &
        \multicolumn{1}{c}{$x$} &
        \multicolumn{1}{c}{Depth} & 
        \multicolumn{1}{c}{Params.} &
        \multicolumn{1}{c}{Function} &
          &  &
        \multicolumn{1}{c}{Term} \\
        \midrule
         $NN_{1}$ & 5 & 3 & 20 & x & 6 & 2283 & swish & LBFGS & He Normal & $1 \cdot 10^{-3}$ \\
         $NN_{2}$ & 5 & 3 & 20 & x & 7 & 2703 & swish & LBFGS & He Normal & $1 \cdot 10^{-3}$ \\
         $NN_{g}$ & 5 & 3 & 40 & x & 6 & 8563 & swish & LBFGS & He Normal & $1 \cdot 10^{-3}$ \\
    \bottomrule
    \end{tabular}
\end{table}

\begin{figure}[t]
\centering
\includegraphics[width=0.8\textwidth]{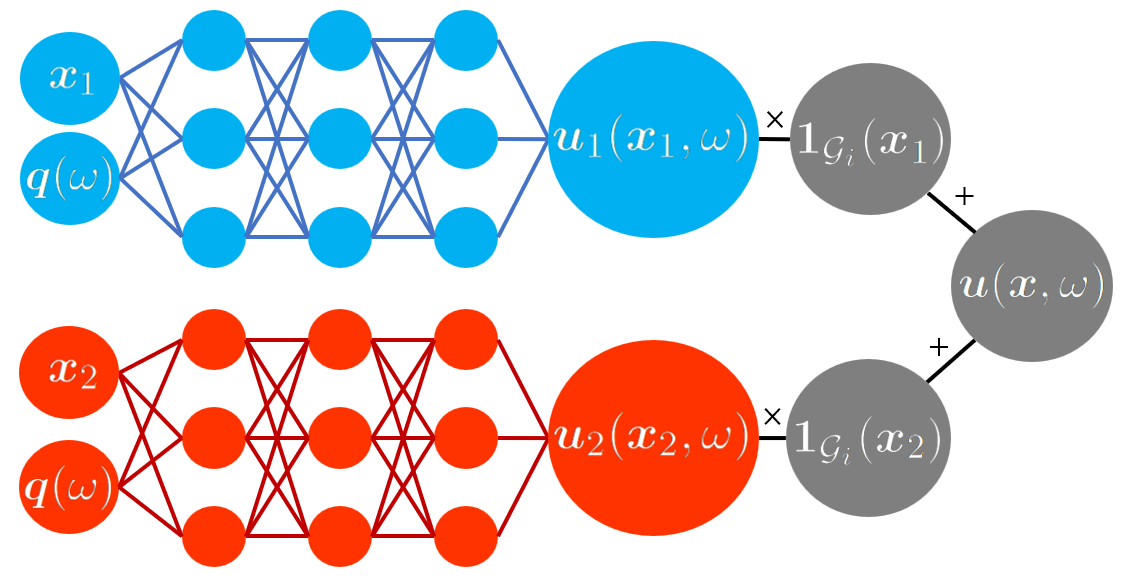}
\caption{Schematic of the decomposed forward prediction model for the spatial split into two domains.}    
\label{Fig:forward_scheme}
\end{figure}

To predict the displacement field, the proposed DDM NN surrogate model is employed, as illustrated schematically in Fig.~\ref{Fig:forward_scheme}. The computational domain is decomposed into two non-overlapping subdomains corresponding to the lower and upper halves of the cylinder, denoted by \(\gebiet_1\) and \(\gebiet_2\), respectively. Accordingly, the displacement field is approximated by two neural networks, \(\hat{u}_{f,1}(\point,\omega)\) for \(\point \in \gebiet_1\) and \(\hat{u}_{f,2}(\point,\omega)\) for \(\point \in \gebiet_2\), as illustrated in the discretized domain shown in Fig.~\ref{Fig:problem}.

Each subdomain is represented by a feed-forward neural network with the hyperparameters listed in Tab.~\ref{Tab:nn_params_3d}. The hyperparameters are optimized independently for each subdomain following the procedure described in Section~\ref{Sec:method}. The network inputs consist of the nodal coordinates and the material parameters, while the outputs correspond to the three-dimensional displacement field.

The global displacement field is reconstructed using the indicator function defined in Eq.~(\ref{Eq:indi}). The displacement vector is expressed as
\(
\boldsymbol{u}=[u_x,u_y,u_z]^T,
\)
where \(u_x\), \(u_y\), and \(u_z\) denote the displacement components along the \(x\)-, \(y\)-, and \(z\)-axes, respectively. The networks are trained using the algorithm proposed in Section~\ref{Sec:method}, which enforces continuity of the predicted displacement field across the interface between adjacent subdomains.

\subsection{Numerical analysis of local neural network approximations}

\begin{figure}[b!]
    \begin{subfigure}{0.5\textwidth}  
    \centering 
    \includegraphics[width=\textwidth]{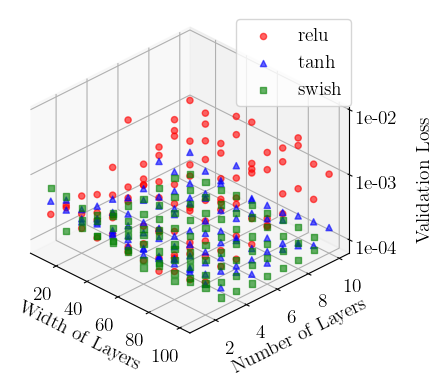}
    \caption{Hyper parameters against loss.}
    \label{Fig:lvnlvwl1}
    \end{subfigure}
    \raisebox{0.1cm}{
    \begin{subfigure}{0.55\textwidth} 
    \centering 
    \includegraphics[width=\textwidth]{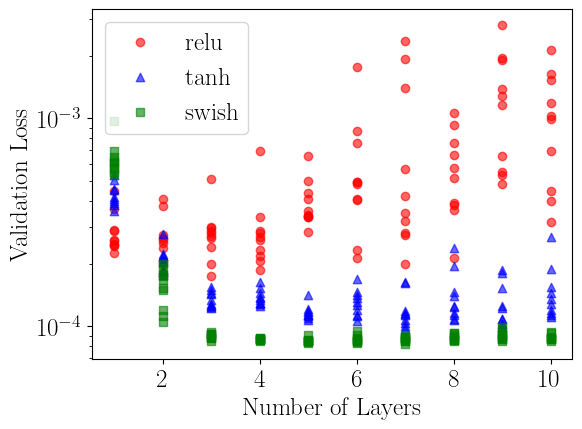}
    \caption{Number of layers against loss.}
    \label{Fig:lvnl1}
     \end{subfigure}
     }
    \caption{The influence of network architecture, e.g., network depth and width on the validation loss of $NN_1$. The corresponding activation functions are indicated in colors and symbols, as depicted in the legends.} 
    \label{Fig:lvnl1f}
\end{figure}

To determine an appropriate network architecture, several hyperparameter configurations are evaluated. The search is restricted to fully connected neural networks with a uniform layer width, i.e., the same number of neurons in each hidden layer. A grid search is performed over the activation function, network width, and network depth.
To ensure a fair comparison, all network configurations are trained for a fixed budget of \(1{,}500\) LBFGS iterations. The resulting validation losses and computational times are then compared under identical optimization settings. During hyperparameter optimization, neither Lagrange multipliers nor interface continuity constraints are incorporated. Consequently, the loss function consists solely of the mean squared error (MSE) evaluated on the training points within each local subdomain.
It is important to note that the reported computational times do not correspond to the time required to achieve a common validation loss. Instead, they represent the wall-clock time required to complete the prescribed \(1{,}500\) optimization iterations for each network configuration.

\begin{figure}[b!]
    \begin{subfigure}{0.5\textwidth}  
    \centering 
    \includegraphics[width=\textwidth]{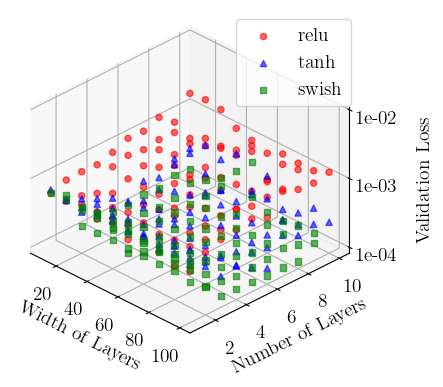}
    \caption{Hyper parameters against loss.}
    \label{Fig:lvnlvwl2}
    \end{subfigure}
    \raisebox{0.1cm}{
    \begin{subfigure}{0.55\textwidth} 
    \centering 
    \includegraphics[width=\textwidth]{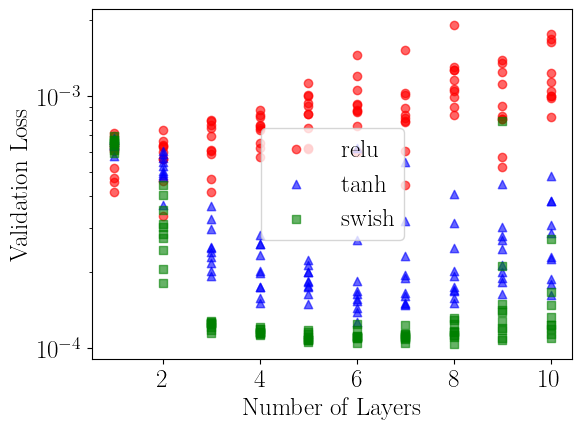}
    \caption{Number of layers against loss.}
    \label{Fig:lvnl2}
     \end{subfigure}
     }
    \caption{The influence of network architecture, e.g., network depth and width on the validation loss of $NN_2$. The corresponding activation functions are indicated in colors and symbols, as depicted in the legends.} 
    \label{Fig:lvnl2f}
\end{figure}

The influence of the network hyperparameters on the validation loss is first examined. Figures~\ref{Fig:lvnl1f} and~\ref{Fig:lvnl2f} present the validation loss as a function of the number of hidden layers (1--10) and the width of the hidden layers (10--100 neurons), with different activation functions distinguished by colors and markers. The analysis is carried out independently for each subdomain.
A first observation is that both subdomains exhibit remarkably similar trends with respect to the selected hyperparameters. Consequently, the following discussion applies to both subdomains. Among the activation functions considered, the swish activation consistently yields the lowest validation loss within the prescribed optimization budget and is therefore selected for the subsequent analyses.
Furthermore, as illustrated in Figs.~\ref{Fig:lvnl1} and~\ref{Fig:lvnl2}, the predictive performance is influenced primarily by the network depth rather than its width. The results indicate that approximately three hidden layers are sufficient to achieve accurate predictions, whereas increasing the depth beyond this point results in only marginal reductions in the validation loss.

\begin{figure}[t!]
    \begin{subfigure}{0.5\textwidth}  
    \centering 
    \includegraphics[width=\textwidth]{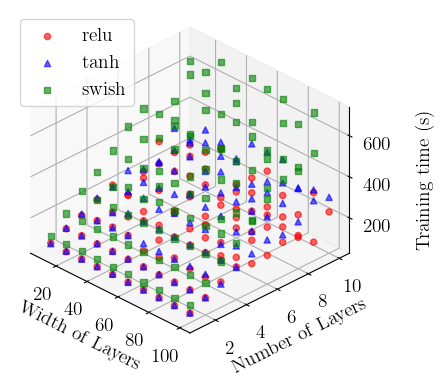}
    \caption{Hyper parameters against time.}
    \label{Fig:tvnlvwl1}
    \end{subfigure}
    \raisebox{0.1cm}{
    \begin{subfigure}{0.55\textwidth} 
    \centering 
    \includegraphics[width=\textwidth]{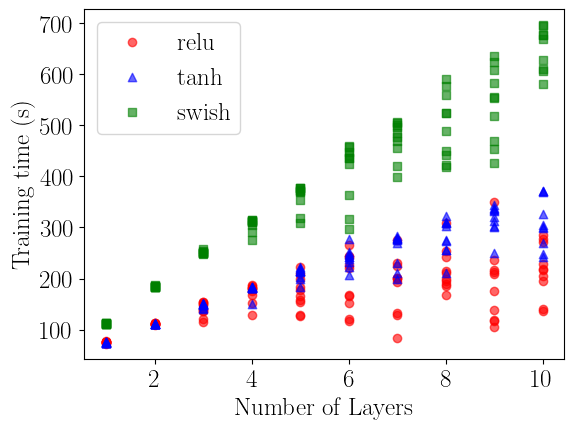}
    \caption{Number of layers against time.}
    \label{Fig:tvnl1}
     \end{subfigure}
     }
    \caption{The influence of network architecture, e.g., network depth and width on the training time of $NN_1$. The corresponding activation functions are indicated in colors and symbols, as depicted in the legends.} 
    \label{Fig:tvnl1f}
\end{figure}
The results also show that the tanh activation function exhibits performance comparable to that of the swish activation function, although the resulting validation losses remain consistently higher than those obtained with swish activation. 
Simultaneously, despite the similarity between the ReLU and swish activation functions, swish yields significantly better performance. 
This behavior is expected since the discontinuity in the ReLU activation function may require more optimization iterations to converge and potentially a larger number of neurons is required to accurately represent these smooth solution fields.

\begin{figure}[t!]
    \begin{subfigure}{0.5\textwidth}  
    \centering 
    \includegraphics[width=\textwidth]{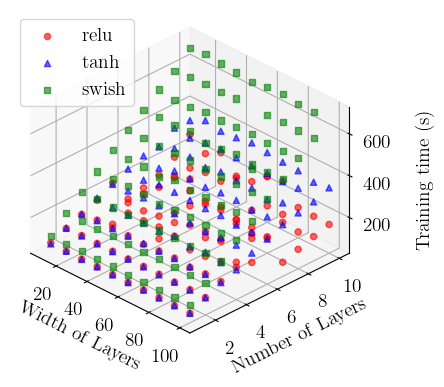}
    \caption{Hyper parameters against loss.}
    \label{Fig:tvnlvwl2}
    \end{subfigure}
    \raisebox{0.1cm}{
    \begin{subfigure}{0.55\textwidth} 
    \centering 
    \includegraphics[width=\textwidth]{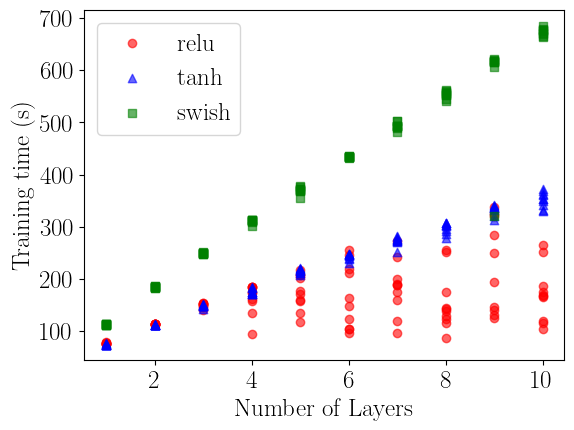}
    \caption{Number of layers against loss.}
    \label{Fig:tvnl2}
     \end{subfigure}
     }
    \caption{The influence of network architecture, e.g., network depth and width on the training time of $NN_2$. The corresponding activation functions are indicated in colors and symbols, as depicted in the legends.} 
    \label{Fig:tvnl2f}
\end{figure}

\begin{table}[t!]
    \centering
    \caption{Neural network hyperparameter combinations after MSE convergence with LBFGS ranked by lowest validation loss values for $NN_1$.}
    \label{tab:hyperparameter_results_local_1}
    
    \small                         
    \setlength{\tabcolsep}{5pt}    
    
    \begin{tabular}{lllrclll}
    \toprule
        \multicolumn{1}{c}{Rank} &
        \multicolumn{1}{c}{Time} &
        \multicolumn{1}{c}{Validation} &
        \multicolumn{3}{c}{Architecture} &
        \multicolumn{1}{c}{Training} &
        \multicolumn{1}{c}{Activation} \\
        \cmidrule{4-6}
        & \multicolumn{1}{c}{(s)} & \multicolumn{1}{c}{loss} &
        \multicolumn{1}{c}{width} &
        \multicolumn{1}{c}{x} &
        \multicolumn{1}{c}{depth} &
        \multicolumn{1}{c}{params.} &
        \multicolumn{1}{c}{function} \\
        \midrule
        1 & 507 & $8.24168 \cdot 10^{-5}$ & 10  & x & 7 &   753 & swish \\
        2 & 377 & $8.31860 \cdot 10^{-5}$ & 90  & x & 5 & 33573 & swish \\
        3 & 435 & $8.32344 \cdot 10^{-5}$ & 20  & x & 6 &  2283 & swish \\
        4 & 442 & $8.34771 \cdot 10^{-5}$ & 100 & x & 6 & 51403 & swish \\
        5 & 378 & $8.38512 \cdot 10^{-5}$ & 100 & x & 5 & 41303 & swish \\
    \bottomrule
    \end{tabular}
\end{table}
\begin{table}[b!]
    \centering
    \caption{Neural network hyperparameter combinations after MSE convergence with LBFGS ranked by lowest validation loss values for $NN_2$.}
    \label{tab:hyperparameter_results_local_2}
    
    \small
    \setlength{\tabcolsep}{5pt}
    
    \begin{tabular}{lllrclll}
    \toprule
        \multicolumn{1}{c}{Rank} &
        \multicolumn{1}{c}{Time} &
        \multicolumn{1}{c}{Validation} &
        \multicolumn{3}{c}{Architecture} &
        \multicolumn{1}{c}{Training} &
        \multicolumn{1}{c}{Activation} \\
        \cmidrule{4-6}
        & \multicolumn{1}{c}{(s)} & \multicolumn{1}{c}{loss} &
        \multicolumn{1}{c}{width} &
        \multicolumn{1}{c}{x} &
        \multicolumn{1}{c}{depth} &
        \multicolumn{1}{c}{params.} &
        \multicolumn{1}{c}{function} \\
        \midrule
        1 & 559 & $1.04571 \cdot 10^{-4}$ & 100 & x & 8 & 71603 & swish \\
        2 & 433 & $1.04726 \cdot 10^{-4}$ & 70  & x & 6 & 25483 & swish \\
        3 & 494 & $1.05519 \cdot 10^{-4}$ & 20  & x & 7 &  2703 & swish \\
        4 & 368 & $1.06424 \cdot 10^{-4}$ & 20  & x & 5 &  1863 & swish \\
        5 & 373 & $1.07533 \cdot 10^{-4}$ & 90  & x & 5 & 33573 & swish \\
    \bottomrule
    \end{tabular}
\end{table}
The influence of the network hyperparameters on the computational cost is also investigated, as shown in Figs.~\ref{Fig:tvnl1f} and~\ref{Fig:tvnl2f}. The results indicate that the training time is governed primarily by the choice of activation function and the network depth, whereas the network width has a comparatively minor influence.
For the continuous activation functions, Figs.~\ref{Fig:tvnl1} and~\ref{Fig:tvnl2} reveal an approximately linear increase in training time with increasing network depth. In particular, networks employing the swish activation function require roughly twice the training time of those using the tanh activation function for the same network depth and fixed number of optimization iterations.
By contrast, networks with the ReLU activation function exhibit a less consistent dependence on network depth, although their training times are generally lower than those of the continuous activation functions. However, this computational advantage is accompanied by inferior predictive accuracy, as demonstrated in Figs.~\ref{Fig:lvnl1f} and~\ref{Fig:lvnl2f}. This behavior is likely attributable to the non-smooth nature of the ReLU activation function, which is less well suited for representing the smooth displacement fields considered in the present study and may adversely affect the convergence of the optimization algorithm. Moreover, continuous activation functions are naturally better suited for constructing smooth approximations, making them more compatible with the continuity requirements imposed across the interfaces between neighboring subdomains.

\begin{table}[b!]
    \centering
    \caption{Neural network hyperparameter combinations after MSE convergence with LBFGS ranked by lowest validation loss values for a global NN approximation.}
    \label{tab:hyperparameter_results_global}
    
    \small
    \setlength{\tabcolsep}{5pt}
    
    \begin{tabular}{lllrclll}
    \toprule
        \multicolumn{1}{c}{Rank} &
        \multicolumn{1}{c}{Time} &
        \multicolumn{1}{c}{Validation} &
        \multicolumn{3}{c}{Architecture} &
        \multicolumn{1}{c}{Training} &
        \multicolumn{1}{c}{Activation} \\
        \cmidrule{4-6}
        & \multicolumn{1}{c}{(s)} & \multicolumn{1}{c}{loss} &
        \multicolumn{1}{c}{width} &
        \multicolumn{1}{c}{x} &
        \multicolumn{1}{c}{depth} &
        \multicolumn{1}{c}{params.} &
        \multicolumn{1}{c}{function} \\
        \midrule
        1 & 496 & $8.24467 \cdot 10^{-5}$ &  60 & x &  7 & 22503 & swish \\
        2 & 405 & $8.35854 \cdot 10^{-5}$ &  40 & x &  6 &  8563 & swish \\
        3 & 690 & $8.35983 \cdot 10^{-5}$ &  80 & x & 10 & 59043 & swish \\
        4 & 570 & $8.36132 \cdot 10^{-5}$ & 100 & x &  9 & 81703 & swish \\
        5 & 668 & $8.36821 \cdot 10^{-5}$ & 100 & x & 10 & 91803 & swish \\
    \bottomrule
    \end{tabular}
\end{table}

Since a large range of feasible solutions exists for the swish activation function, as previously shown in Figs.~\ref{Fig:lvnl1f} and~\ref{Fig:lvnl2f}, the five best-performing architectures with respect to the validation loss are presented for each NN in Tabs.~\ref{tab:hyperparameter_results_local_1} and~\ref{tab:hyperparameter_results_local_2}. 
As already indicated by the figures, swish is the optimal choice, since many different architectures using the swish activation function achieve a similar validation accuracy. 
Overall, the selected hyperparameter configurations are comparable for both local NNs. 
Due to the similar validation performance, the final architecture selection is additionally guided by considerations related to the subsequent DDM NN algorithm. 
It is notable that the objective is not only reducing the number of trainable parameters, but also to retain sufficient representational capacity for the additional interface information introduced by the communication between domains within the domain decomposition method. 
Choosing the smallest possible network may lead to insufficient flexibility once the interface constraints are incorporated into the optimization problem. 
In such a case, the fixed number of activation functions, and therefore network parameters, may no longer be sufficient to accurately represent the solution field. 
It should further be noted that the hyperparameter optimization is performed solely for the unconstrained local problems without Lagrange multipliers in order to keep the optimization procedure computationally feasible. 
Taking these considerations together with the previous analysis into account, the third-best architectures from Tabs.~\ref{tab:hyperparameter_results_local_1} and~\ref{tab:hyperparameter_results_local_2} are selected for both sub-domain networks in the remainder of this work. 
Additionally, it is observed that the selected architectures exhibit lower training times than the best-performing configurations listed in the tables.

\begin{figure}[t!]
    \begin{subfigure}{0.5\textwidth}  
    \centering 
    \includegraphics[width=\textwidth]{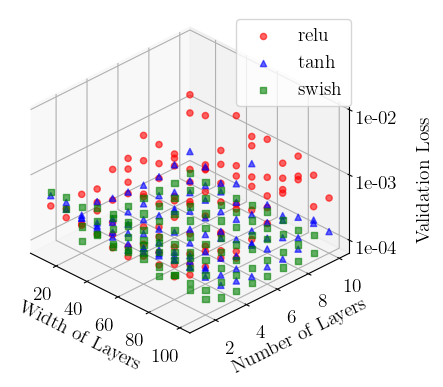}
    \caption{Hyper parameters against loss.}
    \label{Fig:lvnlvwlg}
    \end{subfigure}
    \raisebox{0.1cm}{
    \begin{subfigure}{0.55\textwidth} 
    \centering 
    \includegraphics[width=\textwidth]{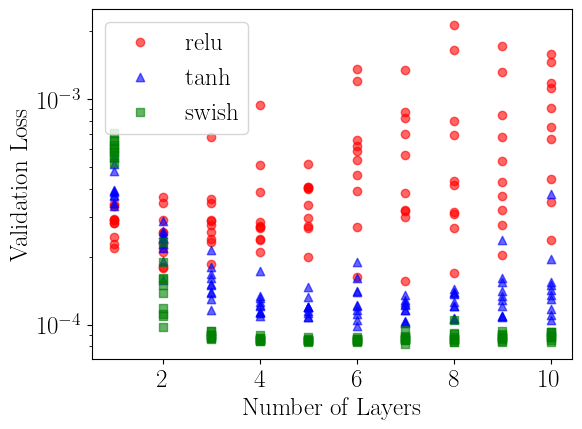}
    \caption{Number of layers against loss.}
    \label{Fig:lvnlg}
     \end{subfigure}
     }
    \caption{The influence of network architecture, e.g., network depth and width on the validation loss of a global NN. The corresponding activation functions are indicated in colors and symbols, as depicted in the legends.} 
    \label{Fig:lvnlgf}
\end{figure}

\begin{figure}[h]
    \begin{subfigure}{0.5\textwidth}  
    \centering 
    \includegraphics[width=\textwidth]{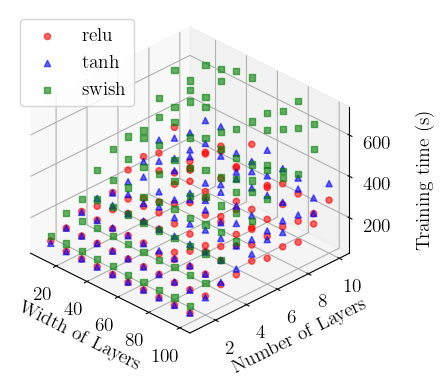}
    \caption{Hyper parameters against loss.}
    \label{Fig:tvnlvwlg}
    \end{subfigure}
    \raisebox{0.1cm}{
    \begin{subfigure}{0.55\textwidth} 
    \centering 
    \includegraphics[width=\textwidth]{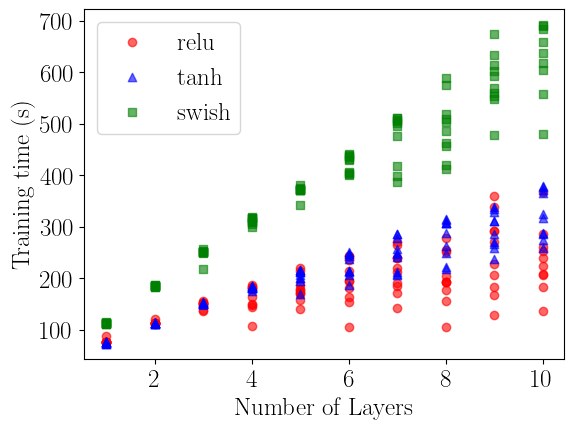}
    \caption{Number of layers against loss.}
    \label{Fig:tvnlg}
     \end{subfigure}
     }
    \caption{Hyperparameter sensitivity analysis: the influence of network architecture, e.g., network depth and width on the training time of a global NN. The corresponding activation functions are indicated in colors and symbols, as depicted in the legends.} 
    \label{Fig:tvnlgf}
\end{figure}

To obtain a reference solution for the presented local NNs, we train full global NN by using LBFGS without the domain decomposition method. In other words, we build a feedforward NN given all spatial points, that is $2,190 \times 10,000$ points. 
The corresponding results are shown in Figs.~\ref{Fig:lvnlgf} and~\ref{Fig:tvnlgf}, which present the validation loss and training time of the global NN after training for different network architectures, while the activation functions are distinguished by different colors. A detailed analysis is omitted here, since the observed trends with respect to the hyperparameters are very similar to those obtained for the local NNs. 
In addition, the five best-performing configurations with respect to the validation loss are listed in Tab.~\ref{tab:hyperparameter_results_global}. 
It can be observed that all selected global architectures contain more trainable parameters than the combined optimized local NN configurations. 
Consequently, the proposed local decomposition approach achieves a substantial reduction in the total number of trainable parameters compared to the global NN that is trained given all points instead of a split set like the decomposed local NNs. 
At the same time, the global NN achieves a slightly lower validation loss of $8.24 \cdot 10^{-5}$ compared to the combined validation loss of the selected local NNs, which is $9.34 \cdot 10^{-5}$. 
This improved accuracy is expected to result from the ability of the global NN to represent the interface region continuously over the entire domain without requiring explicit interface coupling.

\subsection{Numerical analysis of DDM NN surrogate model}
\begin{figure}[b!]
    \begin{subfigure}{0.45\textwidth}  
    \centering 
    \includegraphics[width=\textwidth]{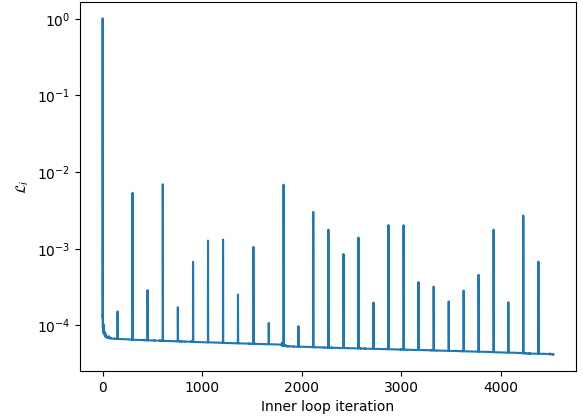}
    \caption{Convergence inner loop algorithm.}
    \label{Fig:conv_inner_3d}
    \end{subfigure}
    \begin{subfigure}{0.49\textwidth} 
    \centering 
    \includegraphics[width=\textwidth]{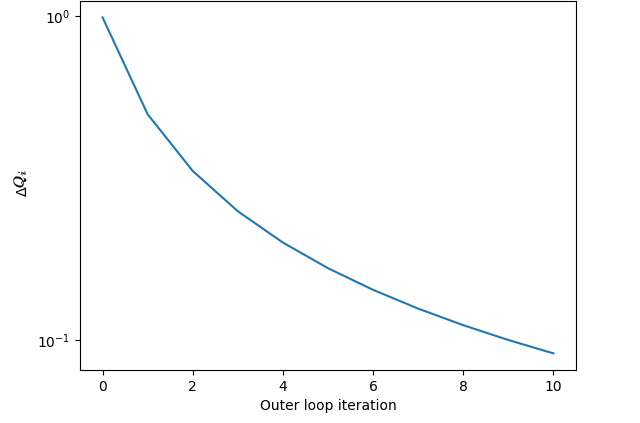}
    \caption{Convergence outer loop algorithm.}
    \label{Fig:conv_outer_3d}
     \end{subfigure}
    \caption{The convergence of the inner and outer loops of the algorithm for the three-dimensional problem.} 
    \label{Fig:3D_conv}
\end{figure}

The convergence of the local NNs  predicting the compressed cylinder is checked, as shown in Fig.~\ref{Fig:3D_conv}. 
Both figures show the corresponding   convergence criteria of the algorithm presented in \cite{Godde2026}, which is the loss function value and the constraint gradient, on the vertical axis and the number of loop iterations on the horizontal axis for the lower NN prediction.
As shown in Fig.~\ref{Fig:conv_inner_3d}, the inner algorithm exhibits a general convergence trend, with several peaks corresponding to instances where the LBFGS algorithm is restarted. Such behavior is characteristic of the optimization procedure, as each restart involves a new initialization of the approximate Hessian. Since this initial approximation is not exact, the subsequent parameter updates may temporarily deviate from the optimal descent direction, resulting in increased loss values before the optimization process resumes convergence.
On the other hand, the convergence of the outer loop is assessed based on the evolution of the Lagrange multipliers, as shown in Fig.~\ref{Fig:conv_outer_3d}. The corresponding loss measure is defined as
\begin{equation*}
    \Delta \Constr=\frac{1}{\ncons}\left(|\Constr(\wi^{l})|-|\Constr(\wi^{l+1})|\right).
\end{equation*}
This quantity measures the change in constraint satisfaction between consecutive outer iterations and indicates the magnitude of the Lagrange multiplier update. As the constraint fulfillment approaches convergence, the improvement between successive iterations diminishes, resulting in a reduction of \(\Delta \Constr\). As observed in Fig.~\ref{Fig:conv_outer_3d}, the loss decreases monotonically with increasing outer iterations until the prescribed convergence tolerance is reached. Combined with the convergence behavior of the inner loop shown in Fig.~\ref{Fig:conv_inner_3d}, these results confirm the successful convergence of the previously proposed algorithm \cite{Godde2026} for the considered problem.

\begin{figure}[t]
    \centering

    \begin{subfigure}[b]{0.32\textwidth}
        \centering
        \includegraphics[width=\textwidth]{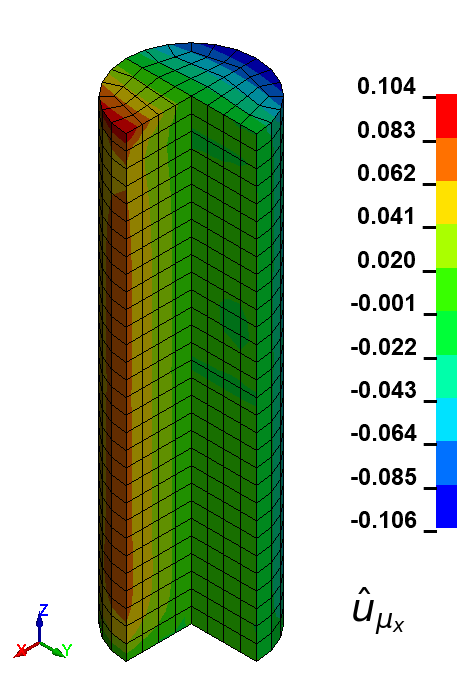}
        \caption{$x$-direction}
    \end{subfigure}
    \hfill
    \begin{subfigure}[b]{0.32\textwidth}
        \centering
        \includegraphics[width=\textwidth]{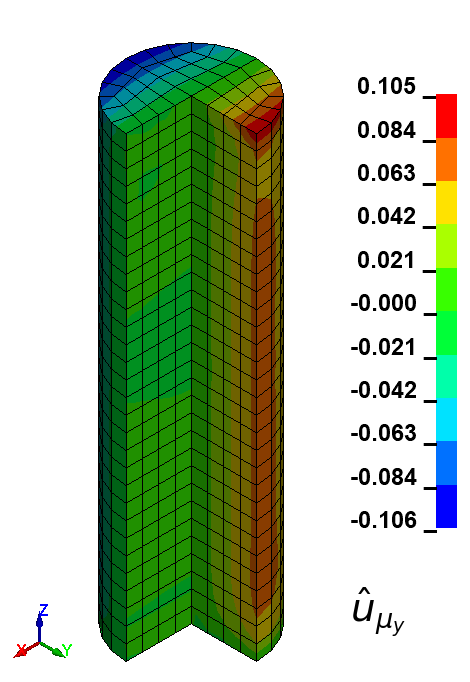}
        \caption{$y$-direction}
    \end{subfigure}
    \hfill
    \begin{subfigure}[b]{0.32\textwidth}
        \centering
        \includegraphics[width=\textwidth]{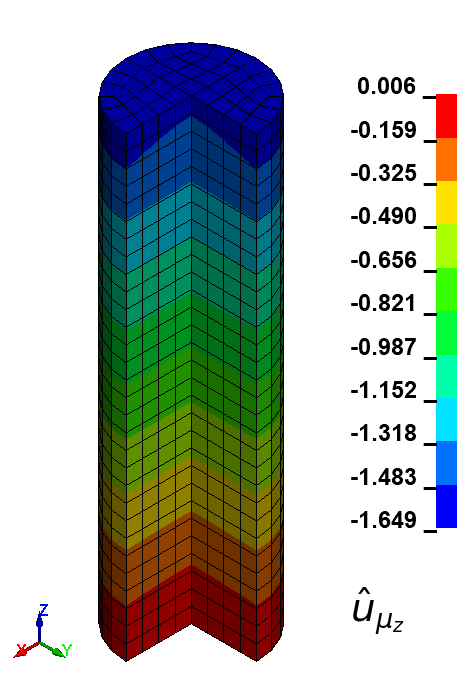}
        \caption{$z$-direction}
    \end{subfigure}

    \caption{Local mean NN DDM predictions of the displacement field $u_\mu$ (test set).}
    \label{fig:pred_m_test}
\end{figure}

\begin{figure}[b]
    \centering

    \begin{subfigure}[b]{0.32\textwidth}
        \centering
        \includegraphics[width=\textwidth]{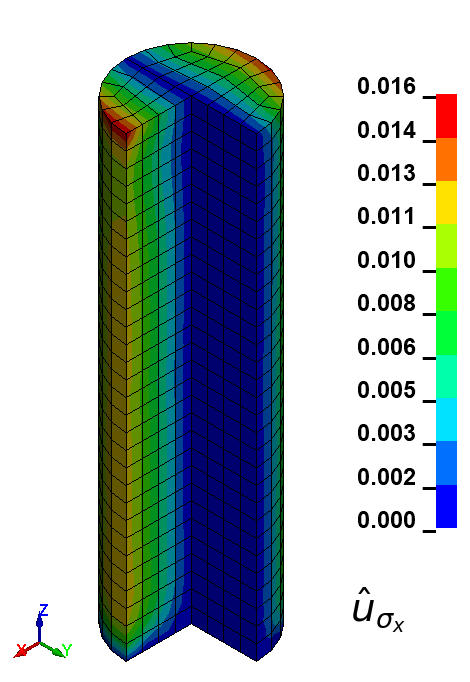}
        \caption{$x$-direction}
    \end{subfigure}
    \hfill
    \begin{subfigure}[b]{0.32\textwidth}
        \centering
        \includegraphics[width=\textwidth]{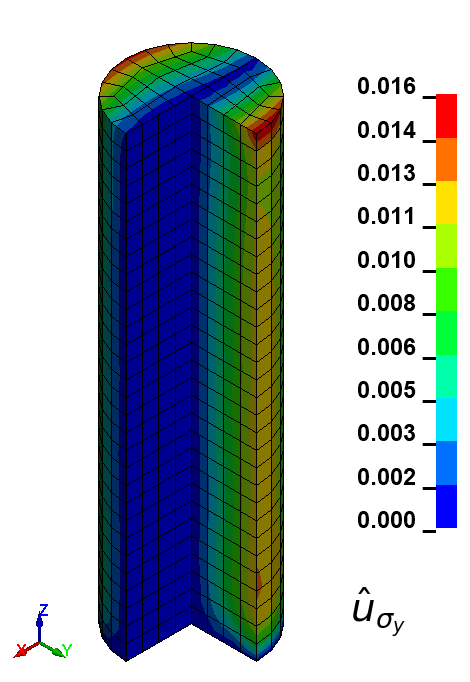}
        \caption{$y$-direction}
    \end{subfigure}
    \hfill
    \begin{subfigure}[b]{0.32\textwidth}
        \centering
        \includegraphics[width=\textwidth]{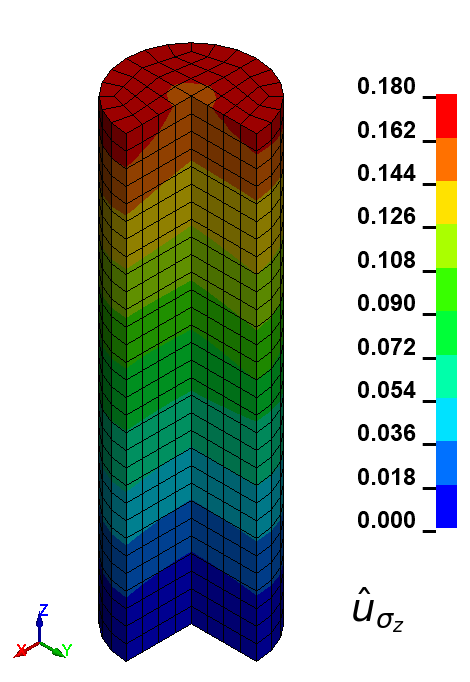}
        \caption{$z$-direction}
    \end{subfigure}

    \caption{Local standard deviation NN DDM predictions of the displacement field $u_\sigma$ (test set).}
    \label{fig:pred_s_test}
\end{figure}

To assess the continuity of the DDM NN predictions, Figs.~\ref{fig:pred_m_test} and~\ref{fig:pred_s_test} present the mean and standard deviation of the predicted displacement field throughout the cylinder domain. Here are shown the displacement field components \([u_x,u_y,u_z]^T,\)
that are reconstructed from the individual subdomain predictions according to Eq.~(\ref{eq:global_solu}).
Each displacement component is visualized separately over the cylinder domain, with a quarter section removed to provide insight into the interior response. The color bars represent the corresponding mean and standard deviation of the predicted displacement components, denoted by \(\hat{u}_{\mu}\) and \(\hat{u}_{\sigma}\), respectively, with all values given in millimeters.
Overall, the predictions exhibit a high degree of continuity across the entire domain. In particular, no visible discontinuity can be identified at the interface between the two subdomains located at the mid-height of the cylinder, indicating that the proposed DDM NN approach successfully enforces continuity between the independently trained networks.
The mean displacement fields correctly capture the expected deformation behavior of the compressed cylinder with a clamped bottom boundary. As shown in Fig.~\ref{fig:pred_m_test}, the \(u_z\) component exhibits the largest displacements at the top surface, with a gradual reduction towards the clamped bottom. A small deviation from the prescribed zero displacement condition can be observed at the bottom boundary, where the predicted displacement reaches approximately \(0.006\,\mathrm{mm}\). This discrepancy can be attributed to the difficulty of neural networks in accurately representing exact zero values, particularly in the presence of uncertainty introduced by the stochastic material parameters.
The in-plane displacement components, \(u_x\) and \(u_y\), do not exhibit any noticeable deviations from the expected symmetric deformation pattern. Both components show equivalent radial behavior, differing only by a rotation of \(90^\circ\), which is consistent with the axisymmetric nature of the loading and geometry. The largest in-plane displacements occur near the outer surface of the specimen, whereas the displacement approaches zero along the central axis and at the bottom boundary.
The standard deviation fields shown in Fig.~\ref{fig:pred_s_test} follow the magnitude of the corresponding mean displacement fields. In particular, larger uncertainties are observed in regions with larger displacements for the \(u_x\) and \(u_y\) components. Similarly, the \(u_z\) component exhibits the highest uncertainty near the top surface, where the accumulated influence of the uncertain material parameters and the applied boundary loading leads to increased variability in the predicted response.

\begin{figure}[htbp]
    \centering

    \begin{subfigure}[b]{0.32\textwidth}
        \centering
        \includegraphics[width=\textwidth]{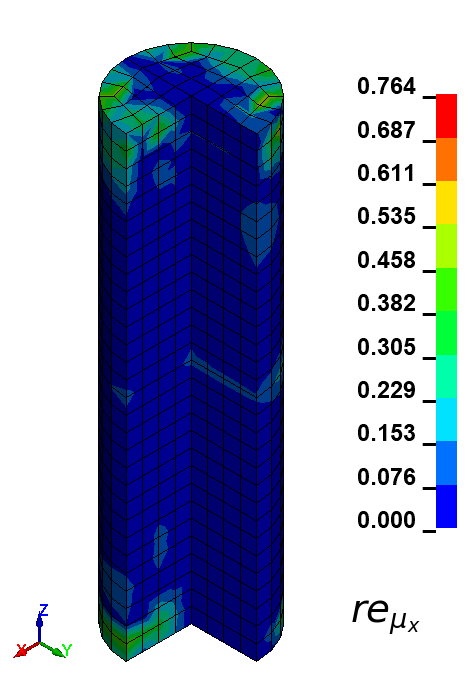}
        \caption{$x$-direction}
    \end{subfigure}
    \hfill
    \begin{subfigure}[b]{0.32\textwidth}
        \centering
        \includegraphics[width=\textwidth]{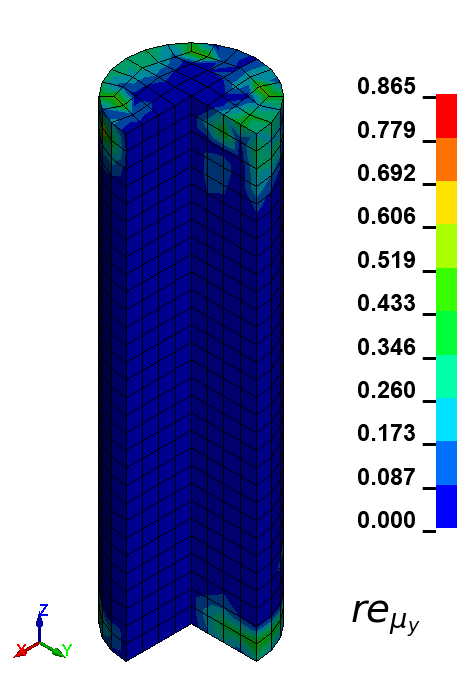}
        \caption{$y$-direction}
    \end{subfigure}
    \hfill
    \begin{subfigure}[b]{0.32\textwidth}
        \centering
        \includegraphics[width=\textwidth]{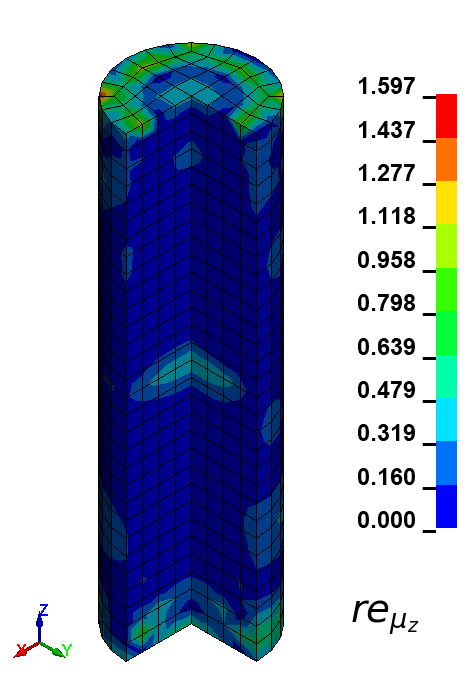}
        \caption{$z$-direction}
    \end{subfigure}

    \caption{Local mean relative errors of the displacement field (test set).}
    \label{fig:rele_m_test}
\end{figure}

\begin{figure}[htbp]
    \centering

    \begin{subfigure}[b]{0.32\textwidth}
        \centering
        \includegraphics[width=\textwidth]{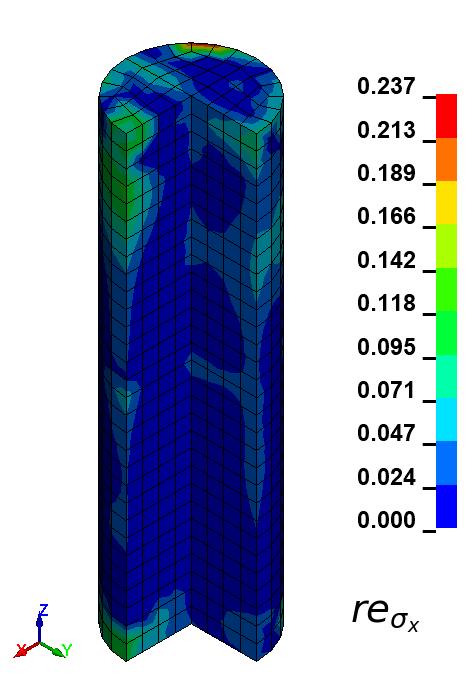}
        \caption{$x$-direction}
    \end{subfigure}
    \hfill
    \begin{subfigure}[b]{0.32\textwidth}
        \centering
        \includegraphics[width=\textwidth]{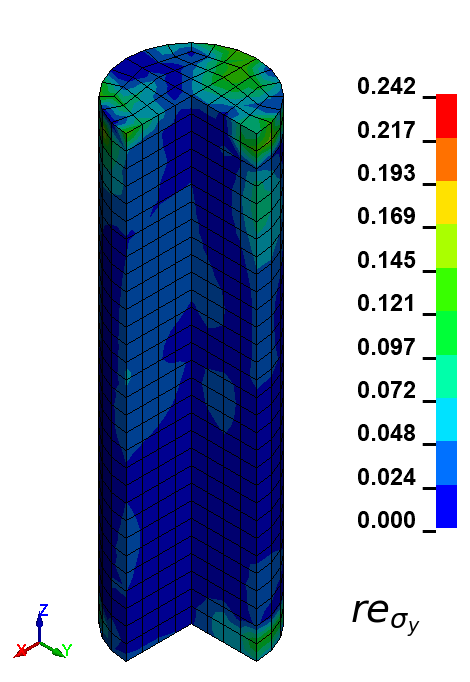}
        \caption{$y$-direction}
    \end{subfigure}
    \hfill
    \begin{subfigure}[b]{0.32\textwidth}
        \centering
        \includegraphics[width=\textwidth]{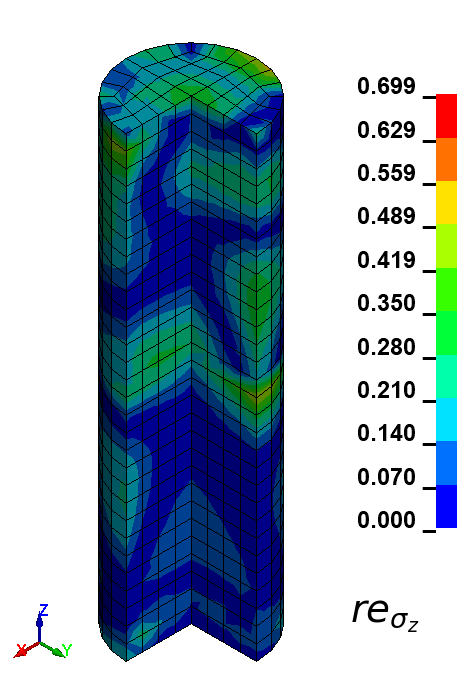}
        \caption{$z$-direction}
    \end{subfigure}

    \caption{Local standard deviation relative errors of the displacement field (test set).}
    \label{fig:rele_s_test}
\end{figure}

\begin{figure}[b]
    \centering

    \begin{subfigure}[b]{0.32\textwidth}
        \centering
        \includegraphics[width=\textwidth]{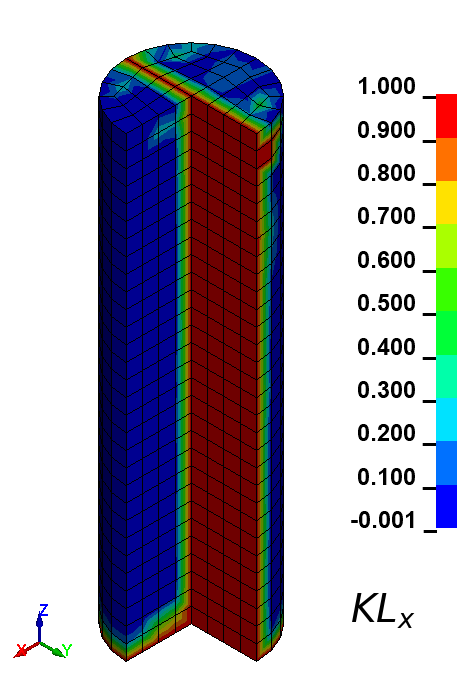}
        \caption{$x$-direction}
    \end{subfigure}
    \hfill
    \begin{subfigure}[b]{0.32\textwidth}
        \centering
        \includegraphics[width=\textwidth]{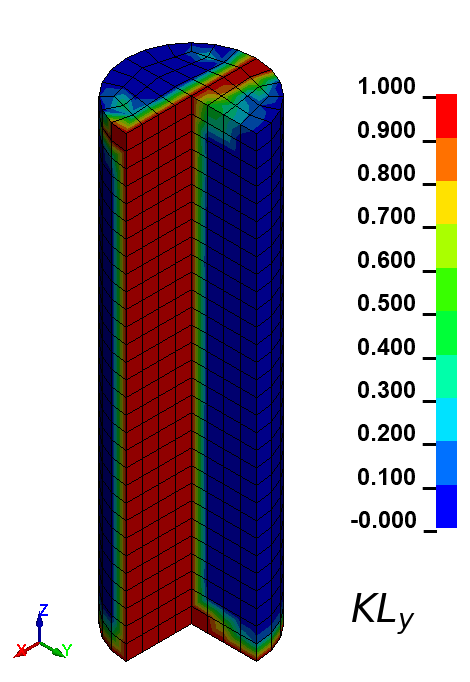}
        \caption{$y$-direction}
    \end{subfigure}
    \hfill
    \begin{subfigure}[b]{0.32\textwidth}
        \centering
        \includegraphics[width=\textwidth]{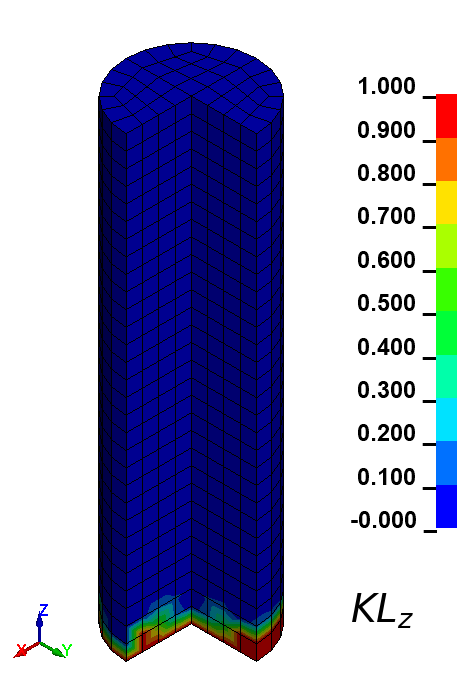}
        \caption{$z$-direction}
    \end{subfigure}

    \caption{Local KLD of the displacement field. Limited to a maximum value of one when the value is larger.}
    \label{fig:kld_test}
\end{figure}

As the previous discussion was qualitative, also the accuracy of the DDM NN surrogate model is checked. The accuracy is checked by comparing the FEM samples as true values to the predictions made by the local NNs.
The corresponding results are shown in Figs.~\ref{fig:rele_m_test},~\ref{fig:rele_s_test},~and~\ref{fig:kld_test}. 
The first two figures present the mean and standard deviation of the relative error, respectively, while the third figure shows the Kullback--Leibler divergence that compares the FEM sample to the local prediction distributions.
Each of the figures shows the error separately for each of the three components across the cylinder domain with a cut-out of one quarter to show the inside of the cylinder.
The relative error $e_{rel}$ of the mean and standard deviations across the cylinder for training and test data, as shown in Figs.~\ref{fig:rele_m_test}-\ref{fig:rele_s_test}, is computed by:
\begin{equation}
    e_{rel} = \frac{|\fwi(x,\omega) - \fvar_m(x,\omega))|}{|\fvar_m(x,\omega))| + 1},
    \label{Eq:rele+1}
\end{equation}
which can account for values at which the measurement $\fvar_m(x,\omega)$ is zero. 
The regions with the largest relative errors are primarily located near the upper and lower boundaries of the local subdomains. 
In particular, the highest errors occur at the top and bottom ends of the cylinder, with slightly elevated errors also appearing in the transition region between the two local domains at the interface. 
Interestingly, the probabilistic characteristics of the displacement field appear to be captured even more accurately than the displacement field itself. That means the mean relative error is slightly higher than the standard deviation relative error of the displacement.
This can be observed from the standard deviation predictions, which exhibit lower relative errors than the corresponding mean predictions. The mean predictions are slightly worse since these mostly represent the displacement field, that is a single FEM simulation, which is the more involved part of the approximation function when compared to spatially constant material parameters representing a lognormal distribution.
For both quantities, the most noticeable errors are located near the outer edge of the upper subdomain, with the largest error occurring in the mean displacement prediction in the $z$-direction. 
Overall, the errors in the first two stochastic moments remain small throughout the computational domain. 
As expected, the relative error is locally increased at the interface between the subdomains, where only communication points between the local NNs are available rather than direct test data. 
Nevertheless, the magnitude of the error remains very low, despite the fact that the solution in this region is effectively interpolated between the local neural networks using the proposed domain decomposition algorithm.

The Kullback--Leibler divergence is computed according to
\begin{equation}
    \mathcal{D}_{KL}(P \| Q)
    =
    \sum_{\omega \in \varOmega}
    P(\omega)
    \log\left(
    \frac{P(\omega)}{Q(\omega)}
    \right),
    \label{Eq:KLD}
\end{equation}
where $P(\omega)$ and $Q(\omega)$ denote the predicted and reference probability distributions, respectively, as shown in Fig.~\ref{fig:kld_test}. 
To obtain these distributions, the displacement predictions and test samples are first represented using Gaussian kernel density estimators, from which samples are subsequently drawn. 
Equation~(\ref{Eq:KLD}) is evaluated independently at each spatial point, allowing the divergence to be visualized throughout the three-dimensional domain.
Compared to the relative error, the Kullback--Leibler divergence reveals an additional limitation of the model in regions where the true displacement is identically zero while the input parameters remain stochastic. 
In such cases, the exact solution corresponds to a discrete zero value. 
The surrogate model, however, still predicts a distribution with finite variance, leading to relatively large Kullback--Leibler divergence values despite the small displacement errors.
For visualization purposes, the values shown in Fig.~\ref{fig:kld_test} are capped at a maximum value of one. 
This scaling makes it possible to distinguish regions with sufficiently accurate distributional predictions from regions where the divergence becomes excessively large. 
Consequently, clear lines can be observed in each displacement component corresponding to locations where the cylinder experiences little or no deformation, such as at the constrained lower boundary. 
These regions exhibit elevated divergence values for the reasons discussed above.
Conversely, in regions where stochastic variability is physically present and the Kullback--Leibler divergence is therefore meaningful, the divergence remains low, indicating that the predicted distributions closely match the reference distributions. 
This suggests that the proposed DDM NN accurately captures the stochastic behavior of the solution field. 
Furthermore, no noticeable increase in the divergence is observed at the interface between the subdomains, indicating that the proposed algorithm produces a smooth and statistically consistent transition across the interface.
In this manner the KLD shows indirectly the values which can be used for the Kalman Filter, since it is high in near zero prediction points, where the prediction model fails. These are points in which the KLD is large in atleast one of the directional components. This is an issue at the boundaries or at locations in which components of the displacement vector are zero since small absolute prediction errors can lead to large relative deviations from the measurement, resulting in unfeasible posterior distributions. This would immediately throw the Kalman filter off and cause divergence.
From a practical standpoint, fixed-point measurements near zero add little information and are therefore less important.
Therefore, any points with magnitudes close to zero are ignored in the Kalman filter to ensure stable posterior distributions.

To evaluate the performance of the local NNs at a spatial point, the local prediction models ability to reproduce the distribution of displacements at an interface point at ($x$, $y$, $z$) = ($5$, $-8,66$, $35$) is examined. 
Specifically, the NN predictions are compared against corresponding displacements from the training data. 
Additionally, test samples are taken and compared to the NN predictions for the samples to assess the generalization capability of the surrogate model in the interface points. 
\begin{figure}[h!]
    \centering
    \begin{subfigure}[t]{0.45\textwidth}
        \centering
        \includegraphics[width=\textwidth]{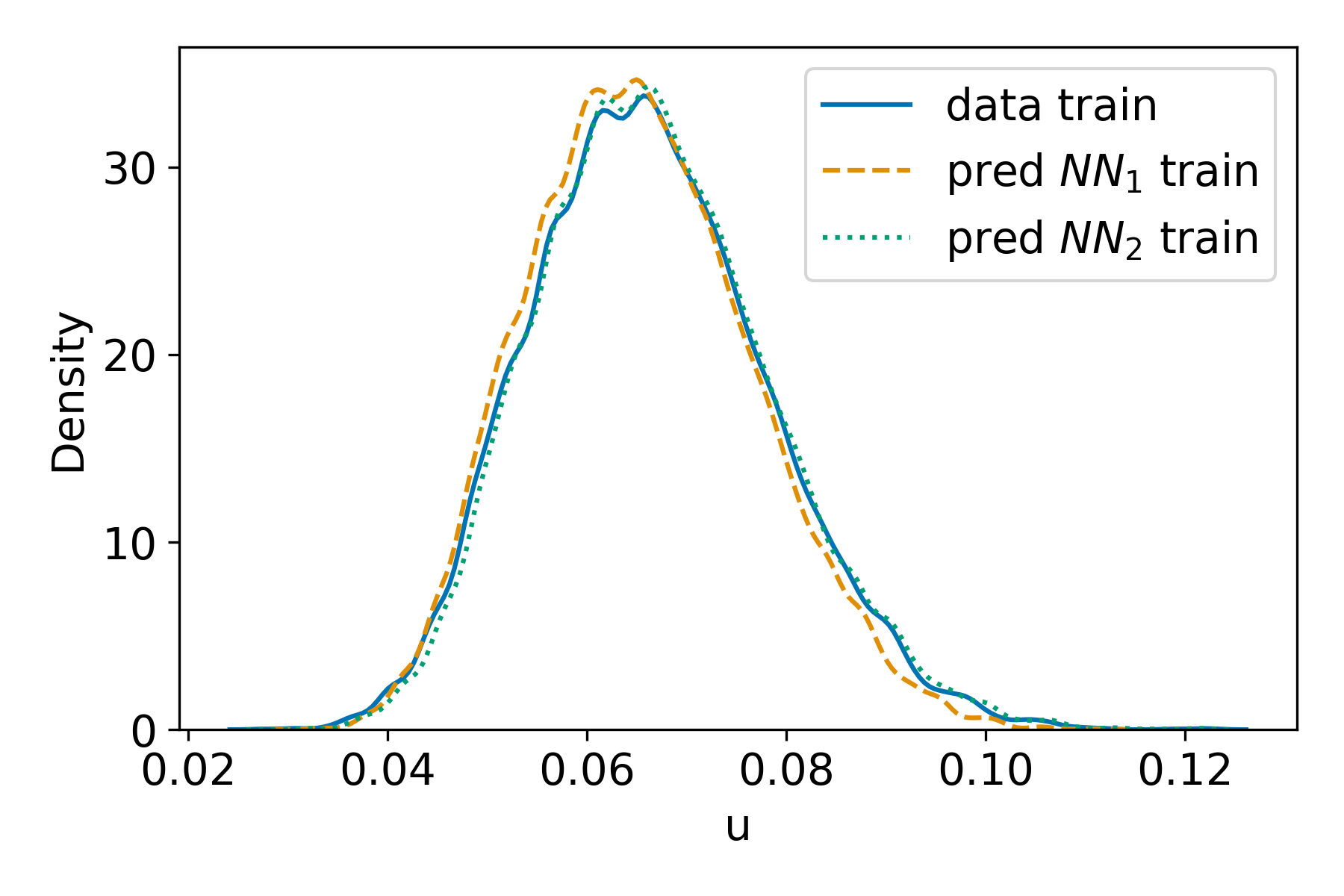}
        \caption{Distribution of $u_x$ at an interface point.}
        \label{Fig:dist_train_u}
    \end{subfigure}
    \hfill
    \begin{subfigure}[t]{0.45\textwidth}
        \centering
        \includegraphics[width=\textwidth]{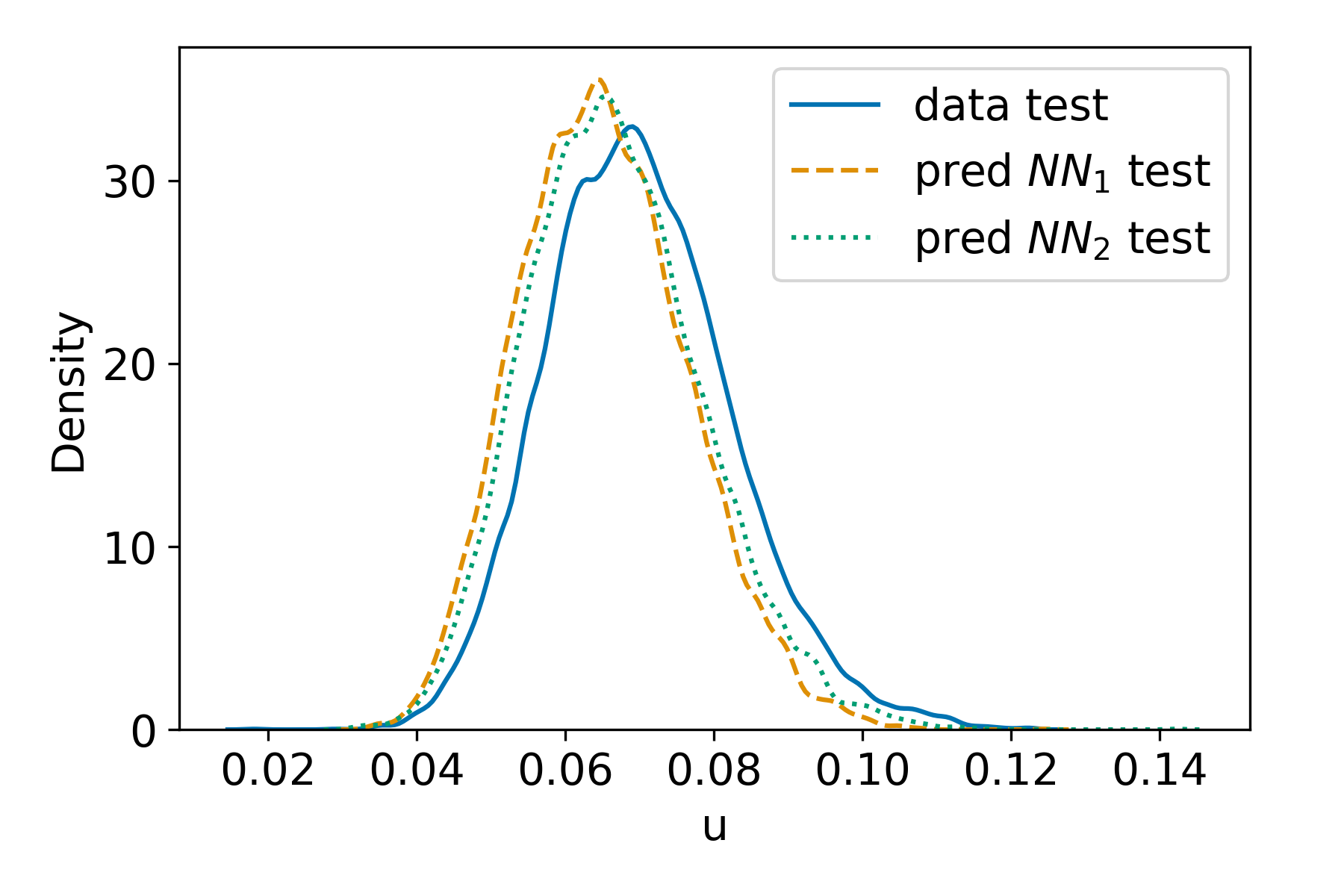}
        \caption{Distribution of $u_x$ at an interface point.}
        \label{Fig:dist_test_u}
    \end{subfigure}

    \begin{subfigure}[t]{0.45\textwidth}
        \centering
        \includegraphics[width=\textwidth]{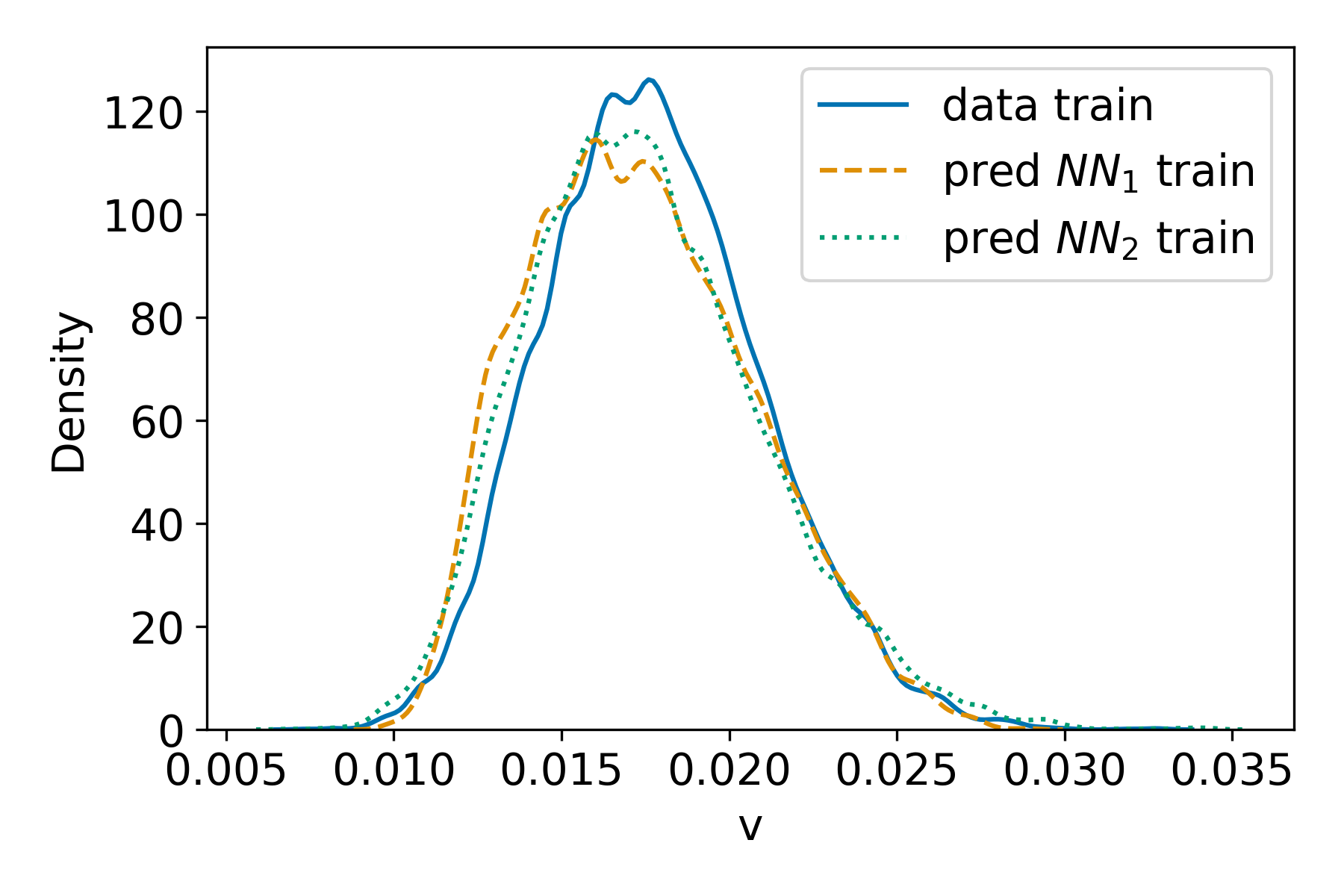}
        \caption{Distribution of $u_y$ at an interface point.}
        \label{Fig:dist_train_v}
    \end{subfigure}
    \hfill
    \begin{subfigure}[t]{0.45\textwidth}
        \centering
        \includegraphics[width=\textwidth]{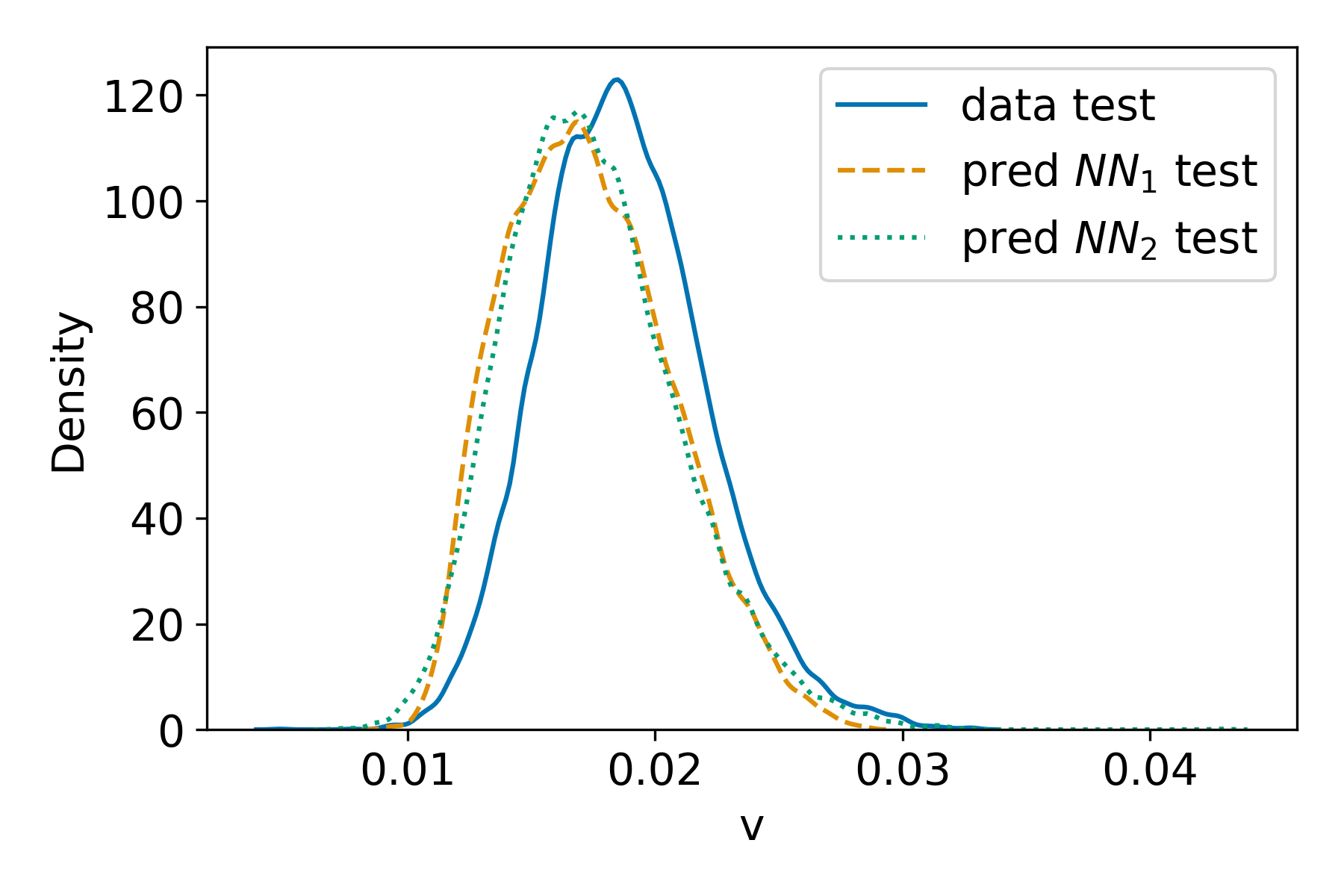}
        \caption{Distribution of $u_y$ at an interface point.}
        \label{Fig:dist_test_v}
    \end{subfigure}

    \begin{subfigure}[t]{0.45\textwidth}
        \centering
        \includegraphics[width=\textwidth]{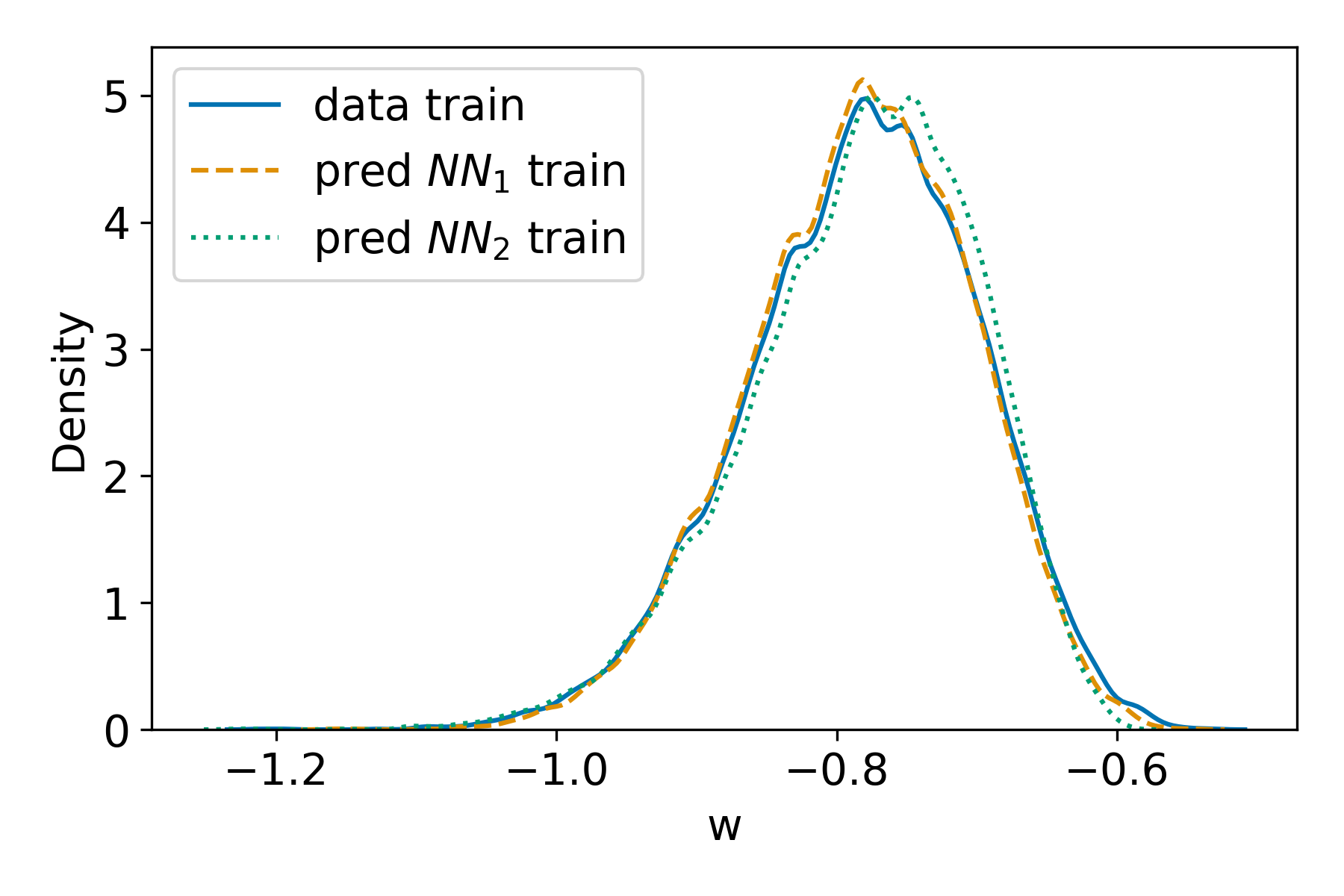}
        \caption{Distribution of $u_z$ at an interface point.}
        \label{Fig:dist_train_w}
    \end{subfigure}
    \hfill
    \begin{subfigure}[t]{0.45\textwidth}
        \centering
        \includegraphics[width=\textwidth]{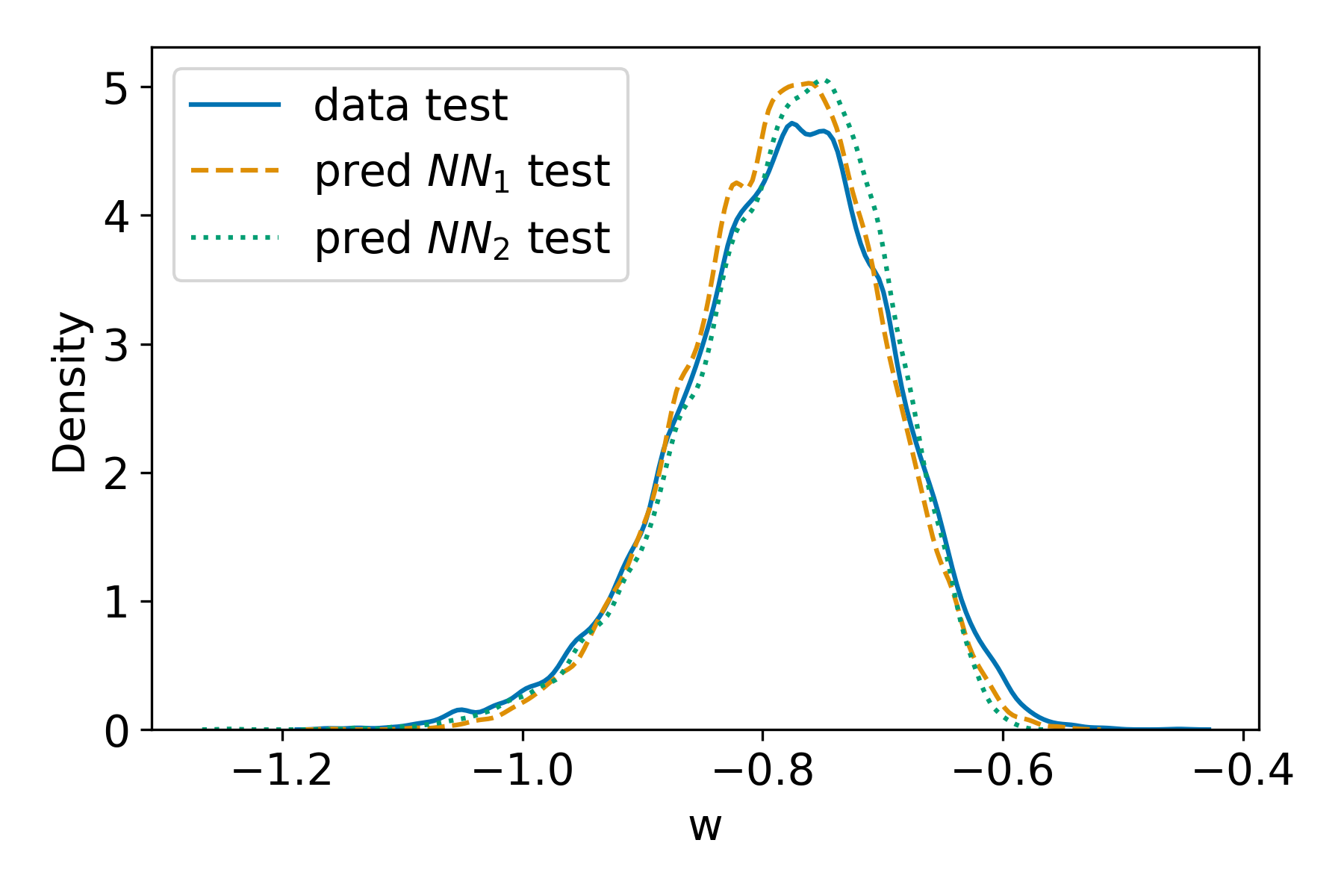}
        \caption{Distribution of $u_z$ at an interface point.}
        \label{Fig:dist_test_w}
    \end{subfigure}
    \caption{Kernel density figures of the displacement distributions on a single interface point between the domains on the cylinder. Each figure includes a blue line for the data, and yellow and green lines for $NN_1$ and $NN_2$ predictions, respectively. From top to bottom each displacement component is shown, and the left side shows predictions with training data, while the right side shows predictions using new test data. The coordinates of the point are ($x$, $y$, $z$) = ($5$, $-8,66$, $35$).}
    \label{Fig:dist_uvw}
\end{figure}
The results corresponding distributions of the displacement components are shown in Fig.~\ref{Fig:dist_uvw}.
On the left side one can see the training sample outputs of data against the predictions, while on the right side the same is shown for the testing sample outputs.
Vertically, the figures are stacked to present each displacement component separately. 
Each of them shows the kernel density estimation of the data, and the predictions for $NN_1$ and $NN_2$, in blue, yellow and green, respectively.
The curves for the 10.000 test samples and training samples indicate that the surrogate model preserves the shape of the original normal distributions accurately, with only minor deviations observed between the training and test sets.
The predictions show strong agreement with data across all three components, with the surrogate model capturing not only the mean behavior but also the tails of the distribution. 
While small discrepancies are visible in the test set, the overall shape remains intact. 
When examining the NN prediction distributions and comparing them with the training data, it can be observed that the predicted heads of the distributions are generally slightly shifted with respect to the reference distributions. 
For the $u_x$ and $u_z$ displacement components, the predictions of both NNs almost perfectly coincide. 
For the remaining displacement component in $u_y$ direction, however, the prediction is slightly off the training data, while the two NN predictions are very similar.
One of the the reasons for this shift might be that the magnitude of the component is closer to zero than the other predictions, which might make the loss function give higher importance to the other directional components.
Also, this shift can be attributed to the fact that both NNs effectively extrapolate at the interface, where no training data are assumed to be available and the information exchange is solely governed by the interface conditions. 
Consequently, small discrepancies between the local predictions may occur in this region.
When comparing the predictions with the test data shown on the right-hand side, somewhat larger differences become visible. 
In particular, the predicted distributions are slightly shifted to the left with respect to the test data and exhibit different height of the peaks than the predicted distributions. 
Nevertheless, the predictions of both NNs remain highly similar to one another with the highest agreement for the highest magnitude of the displacement component in z direction. 
This strong agreement between neighboring local predictions indicates a high degree of continuity across the interface and demonstrates that the proposed algorithm successfully enforces consistency between the subdomains.
Overall, the high similarity between the predicted and reference distributions suggests that the NNs generalize well to previously unseen samples at the interface.

\subsection{Parameter estimation}

After validating the surrogate model and confirming its ability to accurately reproduce the observed displacement data, we employ DDM NN predicted displacements as the forecast model within the Kalman filter framework, see Eq.~(\ref{Eq:enKF}).
Initially, the bulk and shear moduli are represented by broad lognormal prior distributions to account for the uncertainty arising from the lack of direct measurement information while ensuring positive definiteness of the material parameters. These prior distributions, introduced previously and used for training the surrogate model, span a wide range of possible material behaviors. The Kalman update then adjusts the parameter estimates by comparing the predicted DDM NN displacement fields with the corresponding measurements.

Furthermore, the model error between the training data and the surrogate model predictions, see Eq.~(\ref{Eq:mod_err}), is quantified to improve the reliability of the parameter estimation. The resulting error statistics are incorporated into the Kalman filter as an additional model error contribution, as described previously. Specifically, the sampling covariance of the discrepancy between the reference data and the surrogate predictions is computed and used to represent the model error covariance. Furthermore, the full Markov Chain Monte Carlo (MCMC) simulation is run in order to compute the reference solution. 
It utilizes a Metropolis Hastings (MH) algorithm in which a lognormal prior distribution is used for the material parameters. The likelihood is assumed to be Gaussian. The MCMC posterior is computed on the FEM model data directly and thus does not involve the surrogate model. After a burn-in period of 20,000 samples, the later 80,000 MCMC samples are used to generate the posterior distribution using kernel density estimators. Next to this, as another reference solution we use a pure ensemble KF procedure without surrogate model, that is also based on sampling FEM. Both KF representations use 10,000 samples with 600 measurement points on the surface of the cylinder wall.

\begin{figure}[t!]
    \centering
    \begin{subfigure}[t]{0.49\textwidth}
        \centering
        \includegraphics[width=\textwidth]{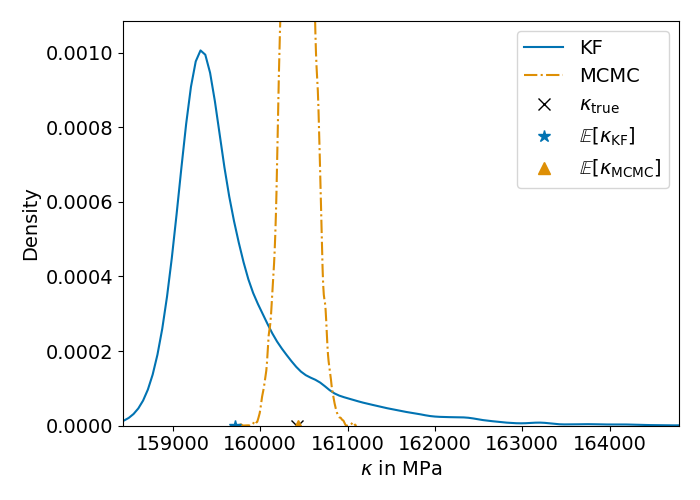}
        \caption{}
        \label{Fig:KF_bulk_MCMC}
    \end{subfigure}
    \hfill
    \begin{subfigure}[t]{0.49\textwidth}
        \centering
        \includegraphics[width=\textwidth]{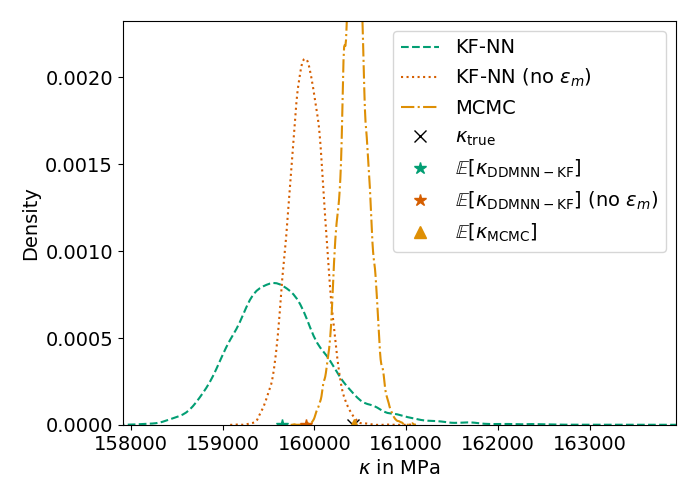}
        \caption{}
        \label{Fig:KF_bulk_MCMC_NN}
    \end{subfigure}

    \begin{subfigure}[t]{0.49\textwidth}
        \centering
        \includegraphics[width=\textwidth]{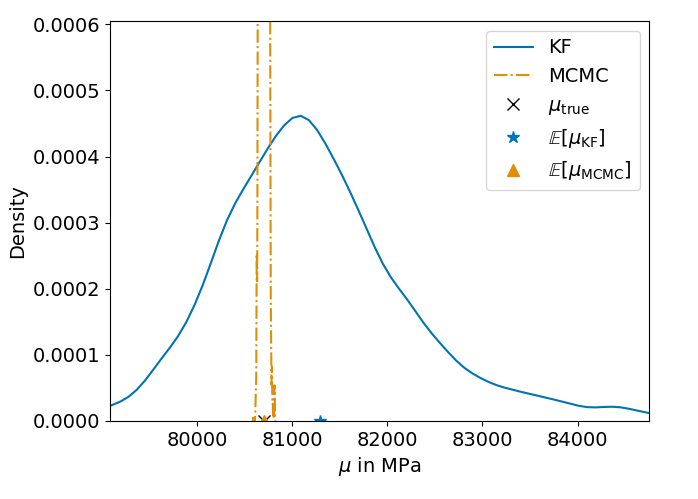}
        \caption{}
        \label{Fig:KF_shear_MCMC}
    \end{subfigure}
    \hfill
    \begin{subfigure}[t]{0.49\textwidth}
        \centering
        \includegraphics[width=\textwidth]{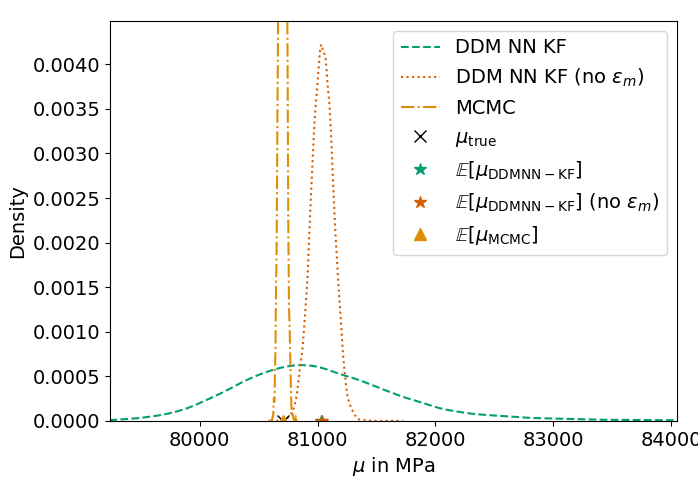}
        \caption{}
        \label{Fig:KF_shear_MCMC_NN}
    \end{subfigure}
    
    \caption{Comparisons of posterior distributions generated with different methods for the bulk modulus $\kappa$ in the top two figures and for the shear modulus $\mu$ in the bottom two figures. In the left figures, the MCMC method is compared to the sample-based KF, while on the right side the MCMC method is compared to the NN-based KF methods with and without an error. The means are denoted at the bottom and the material parameter distributions are shown as kernel density estimators.}
    \label{Fig:KF_vs_MCMC}
\end{figure}

\begin{figure}[t!]
    \centering
    \begin{subfigure}[t]{0.49\textwidth}
        \centering
        \includegraphics[width=\textwidth]{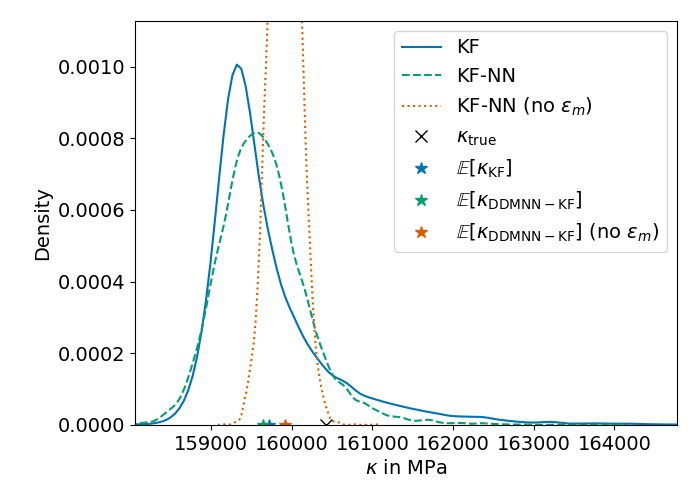}
        \caption{}
        \label{Fig:KF_bulk_full}
    \end{subfigure}
    \hfill
    \begin{subfigure}[t]{0.49\textwidth}
        \centering
        \includegraphics[width=\textwidth]{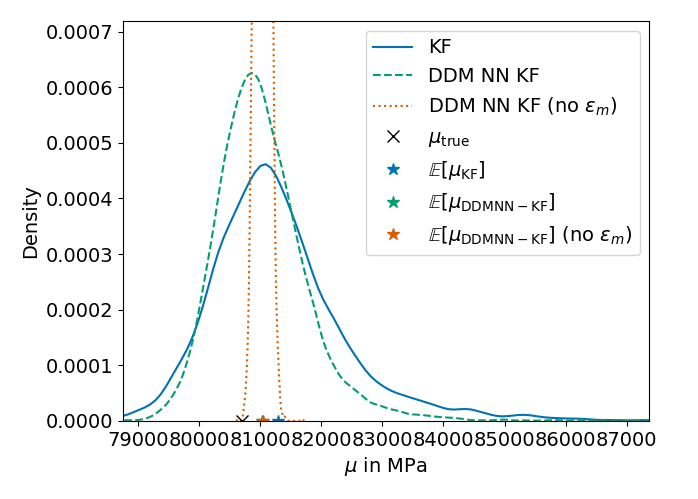}
        \caption{}
        \label{Fig:KF_shear_full}
    \end{subfigure}
    
    \caption{Kalman filter posterior distributions in comparison for $10,000$ samples, a measurement error of $1\%$ and 600 measurement points taken into account along the cylinder wall. The means are denoted at the bottom and the material parameter distributions are shown as kernel density estimators. The left figure shows the bulk modulus and the right figure the shear modulus.}
    \label{Fig:KF_full}
\end{figure}

To validate the proposed approach, the posterior distributions obtained with pure and the DDM NN ensemble based Kalman filter as proposed in this paper are compared with those of the MCMC method. For visualization, all distributions are represented by Gaussian kernel density estimators, with the corresponding posterior means indicated by markers at the base of each distribution and the true  material parameter shown as a black cross.
In Fig.~\ref{Fig:KF_vs_MCMC}a) and Fig.~\ref{Fig:KF_vs_MCMC}c) the posteriorors obtained by the ensemble Kalman filter (blue) without surrogate is compared to the MCMC posterior (yellow) for both the bulk and shear moduli. The MCMC posterior is noticeably narrower, and its mean almost exactly coincides with the true parameter. Although the Kalman filter substantially reduces the uncertainty relative to the prior distribution, its posterior exhibits a larger standard deviation than the MCMC solution together with a slight offset of the mean when compared to the true value. This behavior is observed for both material parameters. The bias in the estimation can be due to slight nonlinearity of the inverse problem. This can also explain the skewness of the bulk modulus posterior.
In Fig.~\ref{Fig:KF_vs_MCMC} b) and c) the MCMC posterior is compared to the DDM NN-based Kalman filter, both with (blue) and without (orange) estimated model error in the update equation. It can be seen that neglecting the model error results in a considerably narrower posterior distribution, whereas incorporating the model error increases the posterior variance by accounting for the uncertainty introduced by the surrogate model. The posterior means of both Kalman filter variants remain similarly close to the true parameter.
To facilitate a direct comparison between the ensemble Kalman filter formulations, Fig.~\ref{Fig:KF_full} compares the ensemble KF without surrogate and those with surrogates. The DDM NN-based Kalman filter with model error correction closely reproduces the sampling-based FEM Kalman filter, yielding a posterior distribution with a similar shape and spread. In contrast, the DDM NN-based Kalman filter without model error correction produces a substantially narrower posterior, indicating a lower estimated uncertainty, while its posterior mean lies slightly closer to the true parameter. However, the true value is located in lower probability region of posterior. 

\begin{table}[b!]
    \centering
    \caption{Parameter configurations and estimation results for the parameter $\kappa$. The bold values indicate the mean estimates that are closest to the target measurement of 160,419~MPa.}
    \label{Tab:kappa_estimation_results}
    
    \footnotesize                 
    \setlength{\tabcolsep}{4.5pt} 
    
    \begin{tabular}{lccccccc}
    \toprule
        \multicolumn{1}{c}{Parameter} &
        \multicolumn{1}{c}{(1)} &
        \multicolumn{1}{c}{(2)} &
        \multicolumn{1}{c}{(3)} &
        \multicolumn{1}{c}{(4)} &
        \multicolumn{1}{c}{(5)} &
        \multicolumn{1}{c}{(6)} &
        \multicolumn{1}{c}{(7)} \\
        \midrule
        Measurement error                               & 1\%   & 1\%   & 1\%   & 1\%   & 1\%   & 5\%   & 10\%  \\
        $N_{\text{samples}}$                          & 10,000 & 10,000 & 10,000 & 1,000  & 100   & 10,000 & 10,000 \\
        $N_{\text{measurements}}$                     & 600   & 300   & 100   & 600   & 600   & 600   & 600   \\
        
        \midrule
        
        $\mathbb{E}[\kappa_{KF}]$                     & 159,715 & 159,786 & \textbf{159,820} & 159,560 & 159,492 & \textbf{159,432} & \textbf{159,430} \\
        $\mathbb{E}[\kappa_{DDMNN-KF}]$ (no $\varepsilon_m$) & \textbf{159,911} & 161,489 & 164,239 & \textbf{159,786} & 159,661 & 158,824 & 158,025 \\
        $\mathbb{E}[\kappa_{DDMNN-KF}]$                   & 159,645 & \textbf{159,987} & 159,568 & 159,515 & \textbf{159,662} & 159,139 & 158,854 \\
        $\mathbb{E}[\kappa_{MCMC}]$                   & 160,435 & -      &  -     & -      &   -    &   -    &   -    \\
        
        \midrule
        
        $\sigma(\kappa_{KF})$                         & 810   & 839   & 904   & 808   & 693   & 1,178  & 1,905  \\
        $\sigma(\kappa_{DDMNN-KF})$ (no $\varepsilon_m$) & 184   & 259   & 441   & 186   & 208   & 882   & 1,716  \\
        $\sigma(\kappa_{DDMNN-KF})$                       & 549   & 584   & 698   & 567   & 468   & 1,062  & 1,874  \\
        $\sigma(\kappa_{MCMC})$                       & 149   &  -     &  -     &   -    &  -     &   -    &  -     \\
    \bottomrule
    \end{tabular}
\end{table}

\begin{figure}[t!]
    \centering
    \begin{subfigure}[t]{0.49\textwidth}
        \centering
        \includegraphics[width=\textwidth]{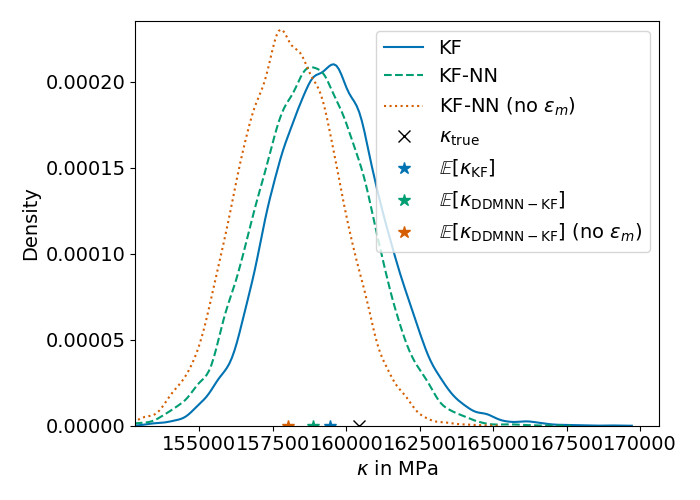}
        \caption{}
        \label{Fig:KF_bulk_herr}
    \end{subfigure}
    \hfill
    \begin{subfigure}[t]{0.49\textwidth}
        \centering
        \includegraphics[width=\textwidth]{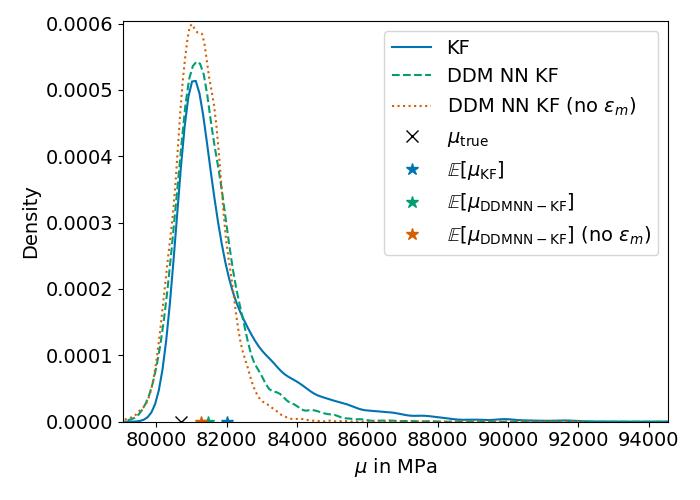}
        \caption{}
        \label{Fig:KF_shear_herr}
    \end{subfigure}
    
    \caption{Kalman filter posterior distributions in comparison for $10,000$ samples, a measurement error of $10\%$ and 600 measurement points taken into account along the cylinder wall. The means are denoted at the bottom and the material parameter distributions are shown as kernel density estimators. The left figure shows the bulk modulus and the right figure the shear modulus.}
    \label{Fig:KF_herr}
\end{figure}

\begin{figure}[b!]
    \centering
    \begin{subfigure}[t]{0.49\textwidth}
        \centering
        \includegraphics[width=\textwidth]{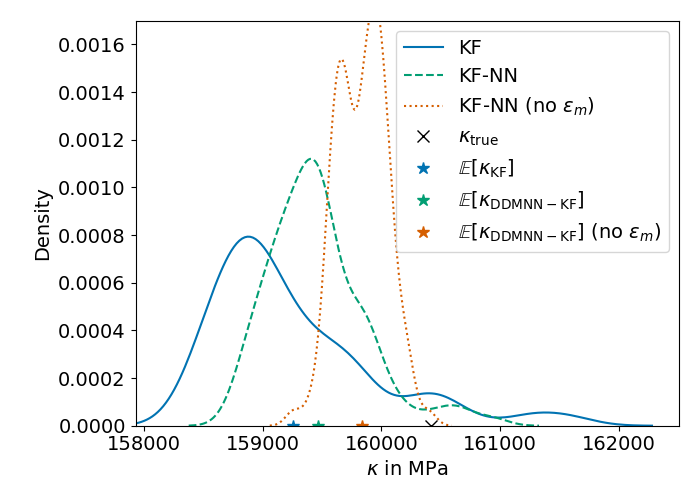}
        \caption{}
        \label{Fig:KF_bulk_lsamp}
    \end{subfigure}
    \hfill
    \begin{subfigure}[t]{0.49\textwidth}
        \centering
        \includegraphics[width=\textwidth]{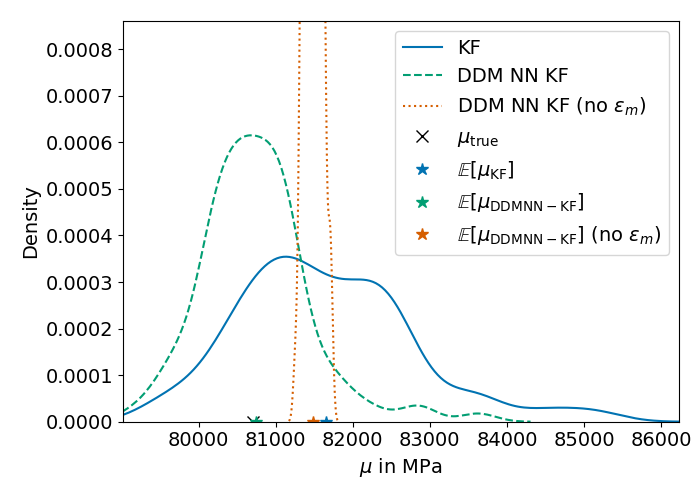}
        \caption{}
        \label{Fig:KF_shear_lsamp}
    \end{subfigure}
    
    \caption{Kalman filter posterior distributions in comparison for $100$ samples, a measurement error of $1\%$ and 600 measurement points taken into account along the cylinder wall. The means are denoted at the bottom and the material parameter distributions are shown as kernel density estimators. The left figure shows the bulk modulus and the right figure the shear modulus.}
    \label{Fig:KF_lsamp}
\end{figure}

\begin{figure}[h!]
    \centering
    \begin{subfigure}[t]{0.49\textwidth}
        \centering
        \includegraphics[width=\textwidth]{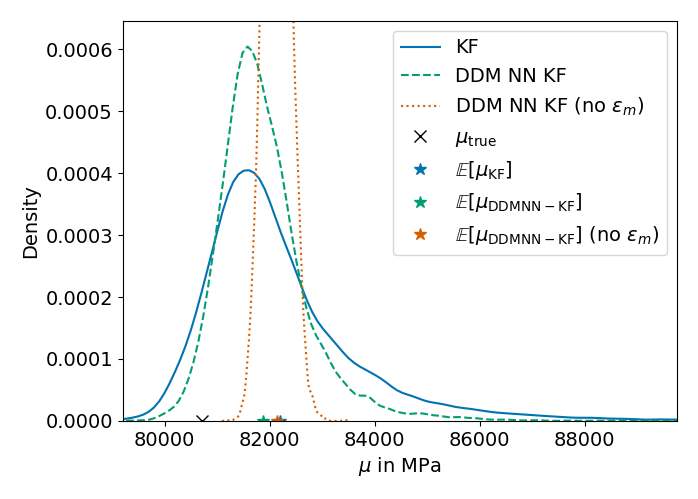}
        \caption{}
        \label{Fig:KF_bulk_lmeas}
    \end{subfigure}
    \hfill
    \begin{subfigure}[t]{0.49\textwidth}
        \centering
        \includegraphics[width=\textwidth]{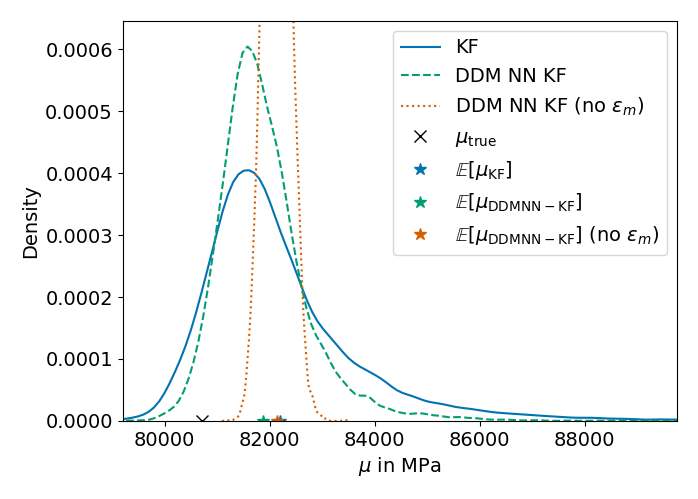}
        \caption{}
        \label{Fig:KF_shear_lmeas}
    \end{subfigure}
    
    \caption{Kalman filter posterior distributions in comparison for $10,000$ samples, a measurement error of $1\%$ and 100 measurement points taken into account along the cylinder wall. The means are denoted at the bottom and the material parameter distributions are shown as kernel density estimators. The left figure shows the bulk modulus and the right figure the shear modulus.}
    \label{Fig:KF_lmeas}
\end{figure}

Since several factors influence the posterior distribution obtained with the ensemble Kalman filter, additional studies were conducted to investigate the effects of the model error, the ensemble size, and the number of measurement points. The corresponding posterior means and standard deviations are summarized in Tabs.~\ref{Tab:kappa_estimation_results} and \ref{Tab:mu_estimation_results} for the bulk and shear moduli, respectively. Furthermore, Figs.~\ref{Fig:KF_herr}--\ref{Fig:KF_lmeas} compare the posterior distributions obtained for increased measurement error, a reduced number of samples, and a reduced number of measurement points. Throughout these figures, the same color scheme is adopted as before: the ensemble KF with is shown in blue, the DDM NN-based Kalman filter with model error correction in green, and the DDM NN-based Kalman filter without model error correction in orange. The tables additionally list the input parameters for each experiment, followed by the posterior means and standard deviations of the different methods. The KF posterior mean closest to the measured material parameter is highlighted in bold. Since the computational cost of MCMC is significantly higher, it is included only for the reference case with 100,000 samples, as discussed previously.
The posterior means demonstrate that no single method consistently provides the closest estimate for all configurations. However, the DDM NN-based Kalman filter without model error correction exhibits the largest variation across the different test cases. This is particularly evident for the case with fewer measurement points, where the estimated mean deviates substantially from the true parameter. In contrast, the DDM NN-based Kalman filter with model error correction produces consistently accurate estimates across all investigated scenarios, while also maintaining good performance when the number of posterior samples is reduced. Overall, the posterior means obtained by all Kalman filter formulations remain relatively similar, whereas the pure ensemble Kalman filter and the DDM NN-based Kalman filter with model error correction consistently predict larger posterior standard deviations than the formulation without model error correction. 
Combined with the larger variation of its posterior means, this indicates that neglecting the model error leads to an underestimation of the posterior uncertainty and therefore to overconfident parameter estimates.
\begin{table}[t!]
    \centering
    \caption{Parameter configurations and estimation results for the shear modulus $\mu$. The bold values indicate the KF mean estimates that are closest to the target measurement of \(80,707\)~MPa.}
    \label{Tab:mu_estimation_results}
    
    \footnotesize                 
    \setlength{\tabcolsep}{4.5pt} 
    
    \begin{tabular}{lccccccc}
    \toprule
        \multicolumn{1}{c}{Parameter} &
        \multicolumn{1}{c}{(1)} &
        \multicolumn{1}{c}{(2)} &
        \multicolumn{1}{c}{(3)} &
        \multicolumn{1}{c}{(4)} &
        \multicolumn{1}{c}{(5)} &
        \multicolumn{1}{c}{(6)} &
        \multicolumn{1}{c}{(7)} \\
        \midrule

        Measurement error                               & 1\%   & 1\%   & 1\%   & 1\%   & 1\%   & 5\%   & 10\%  \\
        $N_{\text{samples}}$ KF                       & 10,000 & 10,000 & 10,000 & 1,000  & 100   & 10,000 & 10,000 \\
        $N_{\text{measurements}}$                    & 600   & 300   & 100   & 600   & 600   & 600   & 600   \\
        
        \midrule
        
        $\mathbb{E}[\mu_{KF}]$                                      & 81,297 & 82,686 & \textbf{80,784} & 81,379 & 81,405 & 81,913 & 82,007 \\
        $\mathbb{E}[\mu_{DDMNN-KF}]$ (no $\varepsilon_m$) & 81,041 & \textbf{80,587} & 77,553 & 81,305 & 81,485 & \textbf{81,340} & \textbf{81,257} \\
        $\mathbb{E}[\mu_{DDMNN-KF}]$                             & \textbf{81,025} & 81,946 & 81,568 & \textbf{80,967} & \textbf{80,972} & 81,384 & 81,478 \\
        $\mathbb{E}[\mu_{MCMC}]$                                   & 80,704 &  -     &  -     & -      & -      & -      & -      \\
        
        \midrule
        
        $\sigma(\mu_{KF})$                        & 1,146  & 1,248  & 1,512  & 1,162  & 1,073  & 1,609  & 1,674  \\
        $\sigma(\mu_{DDMNN-KF})$ (no $\varepsilon_m$)  & 94    & 129   & 226   & 91    & 91    & 419   & 710   \\
        $\sigma(\mu_{DDMNN-KF})$                 & 754   & 800   & 877   & 683   & 700   & 923   & 980   \\
        $\sigma(\mu_{MCMC})$                   & 25    &    -   &   -    &    -   &   -    &    -   &     -  \\
    \bottomrule
    \end{tabular}
\end{table}
The influence of the individual parameters is further illustrated in Figs.~\ref{Fig:KF_herr}--\ref{Fig:KF_lmeas}. Increasing the measurement error, shown in Fig.~\ref{Fig:KF_herr}, causes all posterior distributions to become broader and more similar, since the measurement uncertainty dominates the comparatively small surrogate model error. 
Consequently, the posterior means remain close to one another while exhibiting greater uncertainty. Figure~\ref{Fig:KF_lsamp} presents the results obtained with only 100 posterior samples. As expected, the distributions become less smooth due to the smaller sample size, yet all methods still predict comparable posterior means. Again, only the DDM NN-based Kalman filter without model error correction exhibits a noticeably smaller posterior variance, whereas the other two methods remain in close agreement. A more pronounced difference is observed in Fig.~\ref{Fig:KF_lmeas}, where the number of measurement points is reduced. In this case, the DDM NN-based Kalman filter without model error correction produces an inaccurate posterior distribution, while both the sampling-based Kalman filter and the DDM NN-based Kalman filter with model error correction continue to estimate the material parameters accurately. This behavior suggests that, when only a limited number of measurements are available, the surrogate model error becomes increasingly important and must therefore be explicitly accounted for during the Kalman filter update.
Overall, these results demonstrate that incorporating the surrogate model error enables the DDM NN-based Kalman filter to achieve performance comparable to that of pure ensemble Kalman filter while providing a more reliable estimate of the posterior uncertainty.

%% file: Section/Conclusion.tex
\section{Conclusion}
\label{Sec:final}
In this paper, we proposed a decentralized parameter identification framework that combines a domain decomposition neural network with the ensemble Kalman filter. The DDM NN surrogate is constructed using an augmented Lagrangian domain decomposition algorithm, in which local neural networks are trained independently while interface continuity is enforced through iterative updates of the Lagrange multipliers. The resulting surrogate provides a continuous approximation of the displacement field over the entire computational domain.

The proposed DDM NN is integrated into the EnKF as the forecast model, replacing repeated finite element evaluations during the prediction step. Furthermore, the domain decomposition strategy is extended to the analysis step by decomposing the Kalman gain into local contributions. This avoids the repeated inversion of large global covariance matrices and enabling decentralized parameter updates. Consequently, the proposed DDM NN EnKF framework reduces the computational complexity of both the forecast and analysis stages while preserving the accuracy of the parameter estimation.

Since the EnKF assumes an unbiased forecast model, the surrogate modeling error must be accounted for during the assimilation process. In the present work, this error is represented statistically through the sample covariance of the discrepancy between the finite element solutions and the DDM NN predictions and is incorporated into the forecast error covariance. Although this approach effectively mitigates estimation bias, the development of more rigorous model-error representations remains an important topic for future research.

The proposed three-dimensional DDM NN achieves low prediction errors for both the training and test datasets throughout the computational domain. The largest errors occur in regions where the displacement is zero or nearly zero, such as the clamped boundary. These regions are inherently difficult to approximate because the network must simultaneously satisfy exact boundary conditions while representing the variability induced by uncertain material parameters. This behavior is reflected by locally increased Kullback--Leibler divergence (KLD) values. A possible extension is the incorporation of boundary-condition-preserving ansatz functions, which would enforce the essential boundary conditions exactly.

Nevertheless, these local inaccuracies have only a minor influence on the parameter identification process, since regions with negligible displacement variation provide little information about the unknown material parameters. Consequently, excluding such measurements from the assimilation procedure does not noticeably affect the inferred posterior distributions. Using only informative measurement locations, the proposed DDM NN EnKF framework accurately recovers the reference material parameters while substantially reducing their uncertainty.

Finally, the posterior parameter distributions obtained using the proposed DDM NN EnKF framework are in excellent agreement with those obtained using direct finite element simulations within the EnKF. This demonstrates that the proposed surrogate-based decentralized formulation provides an accurate and computationally efficient alternative to conventional finite element-based ensemble data assimilation.

Future work will focus on extending the proposed methodology to sequential EnKF formulations and time-dependent DDM NN surrogates for parameter identification in nonlinear and transient finite element problems. In addition, adaptive model-error estimation and fully decentralized covariance localization strategies will be investigated to further improve the robustness and scalability of the proposed framework.

%% file: Section/Acknowledgements.tex
\section{Acknowledgements}
This research was carried out under project number T21001a in the framework of the Partnership Program of the Materials innovation institute M2i (www.m2i.nl).